\documentclass[acmtog, nonacm]{acmart}

\renewcommand\footnotetextcopyrightpermission[1]{}
\setcopyright{none}

\usepackage{graphicx}
\usepackage{caption}
\usepackage{subcaption}
\usepackage[ruled]{algorithm2e} 
\usepackage{algorithmicx}
\usepackage{booktabs}
\usepackage{amsmath}
\usepackage{comment}
\usepackage{graphbox}
\usepackage{tabularray}

\begin{document}

\title{Bolt: Clothing Virtual Characters at Scale}

\author{Jonathan Leaf}
\orcid{0009-0004-5822-5074}
\affiliation{%
  \institution{NVIDIA}
  \country{USA}
}

\author{David Sebastian Minor}
\affiliation{%
  \institution{NVIDIA}
  \country{Canada}
}

\author{Gilles Daviet}
\orcid{0000-0003-3154-7423}
\affiliation{%
  \institution{NVIDIA}
  \country{France}
}

\author{Nuttapong Chentanez}
\affiliation{%
  \institution{NVIDIA}
  \country{Thailand}
}

\author{Greg Klar}
\orcid{0000-0002-4569-5956}
\affiliation{%
  \institution{NVIDIA}
  \country{New Zealand}
}

\author{Ed Quigley}
\affiliation{%
  \institution{NVIDIA}
  \country{USA}
}

\renewcommand{\shortauthors}{Leaf et al.}

\begin{abstract}
  Clothing virtual characters is a time-consuming and often manual process.
  Outfits can be composed of multiple garments, and each garment must be fitted to the
  unique shape of a character. Since characters can vary widely in size and shape,
  fitting outfits to many characters is a combinatorially large problem.
  We present Bolt, a system designed to take outfits originally authored on a source
  body and fit them to new body shapes via a three stage transfer, drape, and rig process.
  First, our new garment transfer method transforms each garment's 3D mesh positions to
  the new character, then optimizes the garment's 2D sewing pattern while maintaining key features
  of the original seams and boundaries. Second, our system simulates the transferred
  garments to progressively drape and untangle each garment in the outfit. Finally, the
  garments are rigged to the new character. This entire process is automatic, making it
  feasible to clothe characters at scale with no human intervention. Clothed characters
  are then ready for immediate use in applications such as gaming, animation, synthetic
  generation, and more.
\end{abstract}



\begin{teaserfigure}
  \includegraphics[width=\textwidth]{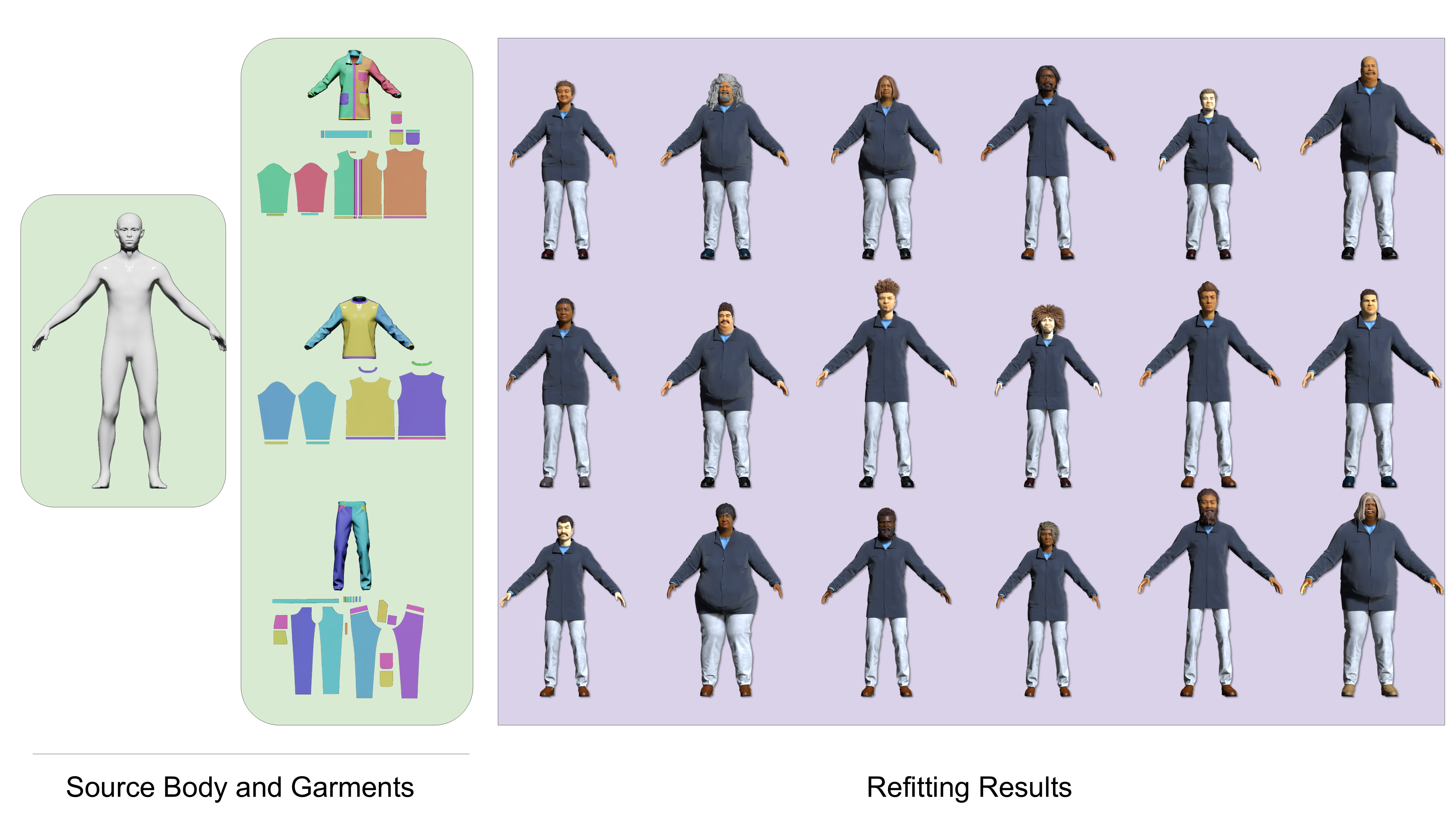}
  \caption{We present Bolt, an automated method for assembling, refitting, draping, and rigging
    outfits from a source character to many new characters.
    On the left is a source character and three garments, each with their corresponding sewing pattern, that fit to its body.
    Using our method we automatically transfer the 3D garments, adjust the 2D sewing pattern,
    untangle and drape the outfit on the new character body, and rig the garments for animation.
    Our method works for a wide range of target body shapes and sizes}
  \Description{Outfits fit onto many body types.}
  \label{fig:teaser}
\end{teaserfigure}

\maketitle

\section{Introduction}

Many industries and applications rely heavily on clothed virtual characters,
including video games, animated films, virtual and augmented reality
applications, social media, fashion, e-commerce, medicine, industrial
applications, and more.

Some applications require large numbers of diverse characters, such as crowds
in a video game. Deep learning applications also need lots of high quality
training data, but collecting comprehensive images of real world scenarios with
clothed humans can be prohibitively costly, difficult, or impossible to
collect. For example, computer vision algorithms for autonomous vehicles need
to observe a significant number of diverse humans in their dataset to reliably
detect the presence of a pedestrian. Some cases are challenging to
collect safely, such as children playing near a street or in a garage, but
having them represented in the data is crucial for reliably detecting children
near an autonomous vehicle. The difficulty of capturing cases like this
motivates the synthetic generation of clothed virtual characters who can take
the place of real humans in these training sets.

For designing clothed digital humans, sewing pattern-based digital content creation (DCC) tools have
emerged as the industry's standard for creating realistic and high-quality 3D
garments\footnote{https://marvelousdesigner.com, https://clo3d.com,
      https://optitex.com, and https://browzwear.com}. The sewing pattern creation
process, which is akin to traditional tailoring methods, involves the creation
of individual pattern pieces (2D panels) that are then stitched together to
form a complete garment. The process begins with the creation of a 2D sewing
pattern with stitching and fine details, which is then transformed onto a 3D
model and draped by simulating the cloth on the character's body mesh.

Sewing pattern garment design allows for a high degree of customization and
precision. However, generating large numbers of clothed characters with DCC tools is
a time-consuming challenge even for expert users. The more complex the outfit, the larger the
number of garments that must be prepared, assembled, simulated, and adjusted to
the character to attain a desirable look. Preparing each garment requires
either manually drafting, sewing, simulating and adjusting the garment's
pattern from scratch, or by refitting an existing garment that has been fit to
a previous character to this new character's body type. Finally, once a new
outfit has been fitted to the character, skinning weights must be painted on the
clothing either manually, or using external tools.

The large numbers of character body shapes and outfit combinations creates a
combinatorial explosion in the number of possible characters that can
be generated. Our tool is designed to specifically address this challenge by
making the refitting and draping of any outfit onto any new character
automatic. For applications that need lots of diverse characters, our approach
can save significant amounts of artist time.

Our contributions are as follows:

\begin{enumerate}
      \item A novel garment transfer algorithm that transfers the 3D cloth mesh positions
            from a source character to a target character and then optimizes the 2D sewing
            pattern to resize and adjust the pattern while preserving style and seams.
      \item An end-to-end pipeline that assembles, refits, drapes, and rigs a multi-garment
            outfit onto a new character automatically.
\end{enumerate}

\section{Related Work}

\subsubsection*{Sewing Pattern Creation}

Recent works have focused on developing improved solutions for creating
garments along with their corresponding 2D patterns. \citet{NeuralTailor2022}
learn patterns and garments from point clouds. \citet{GarmentCode2023} defines
a programmatic method for constructing garments, a modular approach that
provides high-level primitives. \citet{pietroni2022computational} establish how
to create pattern pieces from a 3D garment mesh. \citet{he2024dresscode}
demonstrate a text-to-garment system using a GPT-based architecture. Each of
these approaches can reduce garment creation time, though the issues of
refitting and draping will still remain, making the data produced from these
methods suitable inputs to our system.

\subsubsection*{Cloth Simulation}

To drape garments onto a character, we rely on cloth simulation. Cloth
simulation has been adding physically plausible dynamic motion to cloth models
since the seminal work of \citet{Terzopoulos1987}. \citet{baraff1998large}
introduce implicit integration which improves stability. Many works have
followed since then, such as new methods for collisions
\cite{bridson-02-contact, provot1997collision} and bending strain
\cite{gingold2004discrete} to name a few.

Many approaches like XPBD \cite{macklin2016xpbd}, Projective Dynamics
\cite{bouaziz2014projective}, and others \cite{zeller2005cloth, liu2017quasi,
    tang-siga18} demonstrate improved performance over prior methods. More recent
work reformulates implicit time integration as an energy function
\cite{Gast2015optimization, martin2011example, wang2016descent} to be
minimized. Additionally, \citet{kim-20-fem-cloth} re-represents classic anisotropic material models within a FEM context.

\subsubsection*{Garment Refitting}

Garment retargeting is an area that has been widely studied, primarily
through the lens of capture and virtual try-on \cite{lee2009retargeting}. Many
works rely on parametric body models \cite{pons2017clothcap}, while others
attempt to learn correspondences for the garment-body interface
\cite{naik2024dress}.

\citet{shi2021learning} learn garment transfer from data, specifically from simulation pairs of source and target bodies.
\citet{wang2018rule} precisely optimizes sewing patterns, but only over a narrow range of differing body shapes.
\citet{grigorev2024contourcraft} relies on the SMPL-X body shape system for handling
cloth refitting, by modifying garment nodal positions using the same shape vectors as for the body.

\citet{de2020garment} demonstrate an iterative relax and rebind method for transferring a 3D
garment mesh from one body to another.
\citet{brouet2012design} uses geometric constraints to transfer a garment and the corresponding pattern,
but this method does not include a physically-based drape of the fabric.
\citet{chen2024dress} uses differentiable simulation to jointly
optimize a garment's 2D panels using control cages and 3D positions using differentiable simulation.
This combines the objectives of transfer and draping into a single step,
and is demonstrated on single-garment outfits.

Our approach refits the 3D positions of a garment, then optimizes the 2D
pattern positions based on the transfer. The result of our cloth refitting
process is a simulation ready mesh to be used for draping.

\subsubsection*{Untangling Cloth}

The ability to untangle interpenetrating garment layers is critical for
assembling outfits of multiple garments.

Historically, cloth untangling has focused primarily on untangling
self-intersections. \citet{baraff2003untangling}, \citet{wicke2006untangling},
\citet{ye2017unified} solve this problem by solving garment
self-interpenetration globally. \citet{volino2006resolving} opt for a more
local approach, which is faster but comes with no global guarantees.

Industry tools like Marvelous Designer rely on interactive layer-based
simulation to untangle garments. Users can specify layer-numbers, and the
simulation will order the layers in the collision handling process.

Recent research has focused on untangling multiple garments. Assuming each
garment is assumed to have a unique layer number, the objective is to have the
garments untangle themselves in the user specified ordering.
\citet{buffet2019implicit} use implicit field operators to untangle
multi-layered garments configurations. \citet{lee2023clothcombo} use neural
networks for untangle layers of clothing. \citet{grigorev2024contourcraft} use
graph neural networks to resolve collisions between layers.

Our approach takes a sequential approach to solve this problem, inspired by
practitioners in the VFX community. We simulate each garment one at a time,
progressively draping them on the body. Other recent works like
\cite{li2024isp} also rely on sequentially resolving collisions between layers.

\section{Overview}

With Bolt, we process an outfit composed of one or more garments in three
primary steps.
\begin{itemize}
  \item \textbf{Garment transfer}: We transfer a garment from a source character to a target character,
        adjusting both the 3D positions and the 2D sewing pattern of the garment, which we use
        as the rest state of the garment for simulation.
  \item \textbf{Progressive draping}: We simulate each garment, layer by layer, to untangle and drape each
        garment.
  \item \textbf{Rig transfer}: We transfer the skinning weights from the character's body rig to the corresponding garments.
\end{itemize}

\begin{figure*}
  \centering
  \includegraphics[width=\textwidth]{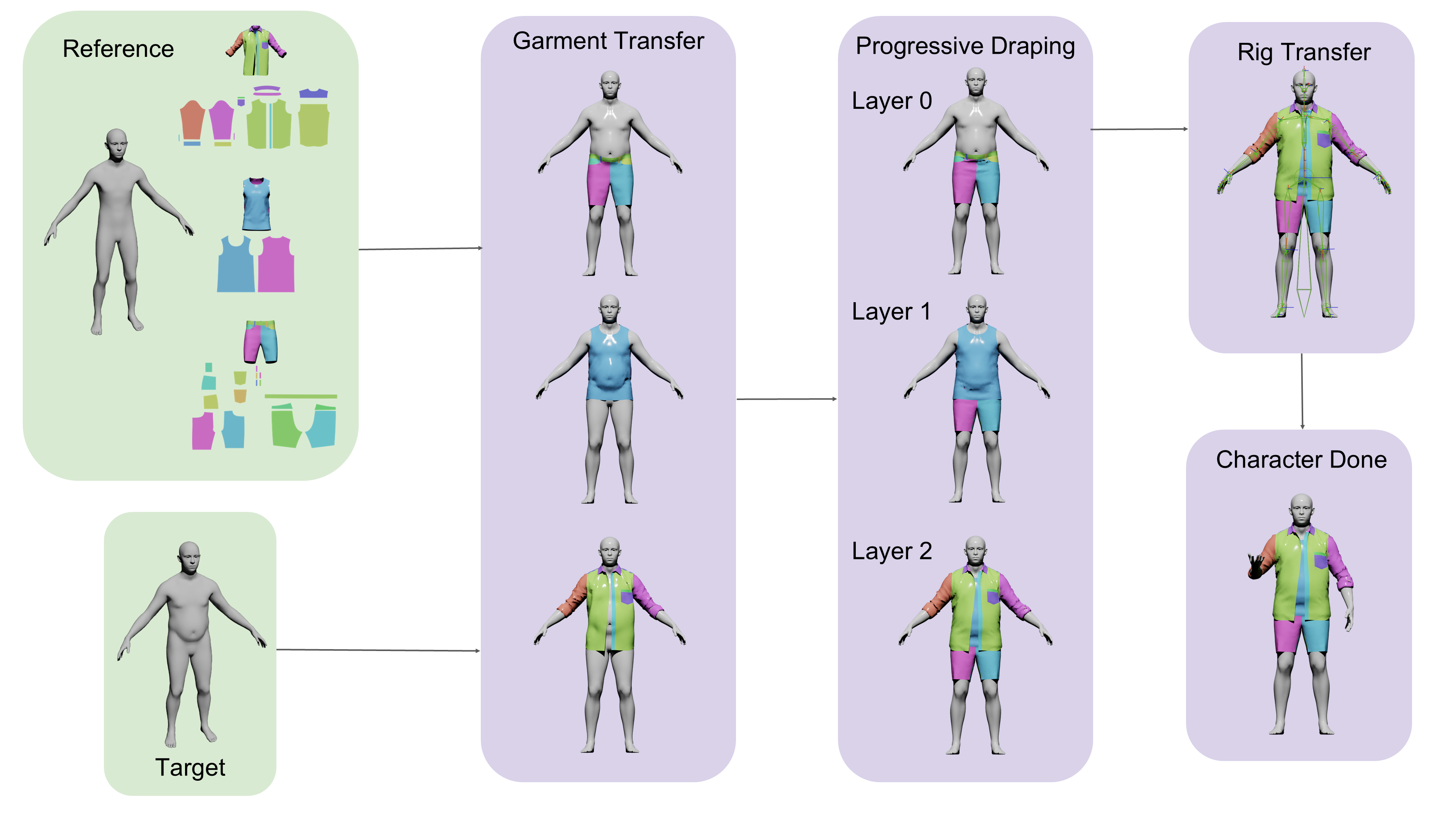}
  \caption{Overview of the Bolt pipeline using an example character.
    In this example, we assemble a 3-layer outfit from a source character's body and garments, and target character's body.
    In the garment transfer step, we fit each garment to the target character separately.
    In the progressive draping step, we simulate the outfit layers one at a time on the character to drape and untangle the clothing.
    Finally, we rig garments by transferring the skinning weights from the target body rig to the garments and finish the character.
  }
  \label{fig:bolt_overview}
\end{figure*}

See Figure~\ref{fig:bolt_overview} for a visual representation of our pipeline.

We develop our system as an offline automated pipeline, prioritizing
flexibility, robustness, and scalability over performance. We wrote the entire
pipeline using Nvidia's Python Warp \cite{warp2022}.

\section{Garment Transfer}

As the first step, we transfer garments from a source character to a target
character. We do this in two parts: 3D Mesh Transfer, then 2D Sewing Pattern
Optimization.

\subsection{3D Mesh Transfer}

We transfer 3D mesh positions from the source body to the target body. Our
algorithm extrapolates a 3D displacement field around the body from a source
to a target mesh. We take inspiration from fluid dynamics to create the
following physically inspired method.
\begin{itemize}
  \item Define a sparse regular grid in a narrow band around the body mesh.
  \item Define displacement boundary conditions on the mesh surface.
  \item Solve for a viscous, weakly compressible flow in the 3D space around the body:
        \begin{itemize}
          \item Viscosity allows diffusion of the displacement away from the surface.
          \item Unilateral incompressibility moves garments away from areas where pinching
                might occur, such as under the armpits.
        \end{itemize}
\end{itemize}

\subsubsection*{Formulation}

We start with the inertialess incompressible viscous fluid equations, inspired
by Stokes' equation. Let $u$ be the mesh displacement and $u_{body}$ be the
target body displacement from the source. We define the domain $\Omega$ as
the outside of the body mesh. Let $\delta\Omega_b$ be the body surface, i.e.
the domain boundary on which we enforce the displacement boundary conditions.
The problem we need to solve is
\begin{equation*}
  \begin{cases}
    -\nu D(u) + \nabla p = 0    & \,                         \\
    \hfill - \nabla \cdot u = 0 & \,                         \\
    \hfill u = u_{body}         & \text{ on } \delta\Omega_b
  \end{cases}
\end{equation*}
where $\nu$ is the viscosity, $p$ is pressure, and $D\left(u\right) =\frac{1}{2}\left(\nabla u +
  \left(\nabla u\right)^T\right)$, the symmetric part of the gradient of $u$.

We do not want or need pure incompressibility, as different body weight classes
will introduce compressing/stretching regions. We instead replace the
incompressibility constraint with an elastic term
\begin{equation}
  - \nabla \cdot u - \gamma p = 0
\end{equation}
where $\gamma$ is a compliance term set to 0.01.

The previous formulation uses a linear approximation of the volume change
($\nabla \cdot u = T r \left( \nabla u \right)$), which does not behave
satisfyingly for large displacements. Moreover, we do not want to penalize
stretching, just compression (to move out cloth from zones like shrinking arm
pits). We switch to a nonlinear elasticity formulation:
\begin{equation}
  - \min \left(0, \det \left(\nabla u + I\right) - 1\right) - \gamma p = 0
\end{equation}

The solve is now a fixed-point iterations loop, where the linearized
formulation is still used for each iteration but with an additional bias term
accounting for the nonlinear part.

\subsubsection*{Implementation}
To avoid having to generate a mesh, we use a non-conforming formulation. We
define a dense regular grid around the body, and mark all cells within a given
distance of the body as active. The solve will be done on this sparse subset of
the regular grid.

This unfortunately means that the grid nodes do not match the body vertices, so
we cannot enforce the boundary condition directly. We instead switch to a weak
formulation,
\begin{equation}
  \int_{\delta \Omega_g}  \,u,v = \int_{\delta \Omega_g}  \,u_{body}\left(\Pi\left(x\right)\right) \cdot v,
\end{equation}
where $\Pi(x)$ denotes the projection of the integration point $x$ onto the body mesh.
In practice, we integrate over all "boundary cells" $\delta \Omega_g$, i.e. grid cells which
intersect the body, and generate quadrature points over those cells. Then for
each quadrature point, we compute the closest point on the surface mesh. If it
is inside the mesh, we enforce the displacement at that point to equal the
prescribed displacement of the closest point on the surface. If it is outside
the mesh, we project the quadrature point onto the surface and enforce
displacement equality there.

As this technique is not guaranteed to generate fully compatible conditions, we
do not use a hard constraint to enforce the displacement boundary condition,
but a compliant penalty instead. This means that at the end of the solve, the
cloth may end up inside the target body because the body has not moved as much
as it should have. To work around this, at each "outer iteration" we displace
not only the cloth, but the body surface mesh as well. We then compare the
displaced body mesh to the target mesh; if we had conforming boundary
conditions they should match exactly, but with our formulation there will be a
discrepancy. If the gap is too big, we restart the transfer process from the
displaced body mesh to the target mesh, and repeat this outer loop until the
delta is small enough.

We compute the various terms using Warp's FEM library \cite{warp2022}. At each
inner iteration, we get a linear system of the form

\begin{equation}
  \begin{bmatrix}
    A & B^T \\
    B & C
  \end{bmatrix}
  \begin{pmatrix}
    \Delta u \\
    \Delta p
  \end{pmatrix} =
  \begin{pmatrix}
    f \\
    k
  \end{pmatrix}
\end{equation}

with $A$ as the matrix corresponding to the viscosity and boundary conditions
bilinear forms, $B$ corresponding to the divergence form, and $C$ to the
elastic compliance. $f$ corresponds to the prescribed boundary displacement
linear form, and $k$ contains the current nonlinear elasticy bias term. As $C$
is diagonal, we can compute the Schur complement of the system, $A+B^TC^{-1}B$,
and solve for the displacement $\Delta u$ using the Conjugate Gradient method.

\subsection{Sewing Pattern Optimization}

We now adjust the 2D sewing pattern based on the already solved for 3D
displacements of the garment. This process optimizes the position of the
panel~\footnote{Note that we use the terms "panel" and "sewing pattern piece"
  interchangeably.} vertices with the following considerations:

\begin{itemize}

  \item Adjust the sewing pattern piece sizes to accommodate for the increase or
        decrease in the target body.

  \item Adjust the sewing pattern positions to minimize the amount of stretch.

  \item Minimize changes to the boundary of the sewing pattern pieces (i.e. make sure
        the edges remain smooth).

  \item Maintain the edge lengths of seams as much as possible on each incident pattern
        piece.

\end{itemize}

Our approach to this is inspired by \citet{pietroni2022computational}, which
generates 2D sewing patterns based on a 3D cloth mesh. The method assigns
orthonormal tangent vectors to each triangle on a 3D cloth mesh representing
the warp and weft directions of the weave, which are computed using
cross-fields. It also "binds" these vectors to the triangle vertex positions so
the tangent vectors can deform with the triangle.

For any 2D panel layout assigned to a 3D cloth mesh, we can compute the
corresponding warp/weft tangent vectors for each triangle in that layout using
this binding information. The goal is to minimize the distortion of each
triangle as it goes from the 3D pose to the 2D layout, so the method defines an
energy that penalizes deviation from orthonormality for all of the triangle
tangents, then finds the 2D layout that minimizes that energy using an
optimization method.

We take a similar approach when reshaping our existing 2D panel layout for the
target body, although our use case is slightly different.

\subsubsection{Defining 3D tangent vectors}

We already have enough information to calculate the ground truth warp/weft
tangent vectors on the 2D sewing pattern for the original character, as they
are just unit vectors in the x and y directions respectively. We can transfer
these vectors onto the original 3D source pose for a given triangle like so:

\begin{itemize}
  \item Find the two edge vectors of the triangle, $\boldsymbol{e}_0 = \boldsymbol{v}_1
          - \boldsymbol{v}_0, \boldsymbol{e}_1 = \boldsymbol{v}_2 - \boldsymbol{v}_0$
  \item Stack these column vectors side by side into a 2x2 matrix and find its inverse
        $\boldsymbol{M}$
  \item Use this matrix to bind the tangent vectors to the triangle, and transfer the
        unit x,y vectors into 3D by finding the 3 by 2 edge matrix of the 3D triangle
        and right multiplying by $\boldsymbol{M}$.
\end{itemize}

Note that if we optimize for orthonormality in the way that we outlined above
we will recover the original 2D panel layout, as the tangent vectors are all
orthonormal in this configuration. To resize the sewing pattern to fit the
clothes on the target character, we need to transfer the vectors onto the
target character and make some modifications.

\subsubsection{Tangent vectors on the target body}

We transfer the tangent vectors onto the deformed garment for the target character
using the previously computed binding information. If we just optimize for
orthonormality again, we will recover the original 2D panel layout without
resizing it to the target character, as the information about the target shape is
encoded in how the tangent vectors are bound to the triangles.

Typically the 3D tangent vectors won't be orthonormal, as among other things,
the transfer between the source character and the target character will scale
and stretch the triangles somewhat. We can make them orthonormal again by
concatenating them into a 3 by 2 matrix $\boldsymbol{F}$, finding the polar
decomposition $\boldsymbol{F} = \boldsymbol{R} \boldsymbol{S}$ and just use the
columns of $\boldsymbol{R}$. We can then recompute the bindings of these
orthogonalized tangent vectors and optimize for a 2D panel layout. We can bind
a 3D tangent vector $\boldsymbol{v}$ to the edge frame by solving the following
minimization problem, where $\boldsymbol{E}$ is the 3 by 2 matrix containing
the triangle edge vectors as columns and $\boldsymbol{b}$ is a 2 element vector
of bind weights. We solve the problem using a pseudo inverse:

\begin{displaymath}
  \begin{split}
    \underset{\boldsymbol{b}}{\mathrm{argmin}} \left|\boldsymbol{E} \boldsymbol{b} - \boldsymbol{v}\right|^2 = \left(\boldsymbol{E}^T\boldsymbol{E}\right)^{-1} \boldsymbol{E}^T \boldsymbol{v} \\
  \end{split}
\end{displaymath}

Using this approach, the optimization will resize the sewing pattern to fit the
target character. The fact that each 3D triangle has orthonormal tangent vectors
means that its optimal shape in the 2D layout is just a rotated version of its
shape in 3D, so the optimization will try and select a 2D layout that matches
all the 3D triangle shapes as closely as possible. When simulated, this will
lead to a loose fitting outfit, as the optimization is trying to avoid all
stretching in the 2D->3D map.

This is not necessarily the desired outcome, since the cloth is often stretched
tight over the body in the source pose and there is already strain on the
triangles. To get a better fit on the target character's body, our method needs
to transfer this strain over to the tangent vectors in the target pose. To do
this, we do the following:

\begin{itemize}
  \item Find the 3D tangent vectors on the source character using the method in the
        previous section.
  \item Find the polar decomposition of each triangle's tangent frame,
        $\boldsymbol{F}_{ref} = \boldsymbol{R}_{ref} \boldsymbol{S}_{ref}$. Even though
        this is the source character, each triangle may be stretched relative to its
        shape in the 2D layout if the clothes are tight fitting, and this stretch
        information is contained in $\boldsymbol{S}_{ref}$, which we save for later.
  \item Find the 3D tangent vectors on the target character in the same way.
  \item Find the polar decomposition of each target triangle's tangent frame,
        $\boldsymbol{F}_{target} = \boldsymbol{R}_{target} \boldsymbol{S}_{target}$
  \item Swap the stretch information in this tangent frame with the stretch information
        in the source pose by building the tangent frame $\boldsymbol{F}_{target} =
          \boldsymbol{R}_{target} \boldsymbol{S}_{ref}$

\end{itemize}

Rebinding these tangent frames to the triangles and optimizing for a 2D layout
preserves the original fit when simulated.

\subsubsection{Optimization}

We use a simpler triangle deformation energy than
\citet{pietroni2022computational}, who have a quadratic term per triangle
forcing the y component of the warp tangent to a target value, a similar term
for the x component of the weft and then an As Rigid As Possible \cite{arap}
term to encourage rigidity. Instead we use an energy that forces the tangent
frames to an identity matrix. This energy is translationally invariant,
yielding an underdetermined problem. To address this, we add a quadratic
penalty term, constraining vertices to their original 2D positions:

\begin{equation} \label{baseenergy}
  E = \frac{1}{2} \sum_{t} A_t |\boldsymbol{F}_t - \boldsymbol{I}|^2 + \frac{1}{2} \epsilon \sum_{v} |\boldsymbol{p}_v - \boldsymbol{p}_{0v}|^2,
\end{equation}

Here, the $t$ index runs over all the triangles, $A_t$ is the area of triangle
t, and similarly $v$ runs over all vertices and $\epsilon$ is a small value,
typically $10^{-8}$. The symbols $\boldsymbol{p}_v$ and $\boldsymbol{p}_{0v}$
refer to the 2D positions of the $v$\textsuperscript{th} vertex in the pose we're optimizing and
the rest pose respectively. We can concatenate these into a 2N component vector
$\boldsymbol{x}$, and write the energy as a quadratic expression using sparse
matrices, as the tangents, $\boldsymbol{F}_t$, are linear functions of the 2D
vertex positions.

\subsubsection{Edges and Seams}

Minimizing this energy gives reasonable results, but often leaves untidy edges,
whose lengths may not match if they're sewn to other edges at seams. To improve
this, we add some extra energy terms that try to force the edges to have the
same direction and curvature as they did in the original pattern, while
allowing them to scale. For this, we consider the vectors from a vertex $i$ on
the edge to its two neighbors, call them $\boldsymbol{z}_{0i}$ and
$\boldsymbol{z}_{1i}$ and find a scale factor $S_i$ that brings their values in
the original configuration, $z_{0i}^r$ and $z_{1i}^r$, as close to their
current values as possible, ie:

\begin{displaymath}
  \begin{split}
    S_i = \underset{S}{\mathrm{argmin}} \left(|\boldsymbol{z}_{0i} - S \boldsymbol{z}_{0i}^{r}|^2 + |\boldsymbol{z}_{1i} - S \boldsymbol{z}_{1i}^{r}|^2\right)
  \end{split}
\end{displaymath}

We then define an energy based on this scaling of the rest pose like so, where
$L_i$ is half the sum of the lengths of vertex $i$'s two edge vectors in the rest
pose:

\begin{equation} \label{edgeenergy}
  \begin{split}
    W_i = \frac{1}{2} L_i \left(|\boldsymbol{z}_{0i} - S_i \boldsymbol{z}_{0i}^{r}|^2 + |\boldsymbol{z}_{1i} - S_i \boldsymbol{z}_{1i}^{r}|^2\right)
  \end{split}
\end{equation}

By ensuring that the vertex neighborhood along the edge is a scaled version of
the neighborhood in the original pose, we can preserve the straightness and
direction of the edges while still allowing them to change size.

These new energy terms are no longer quadratic, so we need a different
optimization method that can handle more complicated energies. For this we use
ADMM \cite{admm}. We formulate the minimization as follows:

\begin{displaymath}
  \begin{split}
    \underset{\boldsymbol{x},\boldsymbol{z}}{\mathrm{argmin}} E(\boldsymbol{x}) + \sum_i g_i(\boldsymbol{z}_i) \\
    s. t. \boldsymbol{z}_i = \boldsymbol{D}_i \boldsymbol{x}
  \end{split}
\end{displaymath}

Where $E(\boldsymbol{x})$ is the energy from equation \ref{baseenergy} and
$g_i(\boldsymbol{z}_i)$ are the edge energy terms in equation \ref{edgeenergy}.
The constraints ensure that the $\boldsymbol{z}_i$ vectors are the vertex
neighbor vectors mentioned above. Following \cite{admm}, we can optimize this
objective by iterating the following optimizations:

\begin{displaymath}
  \begin{split}
    \boldsymbol{x}_{n+1}   & = \underset{\boldsymbol{x}}{\mathrm{argmin}} \left[ E(\boldsymbol{x}) + \frac{1}{2}\sum_i w_i\left|\boldsymbol{D}_i\boldsymbol{x} - \boldsymbol{z}_{n,i} - \boldsymbol{u}_{n,i}\right|^2 \right]   \\
    \boldsymbol{z}_{n+1,i} & = \underset{\boldsymbol{z}}{\mathrm{argmin}} \left[ W_i(\boldsymbol{z}) + \frac{1}{2}\sum_i w_i\left|\boldsymbol{D}_i\boldsymbol{x}_{n+1} - \boldsymbol{z} - \boldsymbol{u}_{n,i}\right|^2 \right] \\
    \boldsymbol{u}_{n+1,i} & = \boldsymbol{u}_{n,i} + \boldsymbol{D}_i \boldsymbol{x}_{n+1} - \boldsymbol{z}_{n+1}
  \end{split}
\end{displaymath}

The first optimization can be performed by solving a linear system and the
optimizations for $\boldsymbol{z}_i$ have straightforward solutions and can be
performed in parallel.

This approach gives cleaner results, although it still allows two edges which
are sewn together to vary in length. To fix this remaining issue we tie the
scale factors together for the neighborhoods of vertices which have been sewn
together.

\subsubsection{Final tidy up}

\begin{figure}
  \centering
  \begin{subfigure}{.49\columnwidth}
    \centering
    \includegraphics[width=\columnwidth]{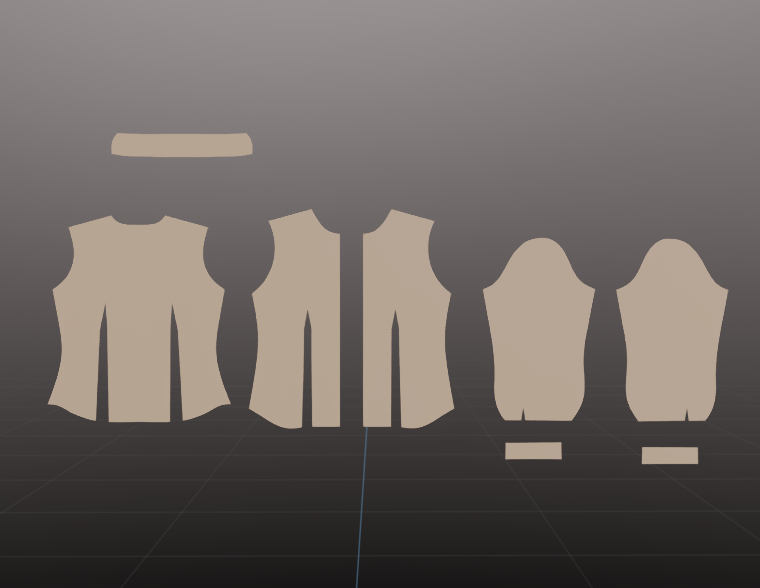}
    \caption{source}
    \label{fig:source_blouse_panels}
  \end{subfigure}%
  \hspace{.01\columnwidth}
  \begin{subfigure}{.49\columnwidth}
    \centering
    \includegraphics[width=\columnwidth]{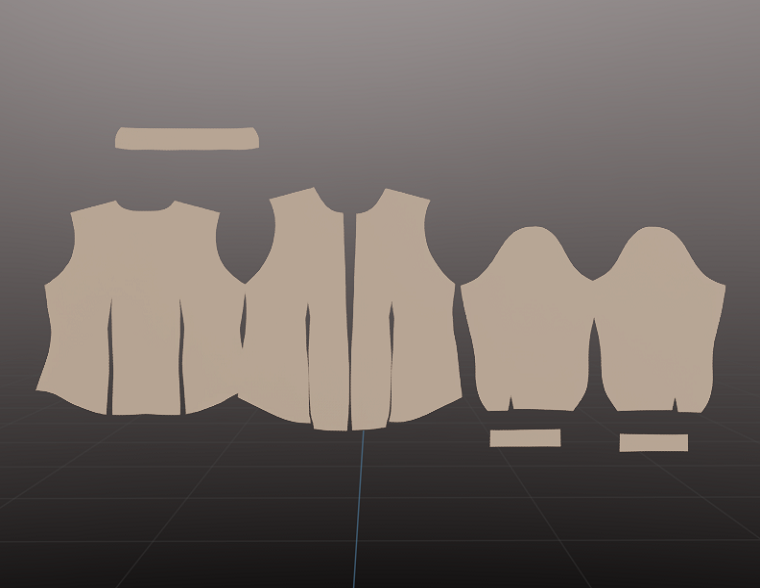}
    \caption{Target}
    \label{fig:target_blouse_panels}
  \end{subfigure}
  \caption{The original sewing pattern adjusted for the size of the target character.}
  \label{fig:blouse_panels}
\end{figure}

By optimizing this energy we achieve good sewing pattern outlines as shown in
Figure~\ref{fig:blouse_panels}. However, the internal vertices can be noisy,
particularly if the cloth is crumpled in the transfer pose. We do a final pass
to clean this up by pinning all vertices involved in seams and borders and
minimizing a Dirichlet energy over the remaining vertices, where the resulting
Laplacian matrix is computed on the original cloth pattern. This smooths the
triangle deformations out and gives them similar shapes to their original ones.

\begin{figure}
  \centering
  \begin{subfigure}{.49\columnwidth}
    \centering
    \includegraphics[width=\columnwidth]{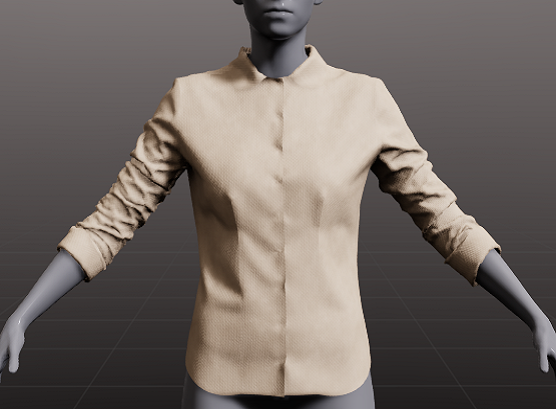}
    \caption{source\label{fig:source_blouse}}%
  \end{subfigure}%
  \hspace{.01\columnwidth}
  \begin{subfigure}{.49\columnwidth}
    \centering
    \includegraphics[width=\columnwidth]{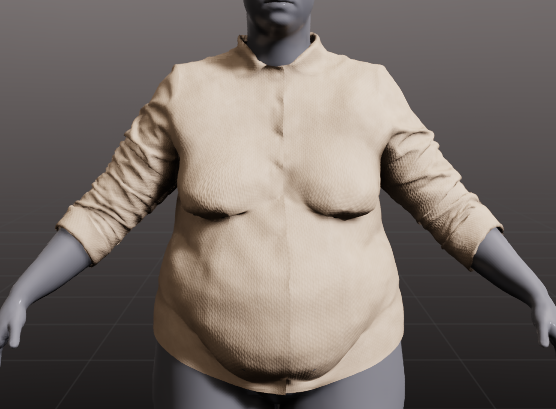}
    \caption{Target\label{fig:target_blouse}}%
  \end{subfigure}
  \caption{The blouse originally fitted to the source character, and transferred onto the target character using our garment transfer method.\label{fig:blouse}}%
\end{figure}

\section{Progressive Draper}

\begin{figure}
  \centering
  \includegraphics[width=\columnwidth]{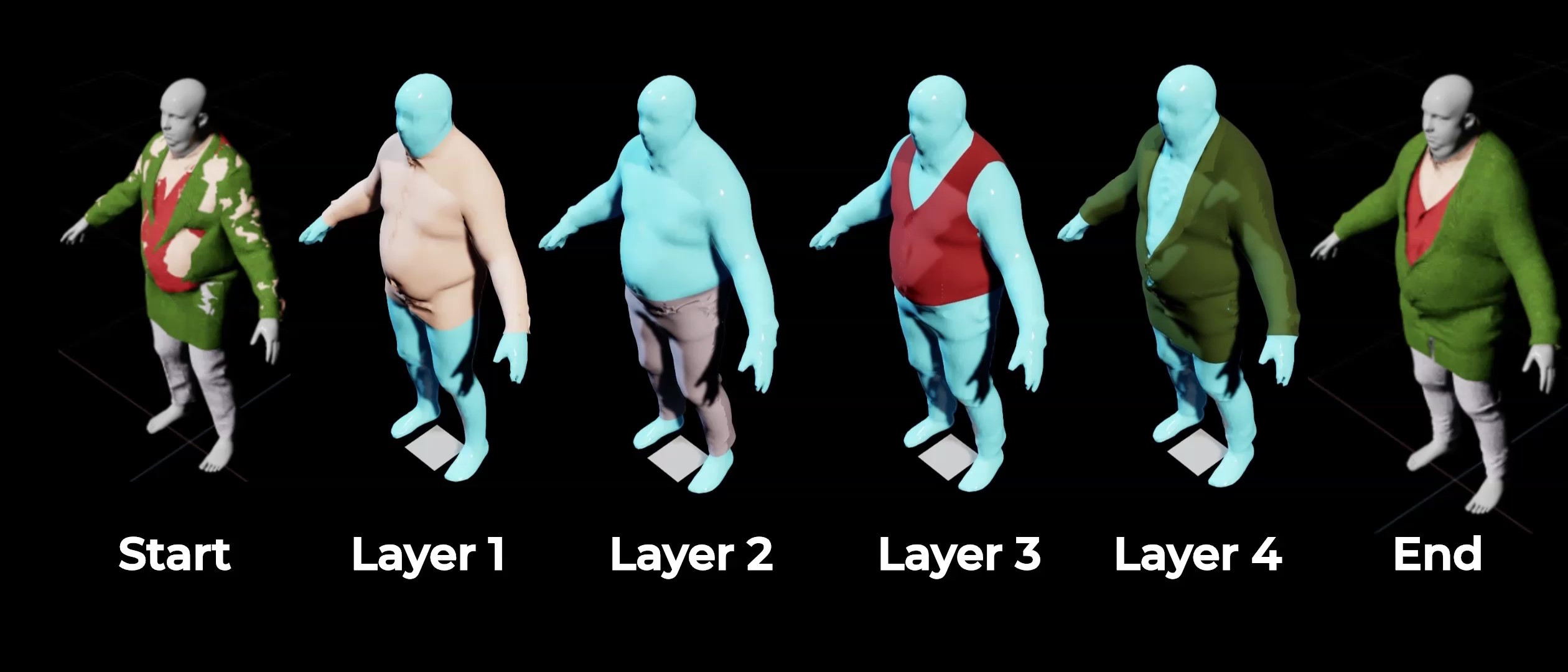}
  \caption{The progressive draper iteratively simulates each layer to untangle the clothing.
    Previous layers are frozen, and their geometry is combined into the collider SDF to
    ensure the next layer successfully pushes it out of the previous layer.}
  \label{fig:progressive_draper_explanation}
\end{figure}

The progressive draper untangles and simulates an outfit on the target
character. The draper runs a simulation once per layer of clothing to achieve
this result. See Algorithm \ref{alg:progressive_draper} for the steps we take
to progressively drape each layer of clothing, and
Figure~\ref{fig:progressive_draper_explanation} for an example of the draping
process on an outfit with four layers.

\begin{algorithm}
  \caption{Progressive Draper Algorithm}
  \label{alg:progressive_draper}
  \bf{Input}: Garment meshes $G$, Layer numbers $L$, collider meshes $C$ \;
  \BlankLine
  \bf{Output}: Garment meshes $G^\prime$ \;

  $G_s = \text{SortByLayer}\left(G, L\right)$

  $s_{total} \gets \emptyset$ \;
  \ForEach{$c \in C$}{
    $s \gets \text{SDF}\left(c\right)$\;
    $s_{total} \gets \text{SDFUnion}\left(s_{total}, s\right)$\;
  }

  $G_{prev} \gets \emptyset$ \;
  \ForEach{$g \in G_s$}{
    $s_{total} \gets \text{SDFUnion}\left(s_{total}, \text{SDF}\left(G_{prev}\right)\right)$\;
    $g\prime \gets \text{Simulate}\left(g, s_{total}\right)$\;
    $G_{prev} \gets G_{prev} \cup g\prime$
  }

\end{algorithm}

\subsection{Unified Collision Signed Distance Functions} \label{unified_sdfs}

A crucial challenge we face for collision detection is to develop a signed
distance function (SDF) for spatial inside/outside queries of a static garment
or mesh collider. Common SDF algorithms cannot provide an accurate value range
on meshes with open holes, as garment meshes lack topological water-tightness.
To address this issue, we employ the generalized winding number method to
determine the inside/outside region around the garment.
\citet{Jacobson-13-winding} demonstrated that generalized winding numbers
robustly determine inside/outside fields, even when meshes contain holes and
artifacts. We set the winding number sampling threshold to 0.25. This threshold
effectively "seals" the holes in garments created from arms, legs, torsos, and
other body parts.

We create a single collision SDF for collision detection between the garment
simulation and the underlying collision geometry, i.e. the body and lower
garment layers (if any). We take the union of all Signed Distance Functions
(SDFs) of each lower garment layer and the body. We combine the SDFs by taking
the minimum and adding a small offset $\epsilon_{sdf}$ to slightly expand the
field. See the blue meshes in Figure~\ref{fig:progressive_draper_explanation}
for a visual example of the SDFs created at each per-layer simulation.

\subsection{Cloth Simulator}

We use a panel-based cloth simulator to resolve collisions and drape the
garment over the character in a physically plausible manner.

\subsubsection{Seams Handling}

We represent each seam as a list of vertex pairs within the overall mesh. Some
vertices may participate in multiple seams, for example at the corners of the
sleeve pattern pieces of a t-shirt. Therefore, we precompute a list of ``seam
groups'': the set of vertices that are all attached together via seam lines.

Artists may sew some panels at a position offset from another panel, such as a
pocket attached to the chest of a t-shirt panel. We store a "seam offset" based
on the initial condition of sewn vertices imported into the tool and enforce
these seam offsets throughout the entire simulation. Most offsets will be zero
for sewing pattern pieces that attach along manifold edges. To enforce the seam
group constraints after a position step, we first find the seam group's new
center by computing the average position and velocity of the seam group. We
then update each vertex position to be offset from the group's center and
update each vertex velocity as the group's velocity.

\subsubsection{Intrinsic Energies}

We use anisotropic constitutive models for stretch and bending, to capture the
features of real fabrics. Industry tools such as Marvelous Designer measure
physical stretch and bending parameters in multiple directions: weft, warp, and
shear. These parameters and their relationship to each other create emergent
material behaviors ~\cite{MD-Fabrics}. To accommodate this for stretch, we use
the energy described by ~\citet{kim-20-fem-cloth}.

We take inspiration from \citet{Gingold-04-thin-shells} for our bending energy.
We derive the bending energy $\psi_{bend}$ from the triangle-level shape
operator, expressed in the coordinate system of the warp and weft directions.
Let $S$ be the 2x2 curvature matrix, where $S_{0,0}$ is the curvature in the
warp direction, $S_{1,1}$ is the curvature in the weft direction. Let $S^r$ be
the rest curvature. We define bending stiffnesses $k$ per panel in each
direction: warp, weft, and shear.

\begin{equation}
  \begin{aligned}
    \psi_{bend} = 1/2 \Bigl(( k_{warp} (S_{0,0}-S^{r}_{0,0})^2 + k_{weft} (S_{1,1}-S^{r}_{1,1})^2 + \\
    k_{shear} (S_{1,0}-S^{r}_{1,0})^2 ) \Bigr)
  \end{aligned}
\end{equation}

We use a spring-based seam bending energy inspired by
\cite{bridson-02-contact}. We measure the angle between two triangles and compare
it to the rest angle to produce a spring force along with an elastic
response and damping term.

\subsubsection{Collisions}

To efficiently detect collisions between cloth triangles, we leverage a BVH to
find nearby triangles. For any contact pair, we verify that the involved
triangles are not topologically adjacent, and are not connected via a seam
vertex. Using this method, we avoid spurious collisions across sewn pattern
pieces.

For collision response, we apply impulses between the closest points between
two colliding triangles. We interpolate impulses into the triangle faces as
described by \citet{bridson-02-contact}. Each impulse is made up of an elastic
portion, a damping portion, and a coulomb friction portion. For collision
between the cloth triangles and all the clothing layers below and the body, we
utilize the SDFs as described in Section~\ref{unified_sdfs}. We detect the
collisions between triangles' vertices and the union signed distance field and
apply impulses to the colliding vertices. The collision normal is simply the
gradient of the SDF.

Typical parameters used in the simulation for collision are available in Table
\ref{tab:params}. Since initial garment configurations frequently start off in
heavily interpenetrated states, we globally damp velocities by $90\%$ at each
step of the simulation to reduce the likelihood of a large energy in the
initial interpenetrated configuration.

\begin{table}
  \begin{tabular}{@{}ll@{}}
    \toprule
    Parameter                     & Value   \\ \midrule
    Collision Force Coefficient   & 2500000 \\
    Collision Damping Coefficient & 25      \\
    Coulomb Friction Coefficient  & 25      \\
    $\epsilon_{sdf}$              & 0.2     \\ \bottomrule
  \end{tabular}
  \caption{Simulation Parameters}
  \label{tab:params}
\end{table}

\section{Automatic Semantic Proxy Generation}

\begin{figure}
    \centering
    \includegraphics[width=\columnwidth]{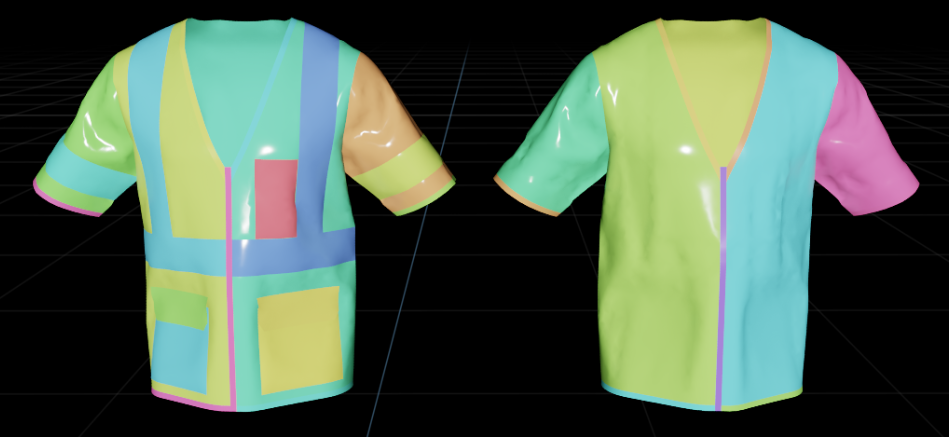}
    \caption{Initial garment (left) and generated proxy mesh (right).}
    \label{fig:proxy_mesh}
\end{figure}

We develop a method to automatically filter out panels based on type, to
produce a simulation-ready proxy mesh. Using this approach, we can address the
simulator's struggle to drape certain garments with complex construction, such
as thin decorative pieces like pockets or straps. For instance, a high
visibility vest may have reflective straps that are tightly constrained to
follow the garment's surface. Improper tuning of the thickness parameter for
this material can cause significant collision forces between the strap and its
attached surface, without adding significant value to the simulation result. We
consistently label each panel in our dataset with semantic meaning, to make
panel filtering simple. We show the proxy mesh of a high visibility vest
generated using our method in Figure \ref{fig:proxy_mesh}. Note that this
approach requires us to make the following assumptions:

\begin{itemize}
    \item We assume that all vertices of a removed panel piece have a nearby part of the
          proxy mesh.
    \item We assume the proxy mesh does not undergo significant rotation. Since we use
          world space offsets, we are not robust to rotations of the garment.
\end{itemize}

We use these generated proxy meshes during garment transfer and the progressive
draper step. Note that for the collisions of the draper to be correct on each
layer, we use the detailed mesh of the lower layers as part of the collision
surface. This ensures we include the thickness of pockets/straps in subsequent
layer simulations.

\section{Rig Transfer}

As the final step to prepare a character's outfit for animation, we transfer
the skinning weights of the character's rig from the body to the cloth.

\begin{figure}
    \centering
    \begin{subfigure}{.49\columnwidth}
        \centering
        \includegraphics[width=\columnwidth]{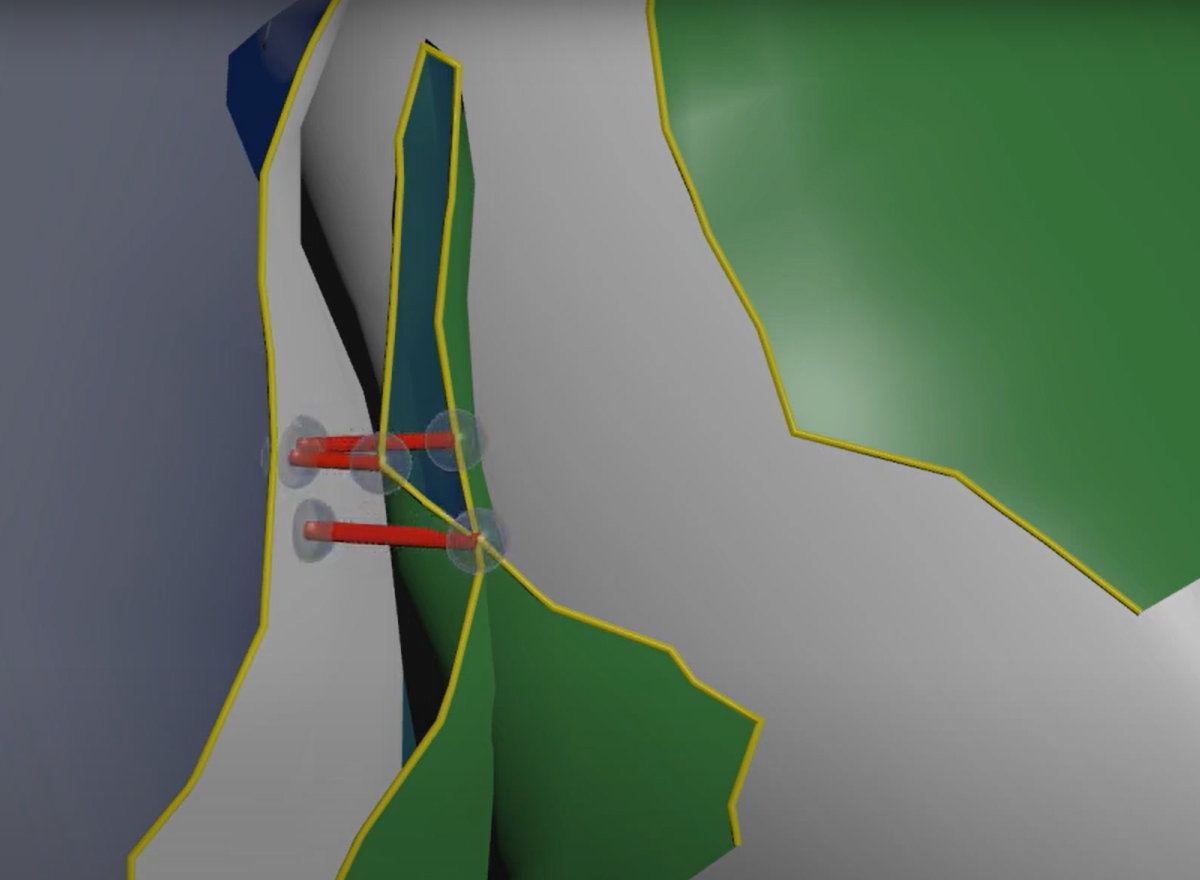}
        \caption{Transfer by positions}
        \label{fig:rig_transfer_xfer_by_pos_cropped}
    \end{subfigure}%
    \hspace{.01\columnwidth}
    \begin{subfigure}{.49\columnwidth}
        \centering
        \includegraphics[width=\columnwidth]{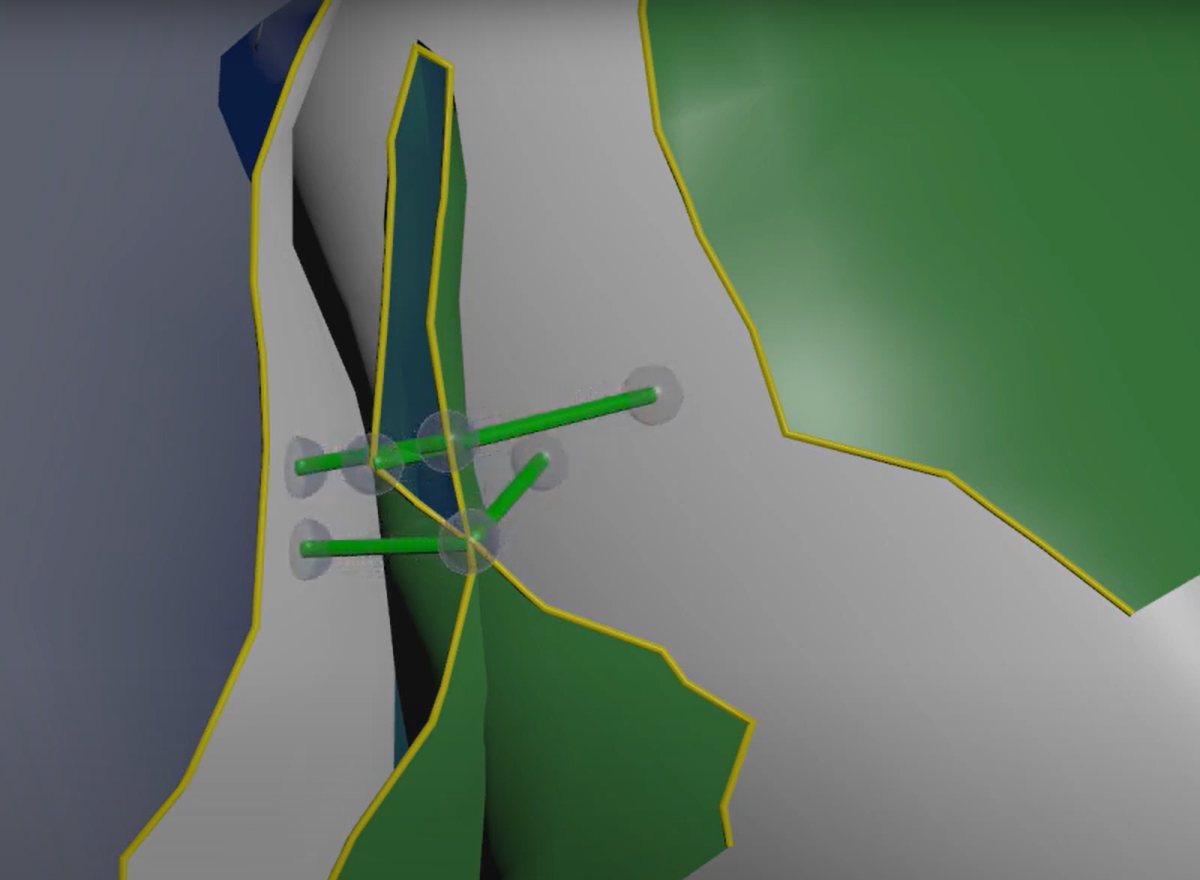}
        \caption{Transfer by normals}
        \label{fig:rig_transfer_by_normals_cropped}
    \end{subfigure}
    \caption{Cross-section of garment layers and their associated body surface locations for
        skinning weights transfer based on different strategies.
        Using garment normals helps each face find the correct body location to inherit skinning weights from.}
    \label{fig:rig_transfer_compare}
\end{figure}

We initially tried a straightforward approach by searching for the closest
point between a cloth vertex and the body surface, then propagating the
skinning weights from the body to the cloth. While this approach may work in
some simple cases, it was not capable of handling situations where cloth layers
can get arbitrarily close or even inverted, since it may associate the cloth
piece with the wrong body part (Fig.
\ref{fig:rig_transfer_xfer_by_pos_cropped}). We observe these problem cases
happen most often in tightly pinched areas, such as the armpit of the
character. To ensure a more robust rig transfer, we use the cloth's normals to
find the best points on the body mesh to transfer skinning weights from (Fig.
\ref{fig:rig_transfer_by_normals_cropped}). We compare these two approaches in
Figure~\ref{fig:rig_transfer_compare}.

We apply normal smoothing to improve the smoothness of the skinning weights
assigned to each cloth vertex. We also constrain skinning weights to be
identical across seam lines, which ensures continuity across seam boundaries.
Finally, we use proximal transfers as a fall-back strategy if the normal based
search path does not find a body vertex to use for skinning weights
association. With all of these strategies together, we create more robust
animated character rigs.

\section{Results}

\begin{figure}
  \centering
  \includegraphics[width=\columnwidth]{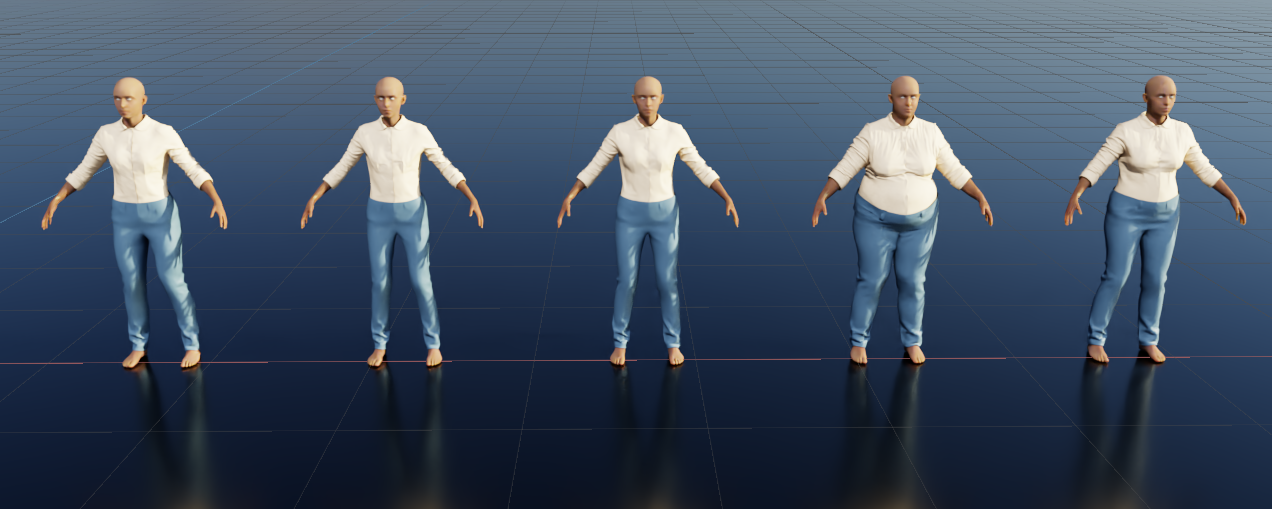}
  \caption{A single outfit fit onto characters of different body types.}
  \label{fig:bolt_results_overview}
\end{figure}

Using our pipeline, we can assemble, fit, drape, and rig a wide range of
outfits onto new characters. As shown in
Figure~\ref{fig:bolt_results_overview}, we can fit a single outfit onto
multiple characters of widely varying body shapes.

We designed the pipeline as an offline process. We measured performance of the
two slowest steps of the pipeline, garment transfer and draping, on a wide
range of outfits using an NVIDIA RTX 5000 Ada Generation Laptop GPU. These
metrics show the scalability of Bolt for generating large character sets
compared to manually transferring outfits.

\begin{figure}
  \centering
  \includegraphics[width=\columnwidth]{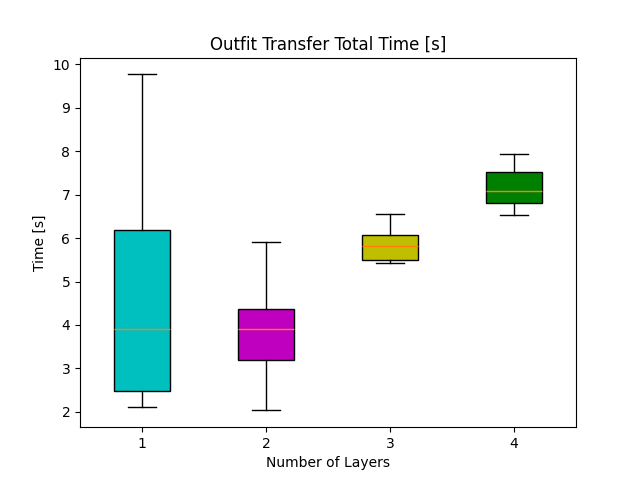}
  \caption{Box plot of transfer times for different outfits
    fit to a single obese male character, organized by number of layers in each outfit.
  }
  \label{fig:garment_transfer_perf}
\end{figure}

Figure \ref{fig:garment_transfer_perf} shows the transfer time for a wide range
of outfits fit to a single obese class-1 male character. The total transfer
time depends on the number of garments, the convergence rate of the conjugate
gradient solver, and the number of source bodies involved in the transfer.
If two garments are fit to two different source bodies, say a skirt fit to a
female source character and a button-up shirt fit to a male source
character, the total transfer time includes the time to transfer both garments
from their respective source bodies to the target body.

\begin{figure}
  \centering
  \includegraphics[width=\columnwidth]{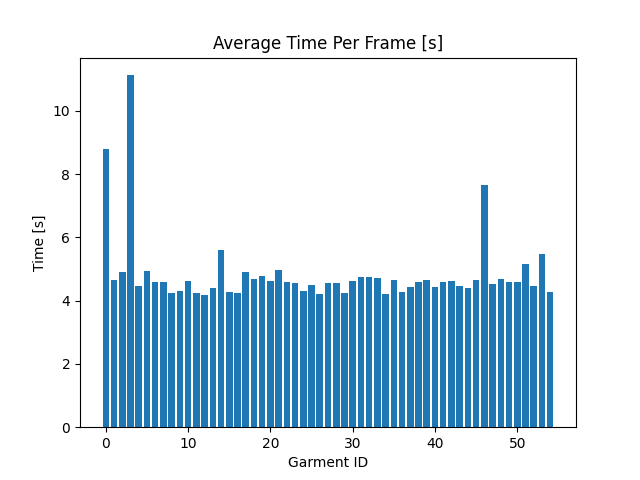}
  \caption{Histogram of average simulation times per frame across 55 different garments.
    Since the draper simulation only simulates one garment at a time, an outfit's total
    simulation time is the sum of the N layers involved in the outfit. All garments are
    simulated for only 6 frames.}
  \label{fig:garment_perf}
\end{figure}

Figure \ref{fig:garment_perf} shows the average draping time per frame for a
range of garments. The total time to assemble an outfit will depend on the
combination of garments, and their individual simulation times. For our
results, we run each garment for 6 frames of the simulation. Across our dataset
of garments, the average frame time is 4.8 seconds.

\begin{figure}
  \centering
  \includegraphics[width=\columnwidth]{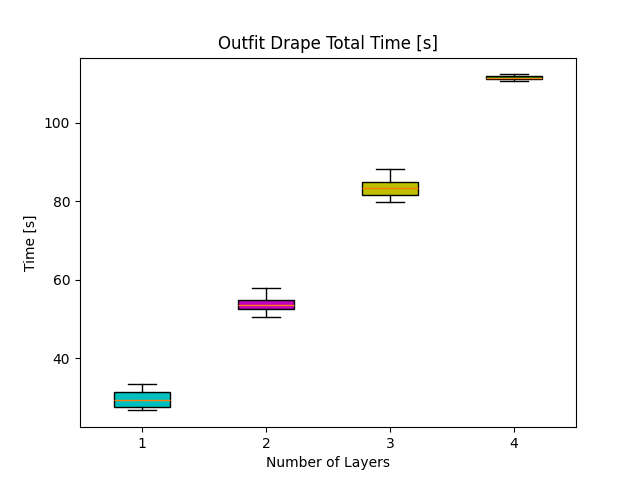}
  \caption{Box plot of draping times across a collection of outfits fit to the obese class 1 individual, sorted by number of layers in the outfit. }
  \label{fig:outfit_draping_perf}
\end{figure}

\begin{figure}
  \centering
  \includegraphics[width=\columnwidth]{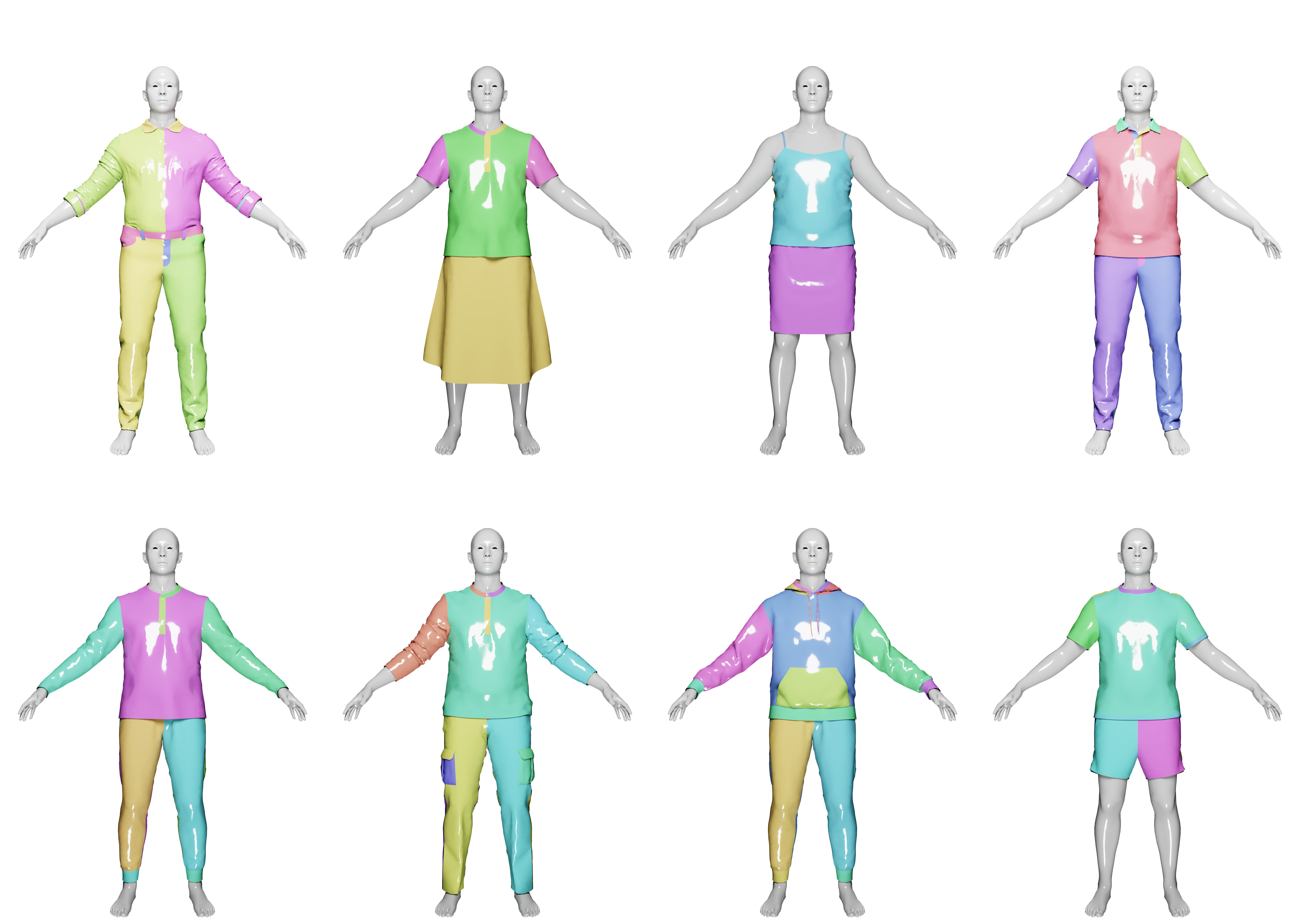}
  \caption{Multiple outfits on the same character. In this visualization each panel is assigned a different random color.}
  \label{fig:outfit_gallery}
\end{figure}

Figure \ref{fig:outfit_draping_perf} shows the draping simulation time for an
outfit collection fit to an obese male character. We observe that the
progressive draper step is the main performance bottleneck of the system, since
we simulate each layer separately. We demonstrate successful fits on large
characters of up to 4 layers, as previously shown in Figure
\ref{fig:progressive_draper_explanation}. For our 4-layer outfits, the slowest
completes in 112 seconds. Figure \ref{fig:outfit_gallery} shows a subset of
outfits on a single class one obese character. See Table
~\ref{tab:male_outfits_grid}, and Table ~\ref{tab:female_outfits_grid} for
examples showing different characters wearing a variety of outfits.

Note that these performance numbers do not include file I/O, rigging, or any of
the transfer initialization and simulation initialization processes. We ran an
experiment to generate 220 characters, and observed an average wall-clock time
of approximately 6 minutes per character (run on a cloud instance with an A40
GPU). Using cloud resources, and with a budget of 64 A40 GPUs, we were able to
generate all 220 in under 30 minutes. We have used our system to produce
characters in practical applications. At the time of writing, we have generated
and published just over ~100 characters for demos or public releases. As part
of our submission video, we show images of nearly 1000 characters we generated
using this tool.

\section{Conclusion}

We have demonstrated a tool for the automatic assembly, fitting, draping, and
rigging of many garments onto many characters. Using this system, we can
assemble garments into new outfits, and fit them onto new characters in a
scalable way. As the demand for large numbers of digital humans continues to
grow, our approach can help keep up with that demand.

\subsection{Limitations}

In the garment transfer step, we assume a per-vertex correspondence between the
source body and target body. To support general character refitting, we
would need a spatial correspondence map between any two characters. We derive
all of our characters from a single parametric model, which ensures they all
share the same topology and are compatible in our pipeline.

In the progressive draping step, we cannot propagate information from later
simulation layers back to earlier layers. For example, with a collared shirt on
layer 0 and a sweater on layer 1, we cannot achieve a look where the collar is
above the sweater since the sweater's simulation will see the final ``frozen''
dress shirt positions as a collision surface. To address this issue, we would
need to untangle all garments simultaneously as part of the same simulation,
and/or carefully break down the garment sewing pattern pieces into "sub-layers"
that could interleave with one another. For example, we could put the body of
the shirt on layer 0, the sweater on layer 1, and the collar of the shirt onto
layer 2, as long as we add constraints to ensure the collar remains connected
to its counterpart pattern piece on layer 0.

Our cloth simulator cannot be used in full scale animations because our seam
handling approach does not allow for seam group rotations. The simulator also
does not have a robust self-collision untangling method, so if the cloth starts
off in a tangled state, due to an artifact introduced in an earlier stage of
the pipeline, the simulator will not fix the tangling artifact, and it will
persist even after the draping stage.

Our system may introduce artifacts for characters with extreme body shape
differences. For example, if a class 3 obese character's body has a region that
causes the surface to fold back onto itself, the transfer process will place
the cloth close to the surface of the body mesh, essentially ``pinching'' the
fabric underneath the body region. Once the cloth starts off in a pinched
configuration, it rarely succeeds in pulling the cloth out from the pinched
section in a way that is nicely untangled.

\subsection{Future Work}

In addition to addressing the limitations mentioned previously, we would like
to investigate other approaches to improve the robustness of our pipeline. We
would like to investigate untangling approaches that can resolve
self-collisions, especially in tight/pinched areas, to help the system recover
from poor garment initialization or issues during garment transfer.

Our garment transfer method resizes the garment solely on the character's
differences in geometry. Sometimes this approach can make garments too
form-fitting on larger characters, and place extra material in ways that modify
the original style or design. Additionally, we are inspired by the traditional tailoring process where a tailor takes specific measurements of the body for fitting a suit. We would like to investigate using similar measurements to achieve smarter garment refitting.

\begin{acks}
  We thank Nvidia for supporting this project and in particular Vanni Brighella, Miguel Guerrero, Adeline Aubame, and Qiao Wang for help with creating the input assets and software support.
  We also thank Michael Honke for proof-reading our paper, and Ken Museth, Simon Yuen, and Miles Macklin for helpful discussions and feedback.
\end{acks}

\bibliographystyle{ACM-Reference-Format}
\bibliography{bolt.bib}

\appendix

\newlength{\imagewidth}
\setlength{\imagewidth}{0.15\textwidth}

\def\bodyA{char_male_age16_body_mid_model_published}
\def\bodyB{char_male_age65_body_mid_model_published}
\def\bodyC{char_male_bodybuilder_body_mid_model_published}
\def\bodyD{char_male_obesityClassI_body_mid_model_published}
\def\bodyE{char_male_atrophied_body_mid_model_published}

\def\outfitA{outfit_unisex_casual_0006}
\def\outfitB{outfit_unisex_sportive_0001}
\def\outfitC{outfit_unisex_sportive_0003}
\def\outfitD{outfit_unisex_formal_0005}

\begin{table*}
  \begin{tblr}{
      colspec = {X[c,m] | X[c,m] X[c,m] X[c,m] X[c,m] X[c,m]},
    }

     &
    \includegraphics[width=\imagewidth, align=c]{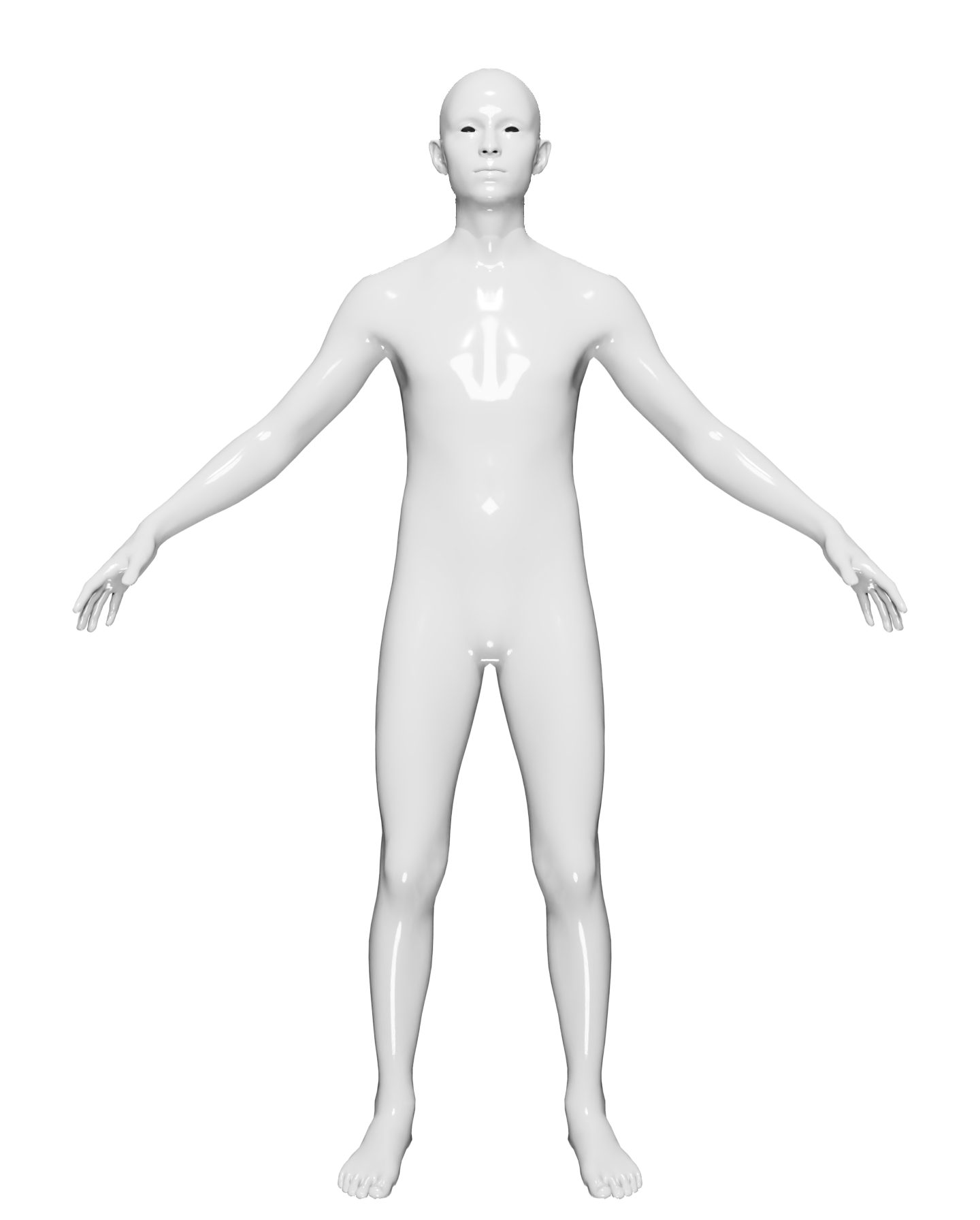}
     &
    \includegraphics[width=\imagewidth, align=c]{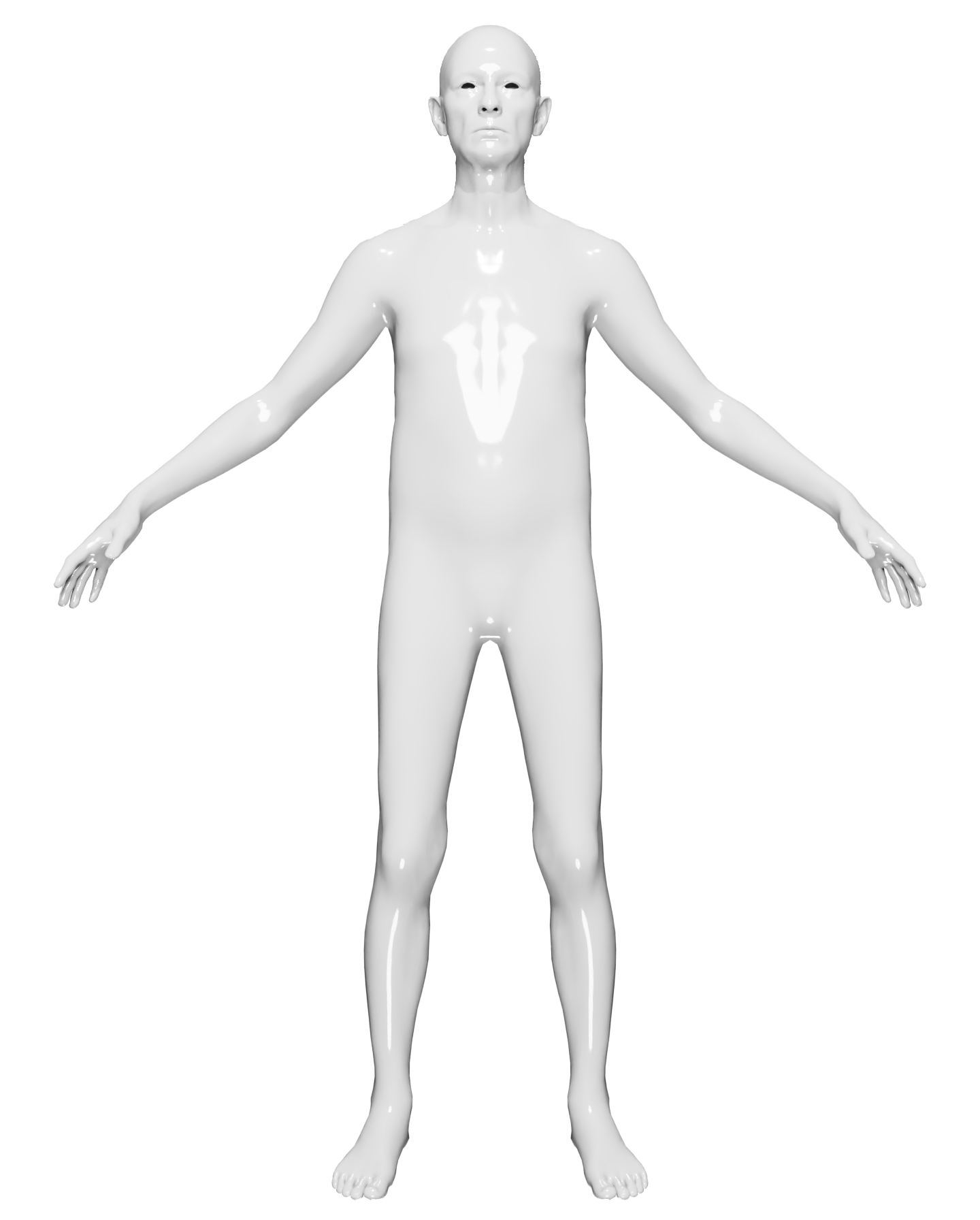}
     &
    \includegraphics[width=\imagewidth, align=c]{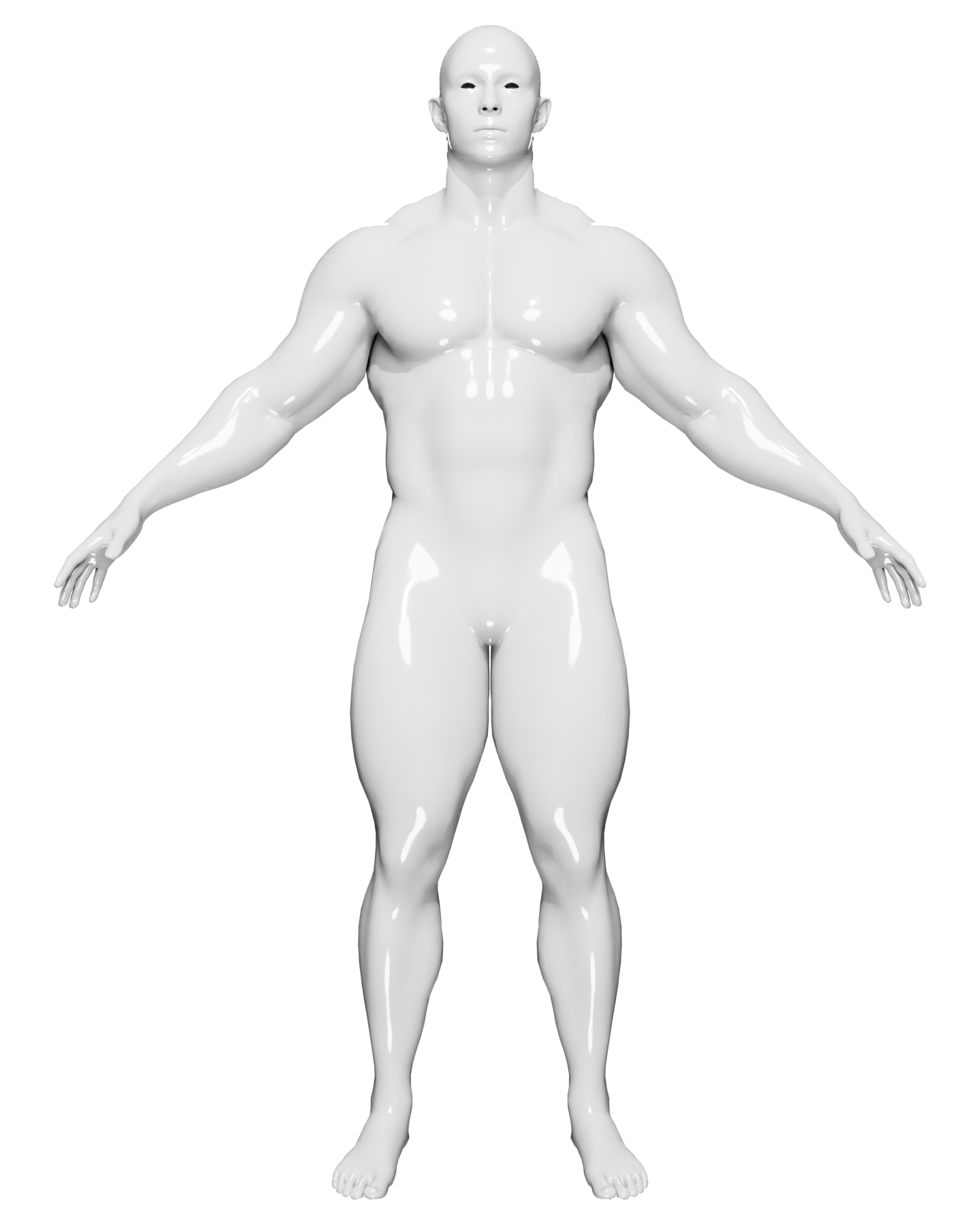}
     &
    \includegraphics[width=\imagewidth, align=c]{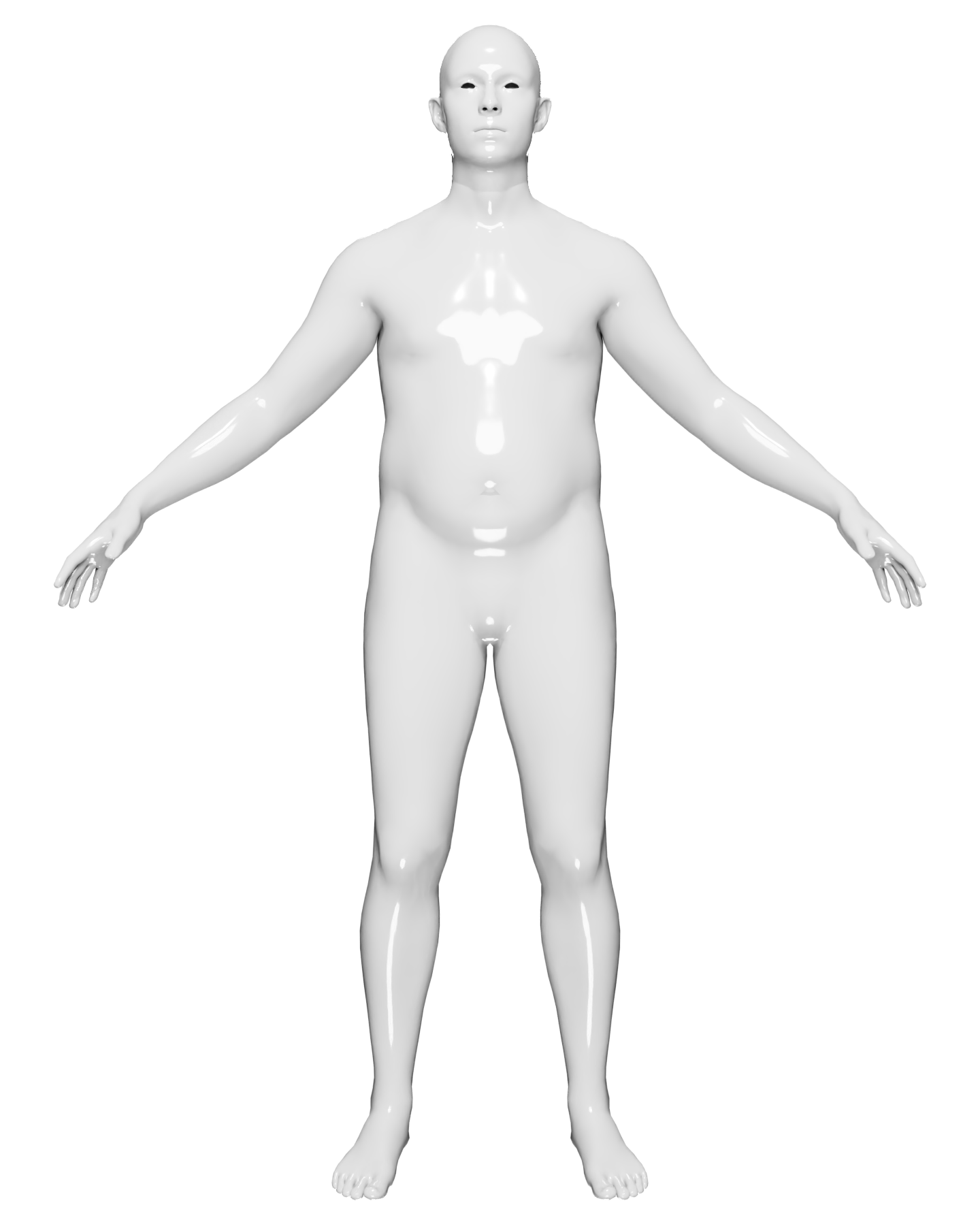}
     &
    \includegraphics[width=\imagewidth, align=c]{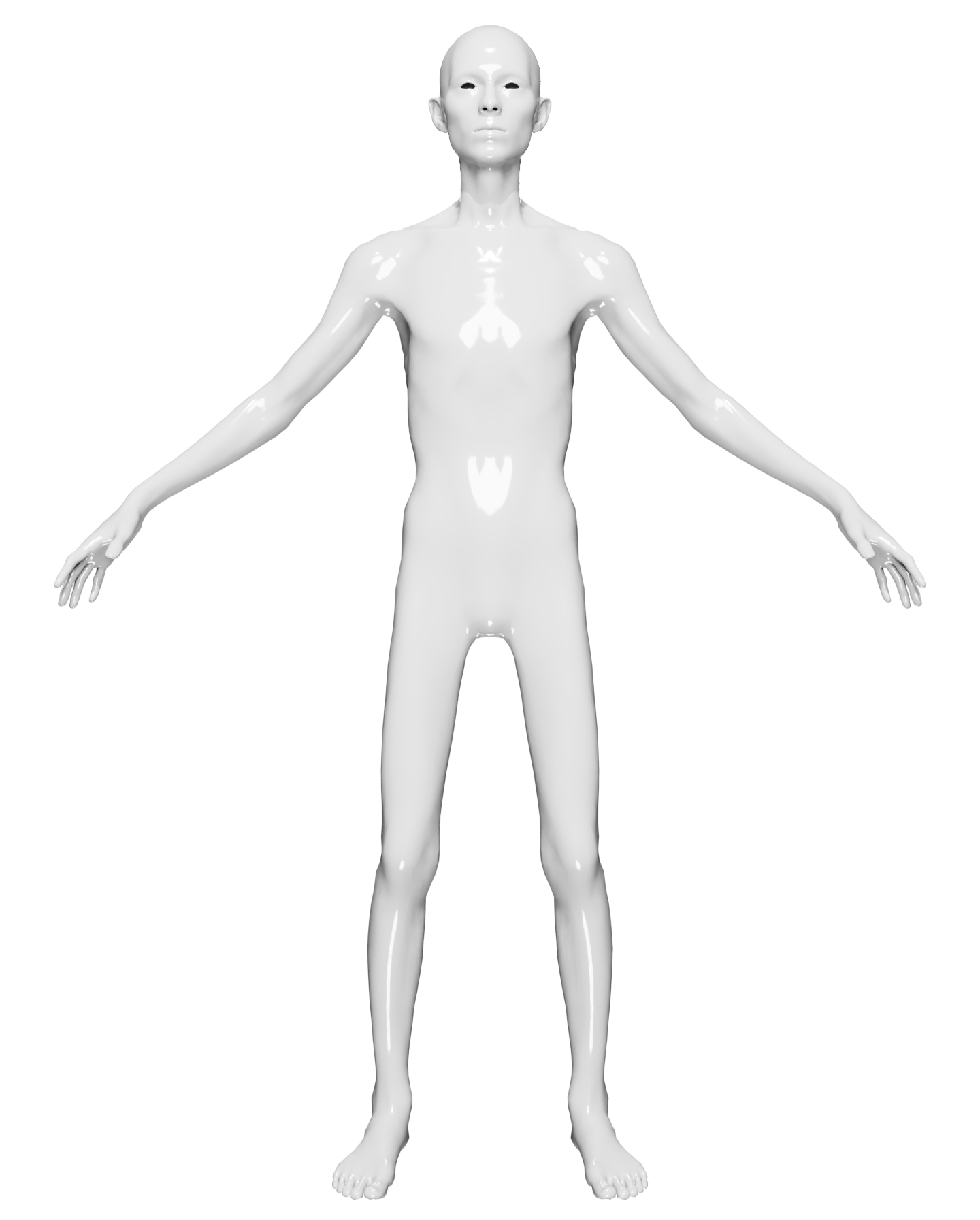}

    \\

    \hline

    \includegraphics[width=\imagewidth, align=c]{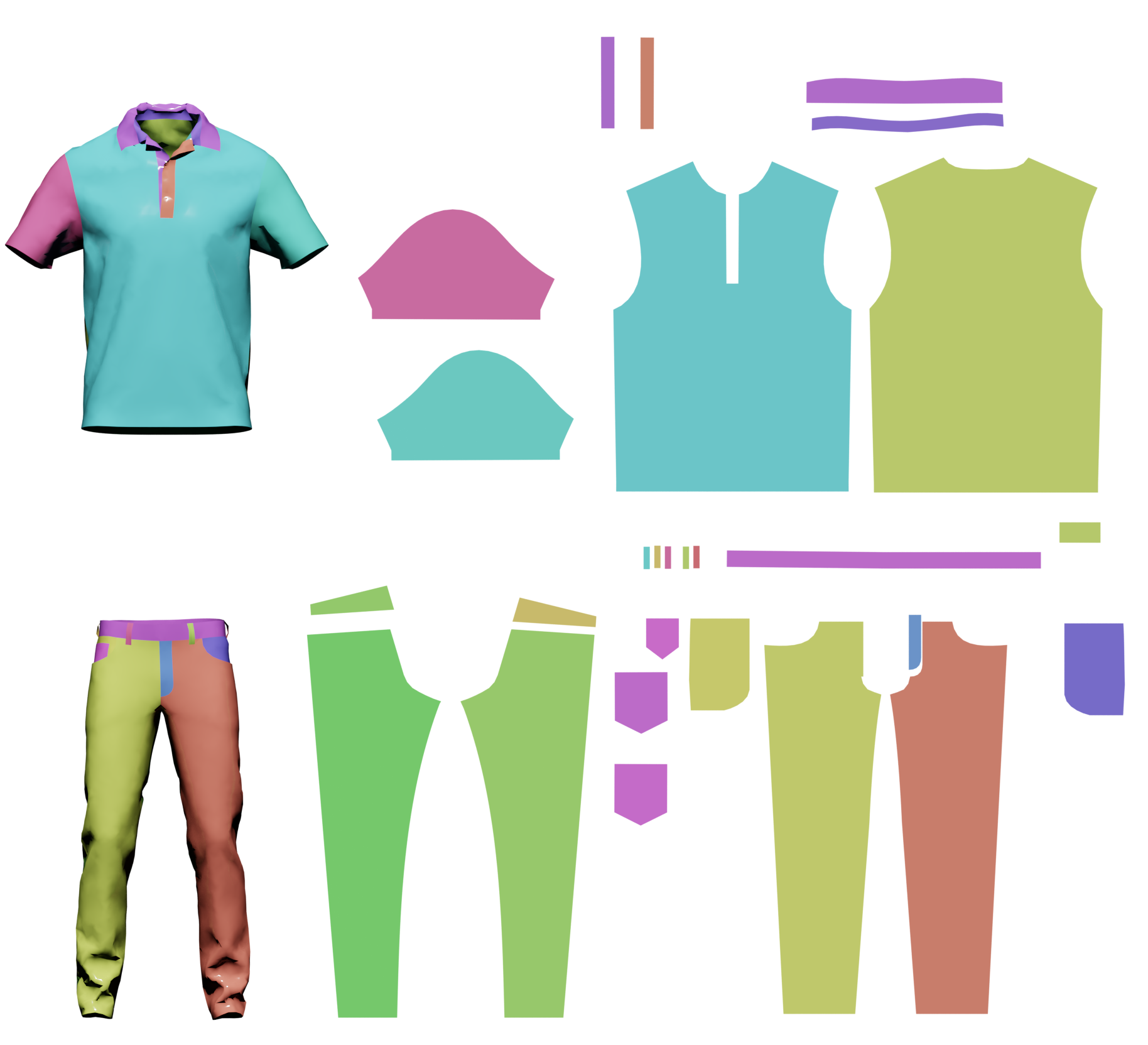}
     &
    \includegraphics[width=\imagewidth, align=c]{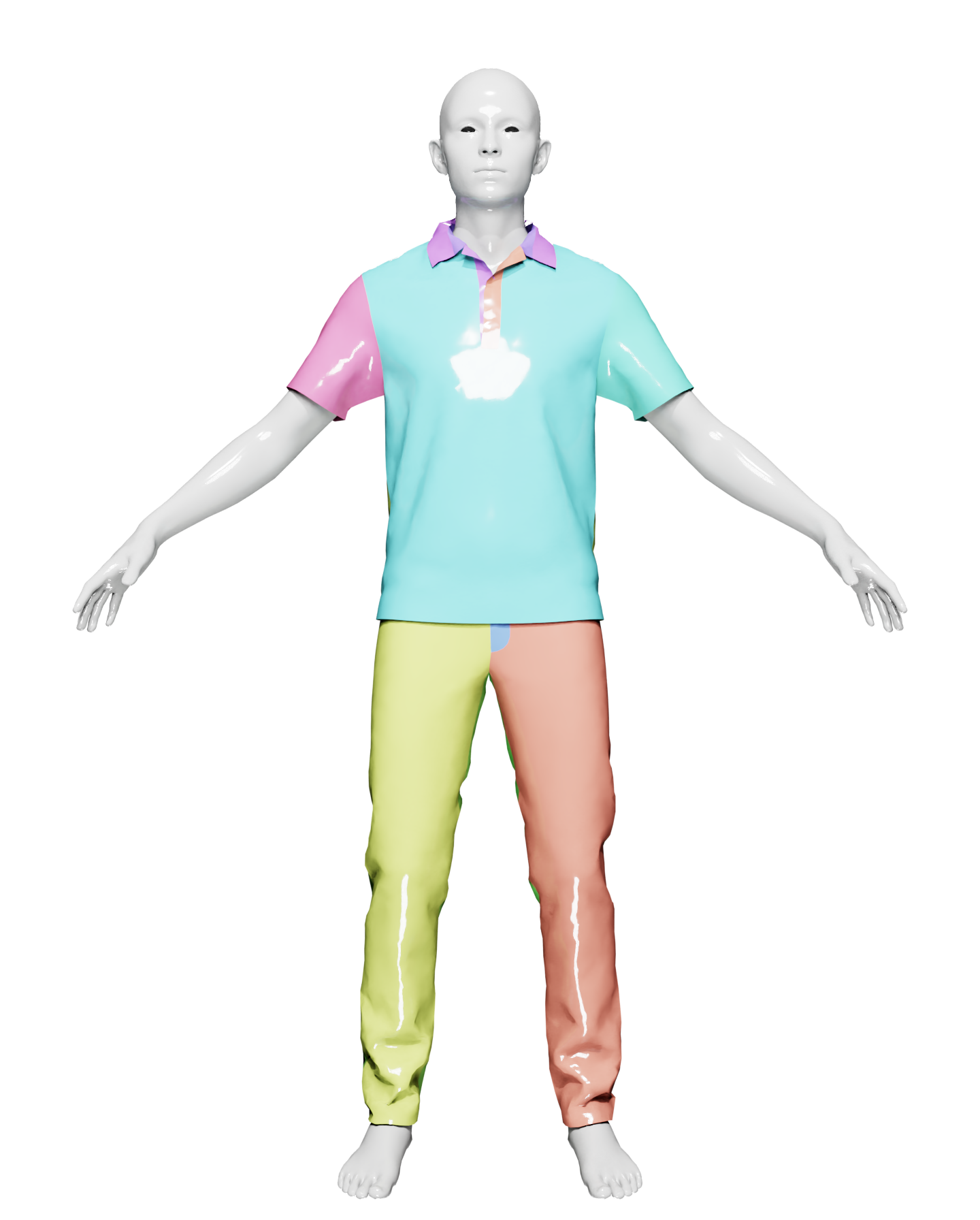}
     &
    \includegraphics[width=\imagewidth, align=c]{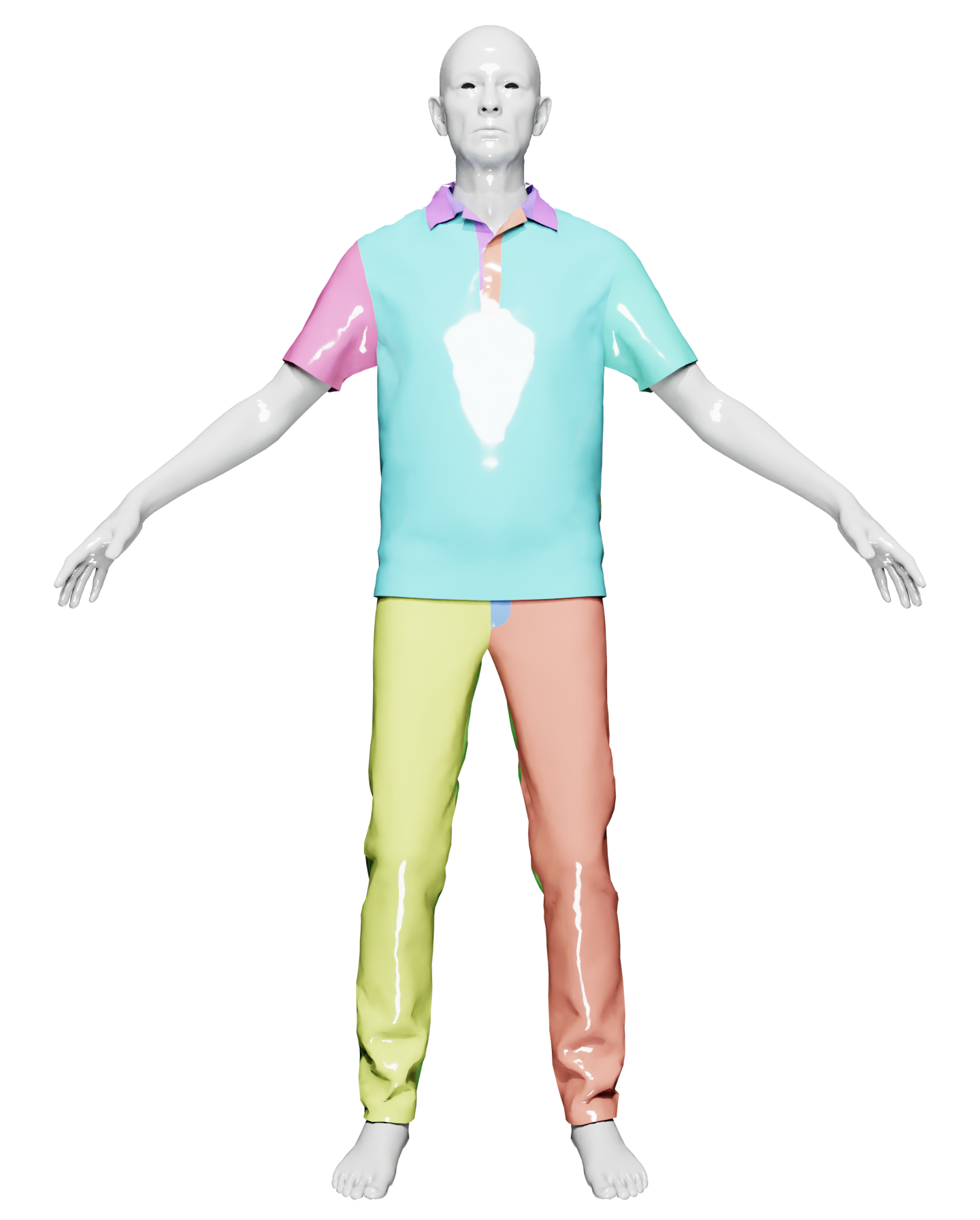}
     &
    \includegraphics[width=\imagewidth, align=c]{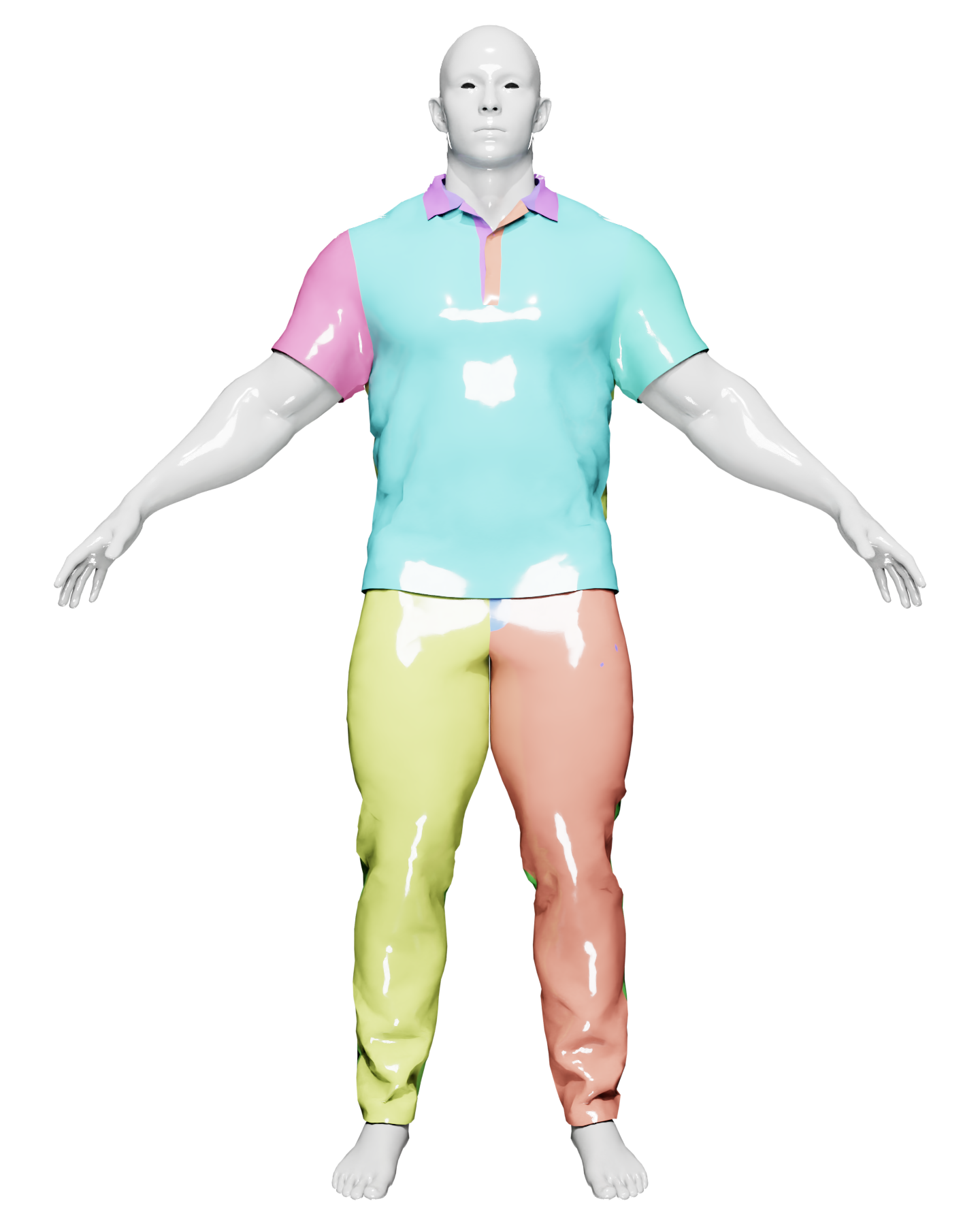}
     &
    \includegraphics[width=\imagewidth, align=c]{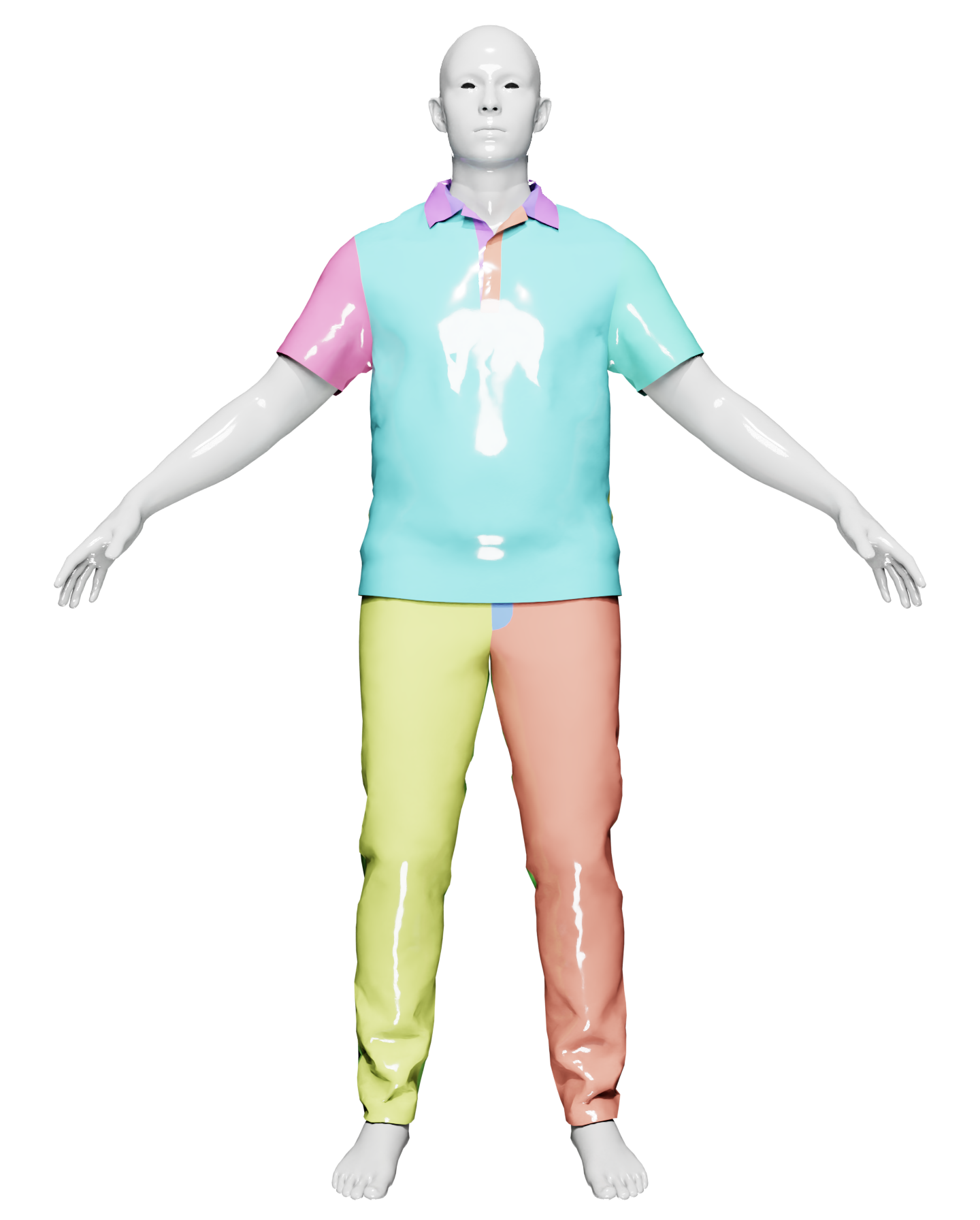}
     &
    \includegraphics[width=\imagewidth, align=c]{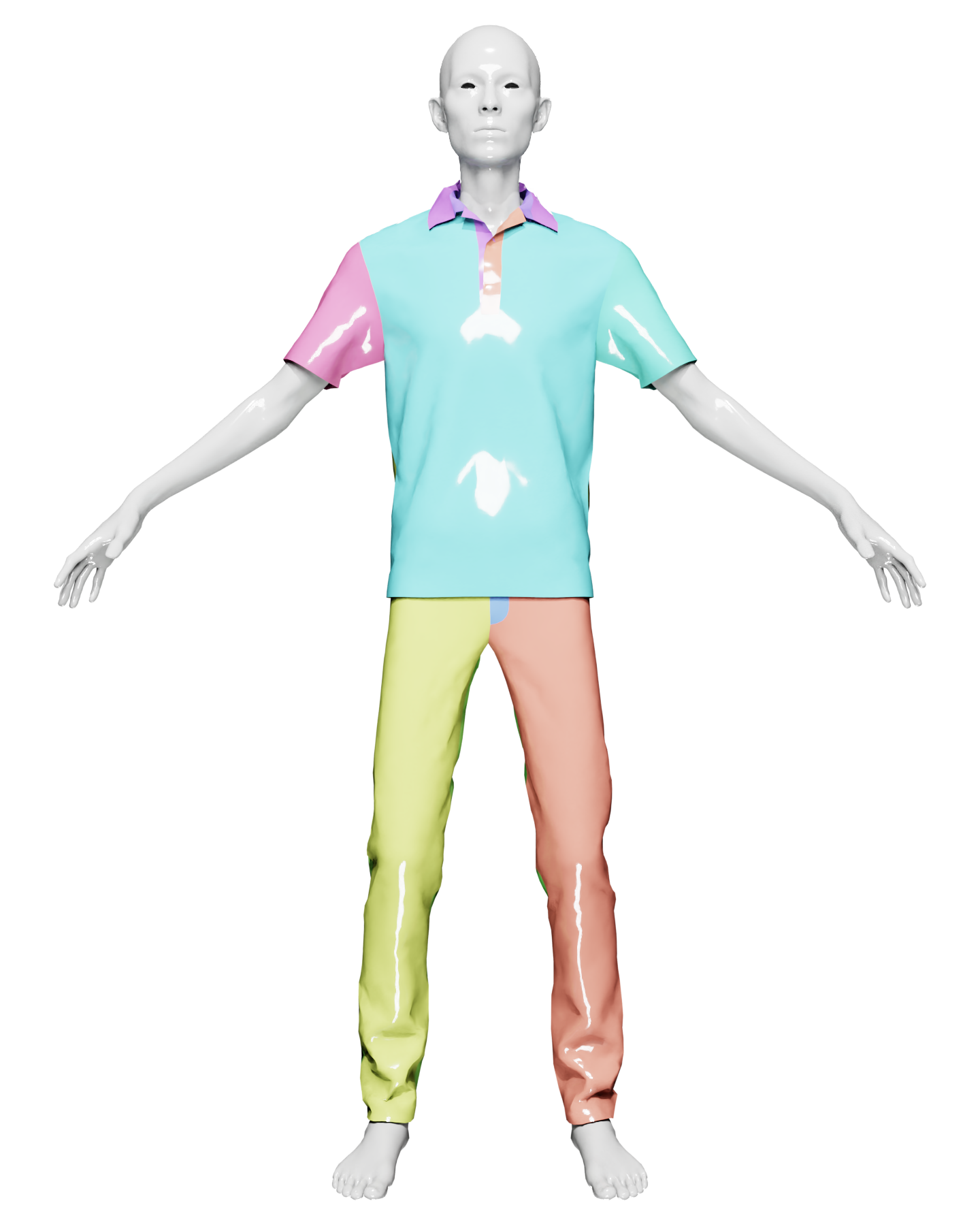}

    \\

    \includegraphics[width=\imagewidth, align=c]{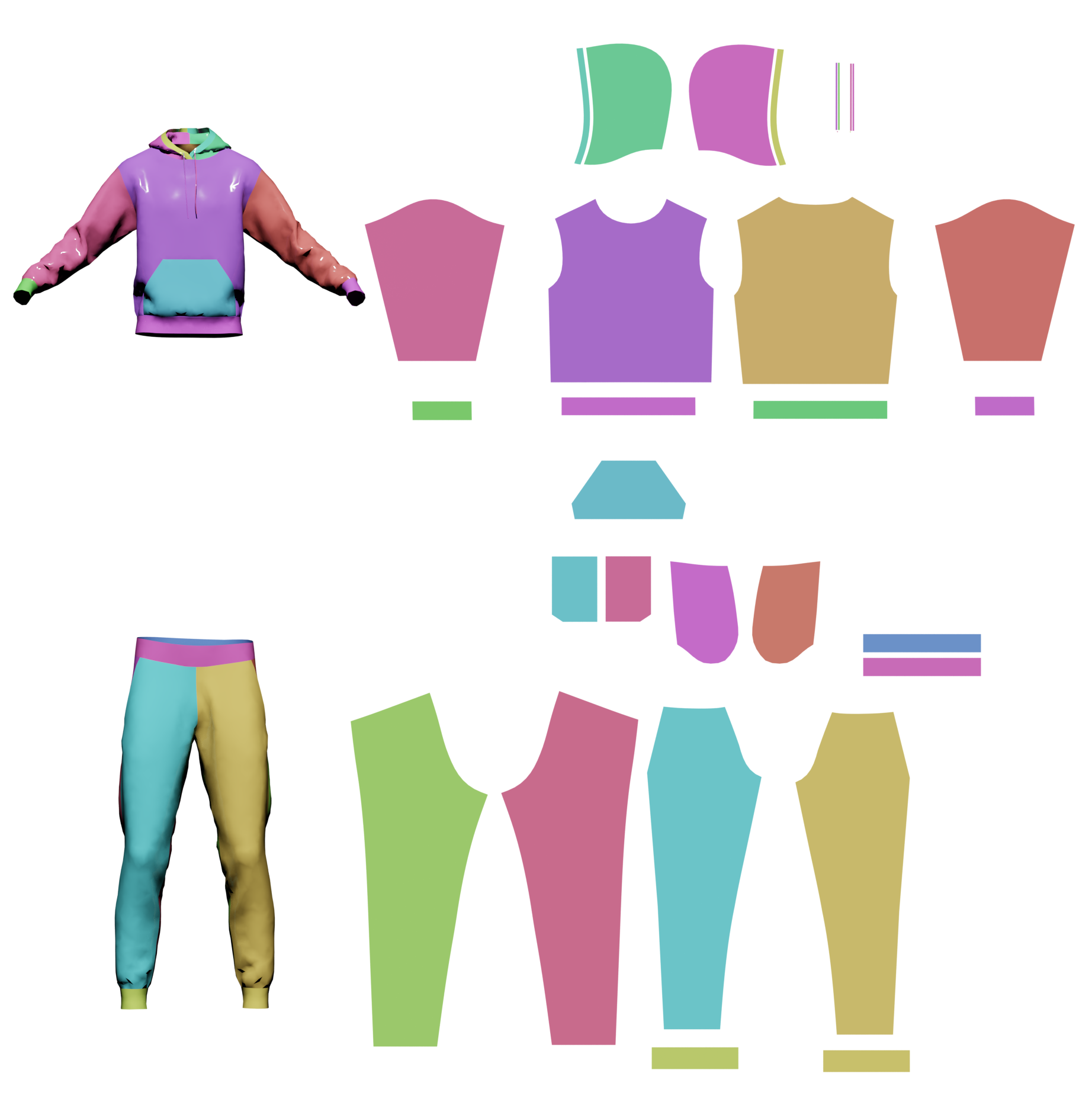}
     &
    \includegraphics[width=\imagewidth, align=c]{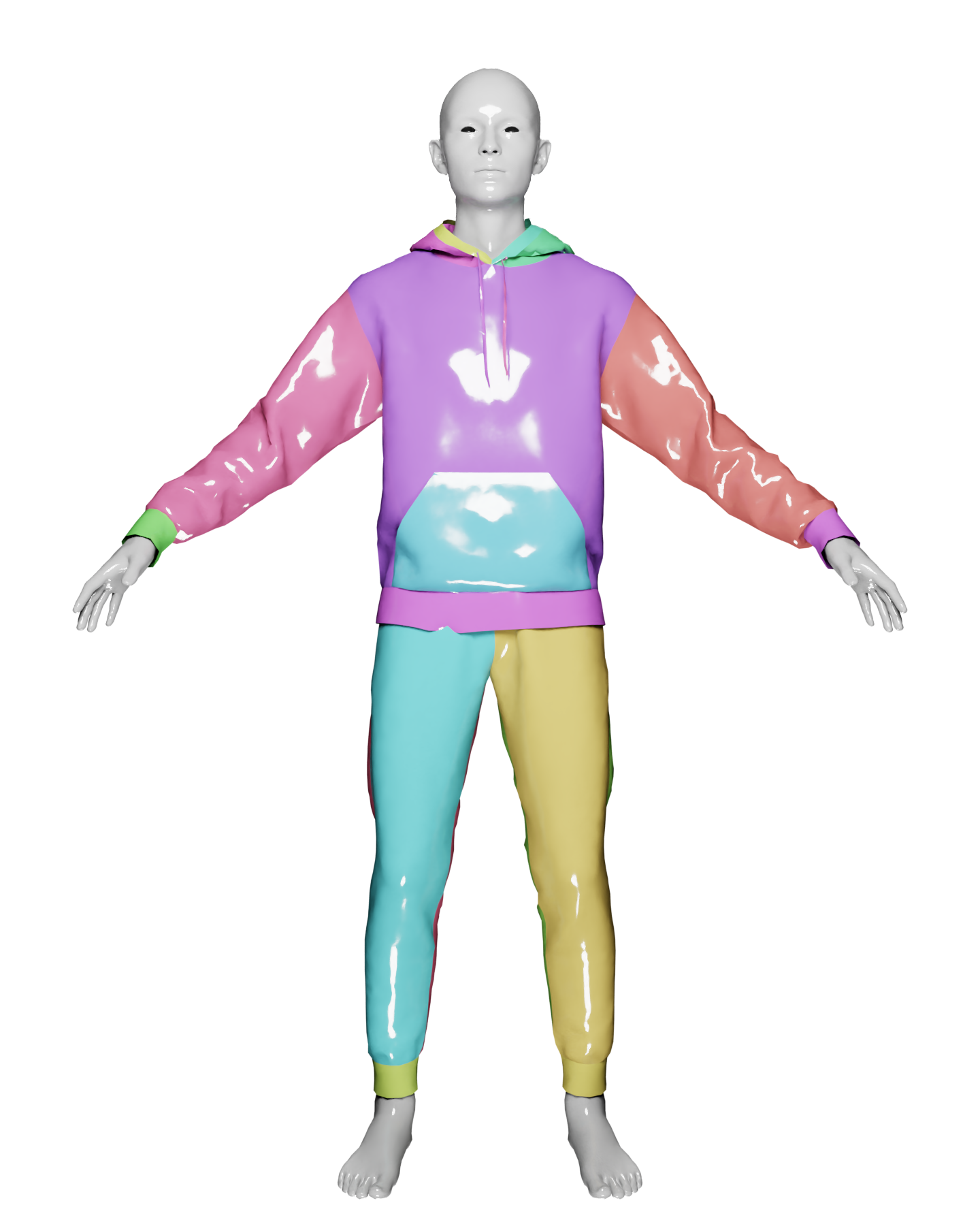}
     &
    \includegraphics[width=\imagewidth, align=c]{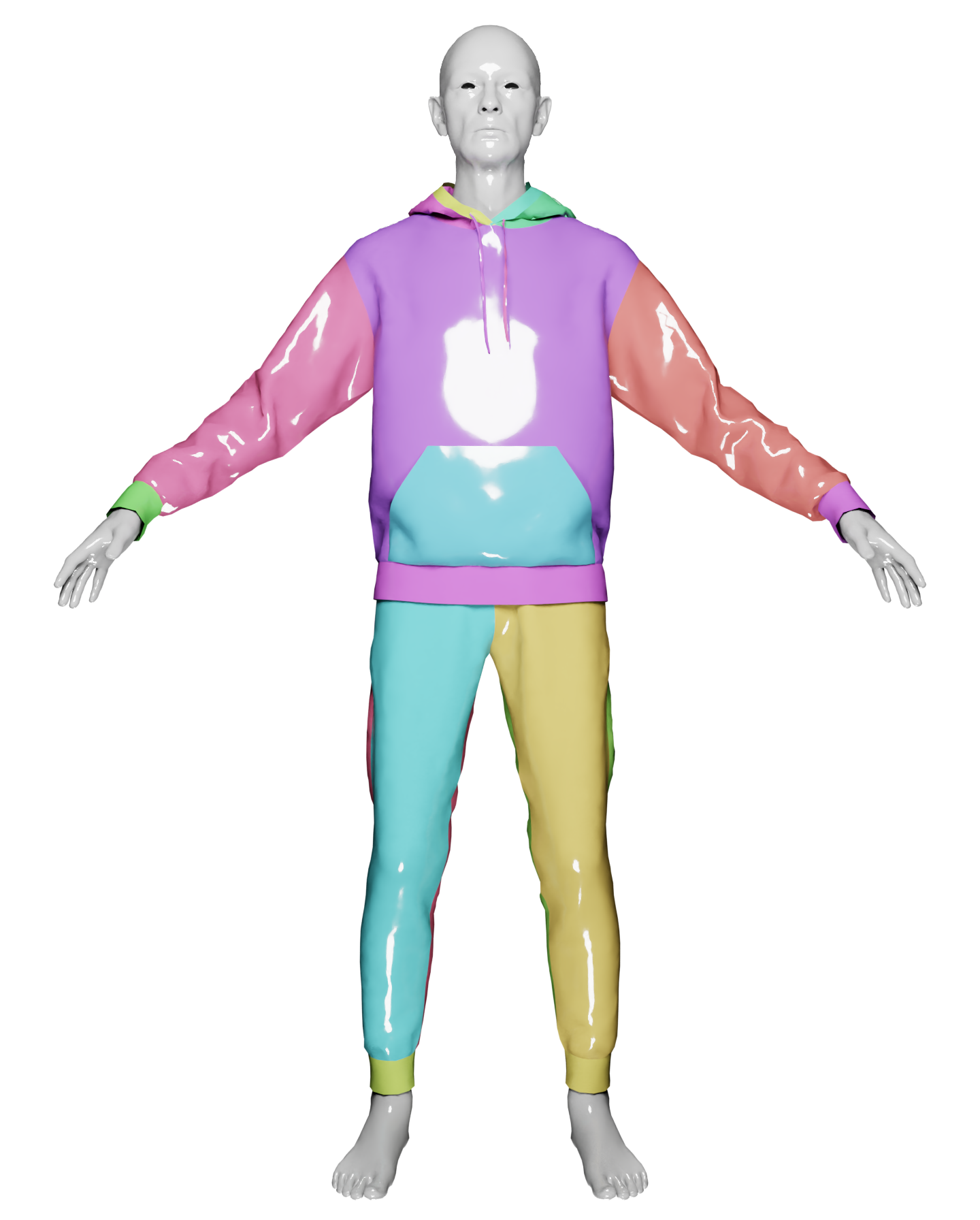}
     &
    \includegraphics[width=\imagewidth, align=c]{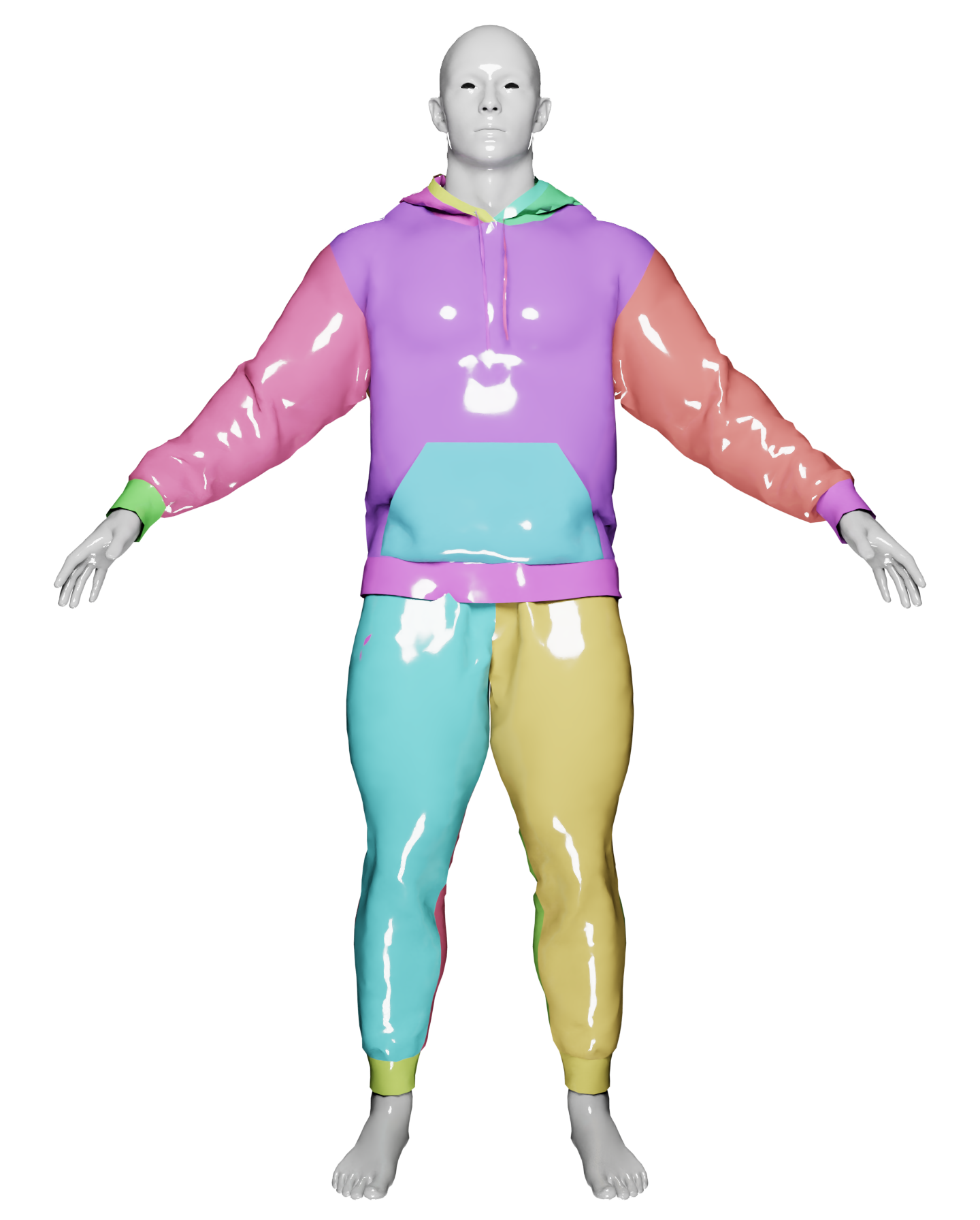}
     &
    \includegraphics[width=\imagewidth, align=c]{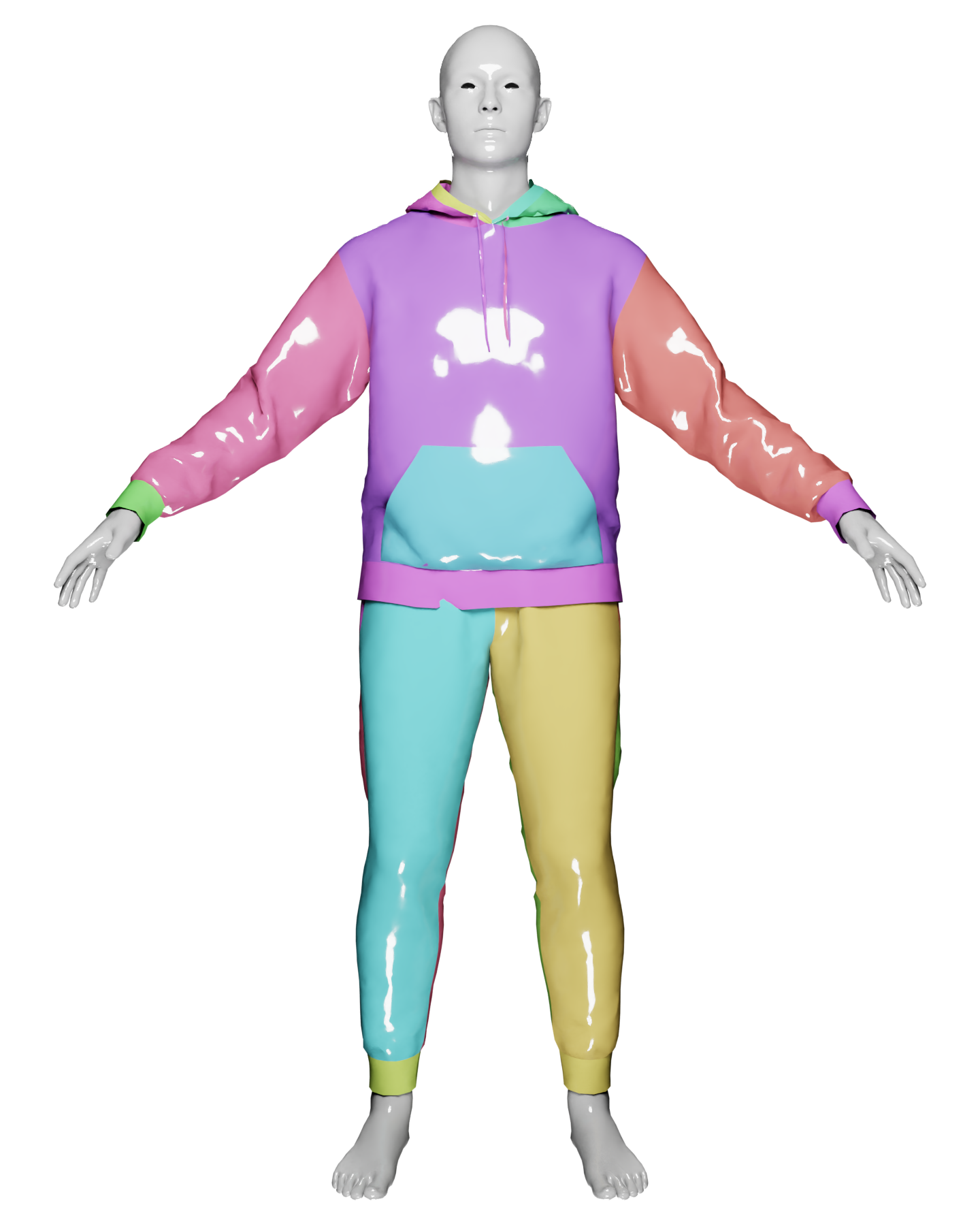}
     &
    \includegraphics[width=\imagewidth, align=c]{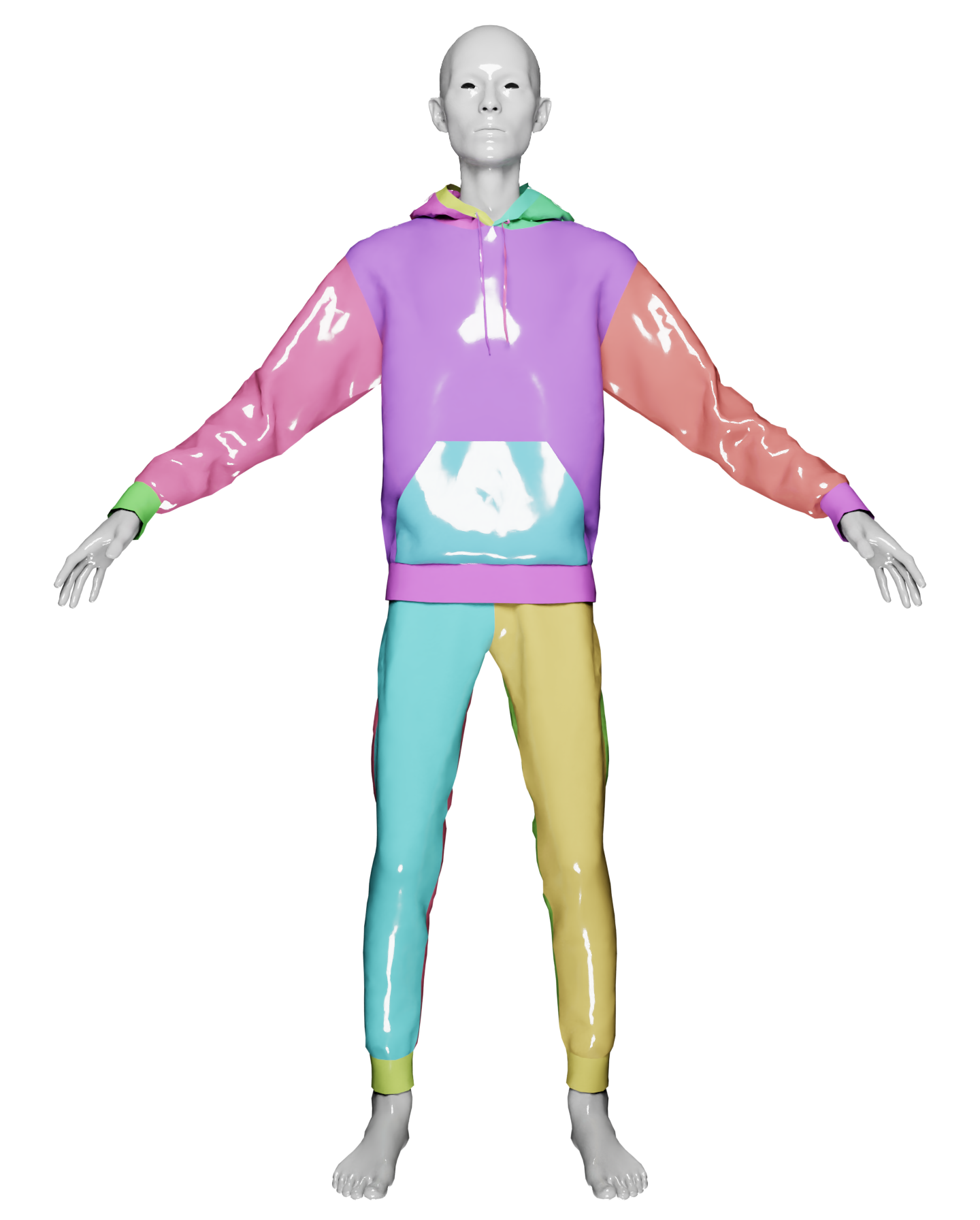}

    \\

    \includegraphics[width=\imagewidth, align=c]{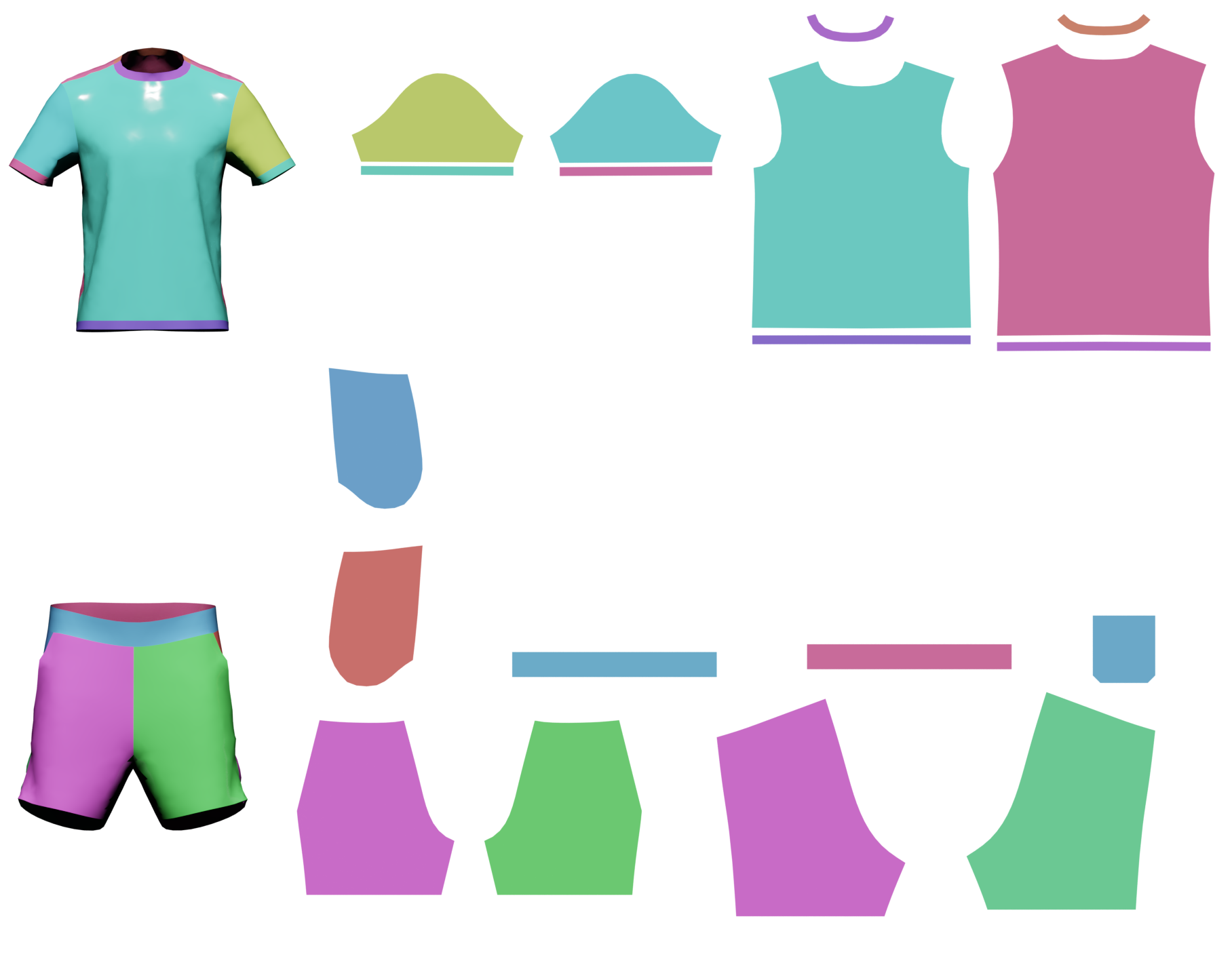}
     &
    \includegraphics[width=\imagewidth, align=c]{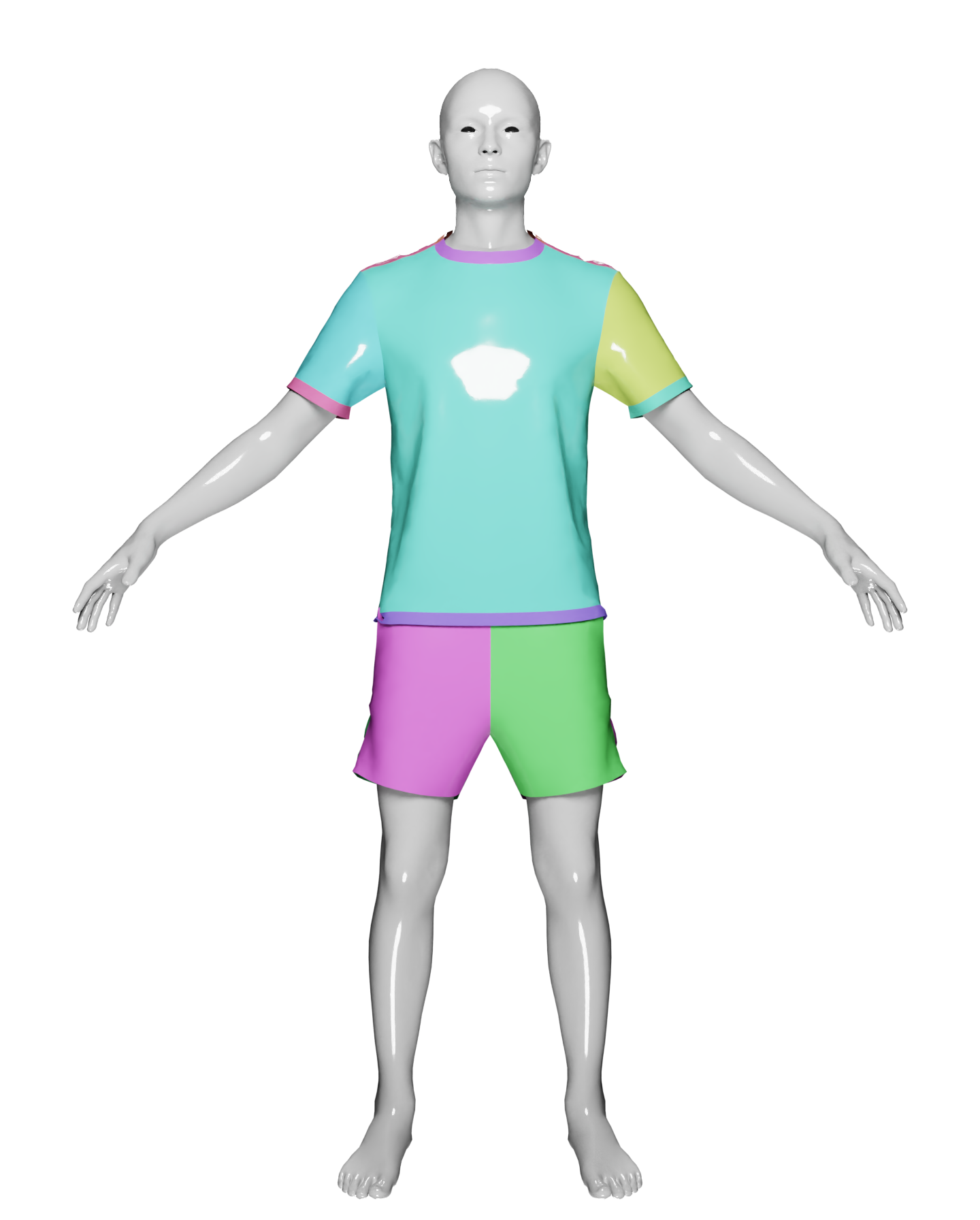}
     &
    \includegraphics[width=\imagewidth, align=c]{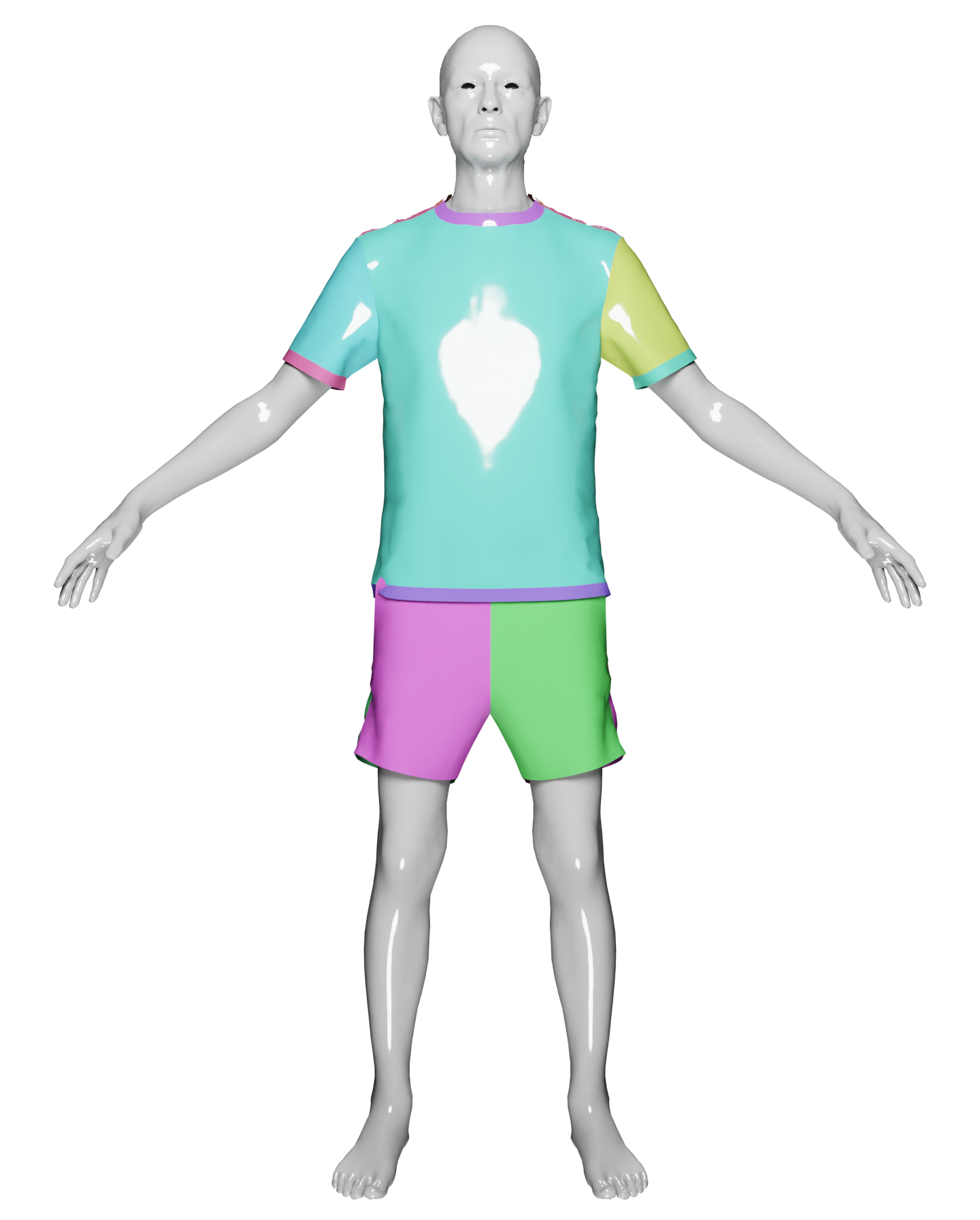}
     &
    \includegraphics[width=\imagewidth, align=c]{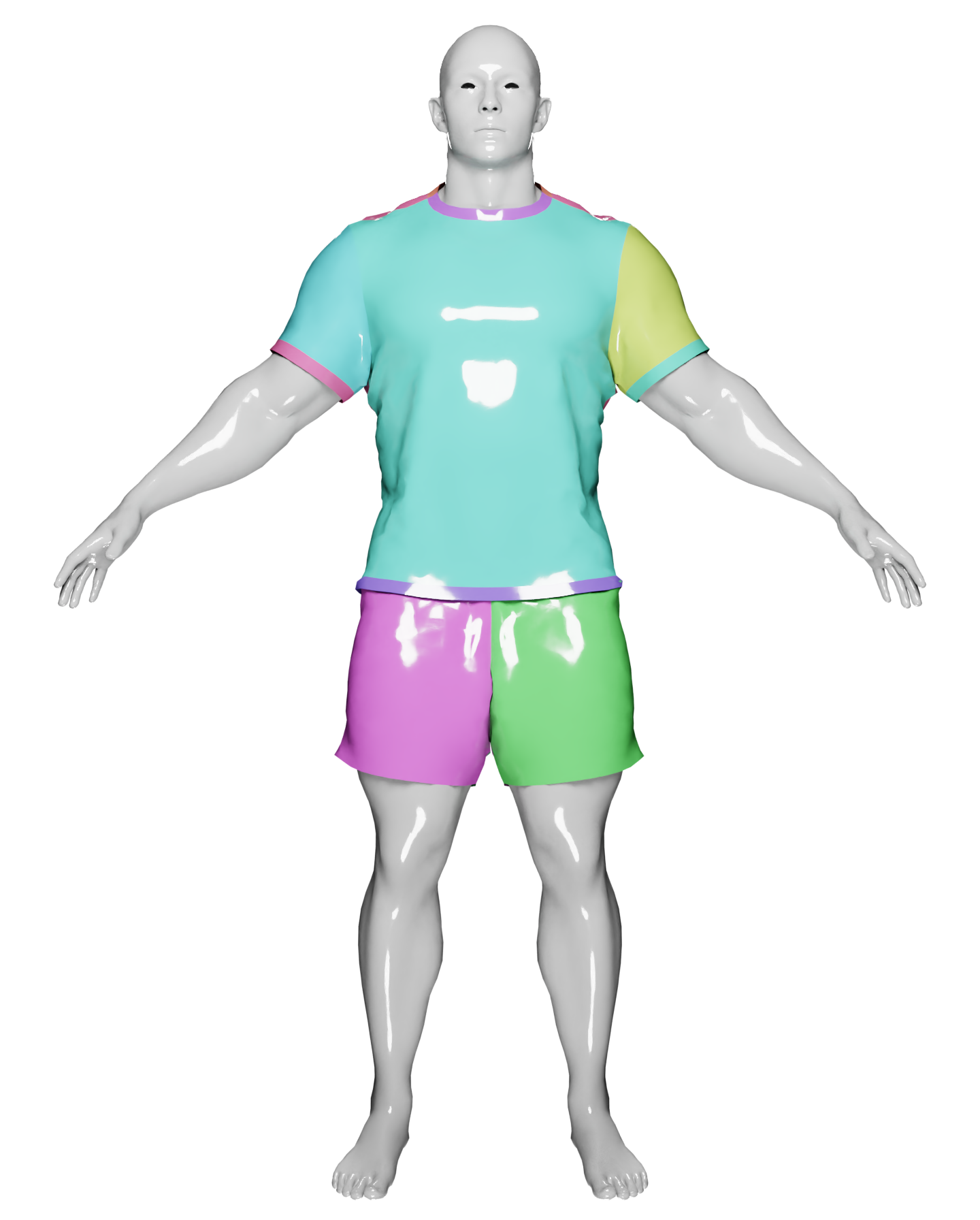}
     &
    \includegraphics[width=\imagewidth, align=c]{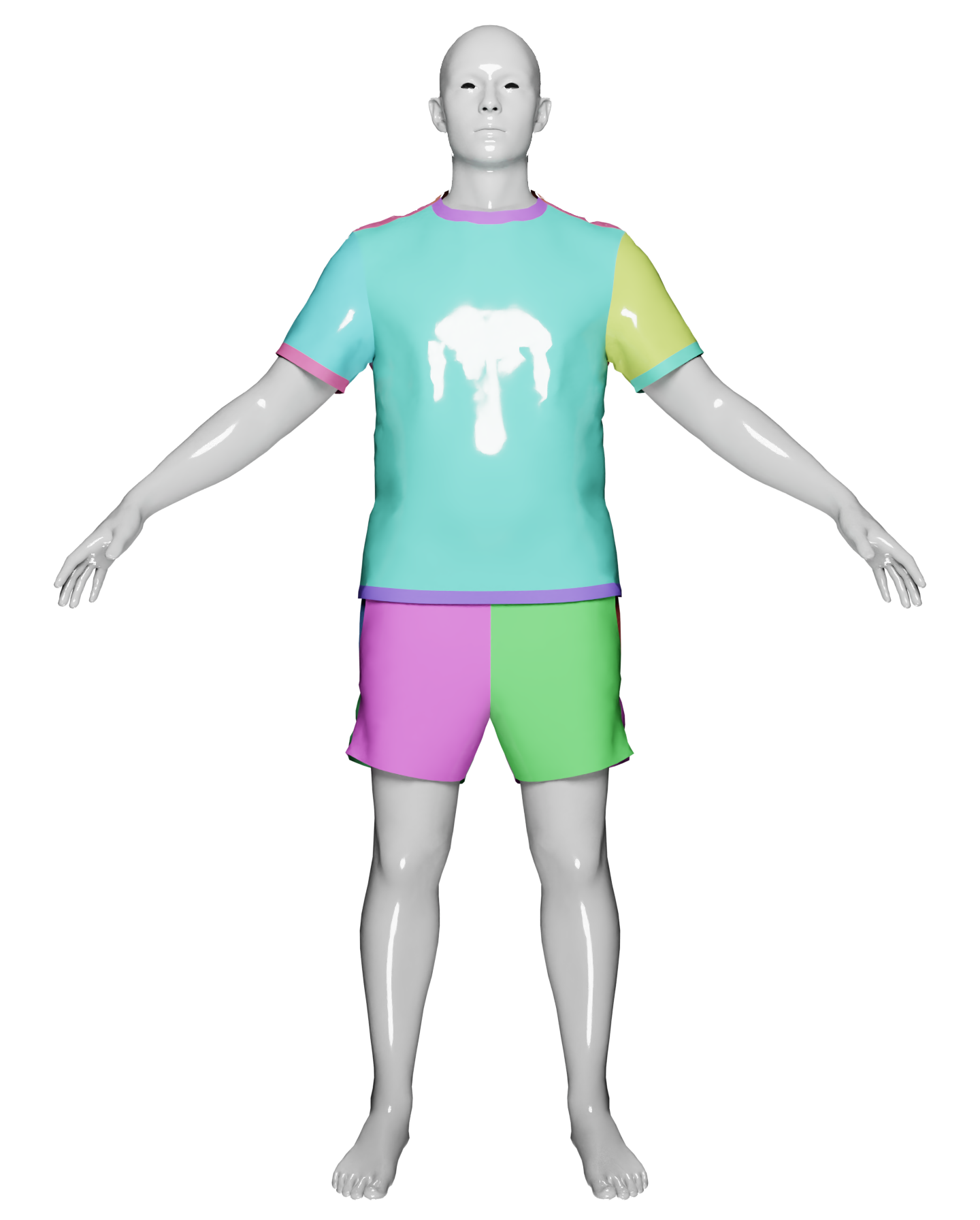}
     &
    \includegraphics[width=\imagewidth, align=c]{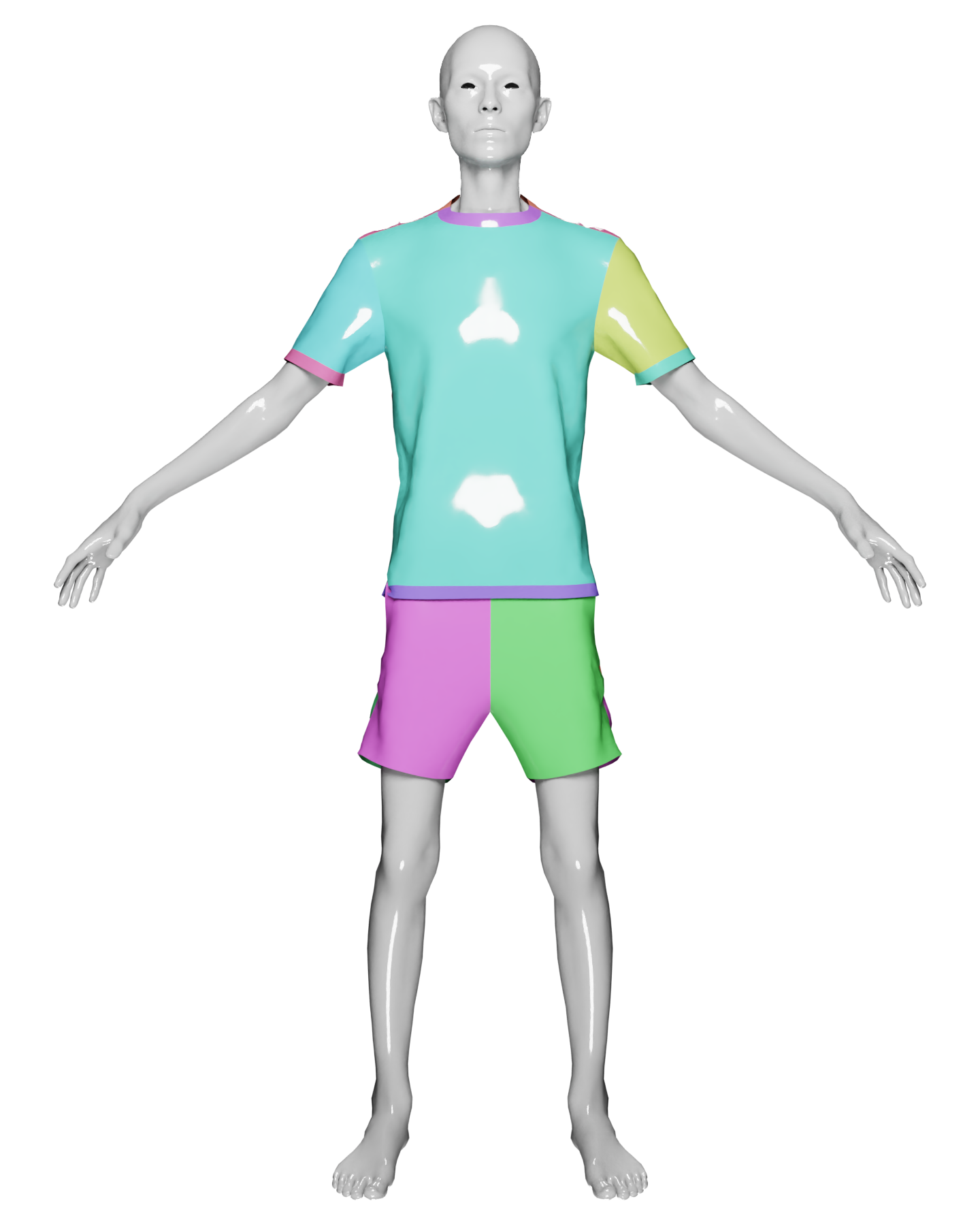}

    \\

    \includegraphics[width=\imagewidth, align=c]{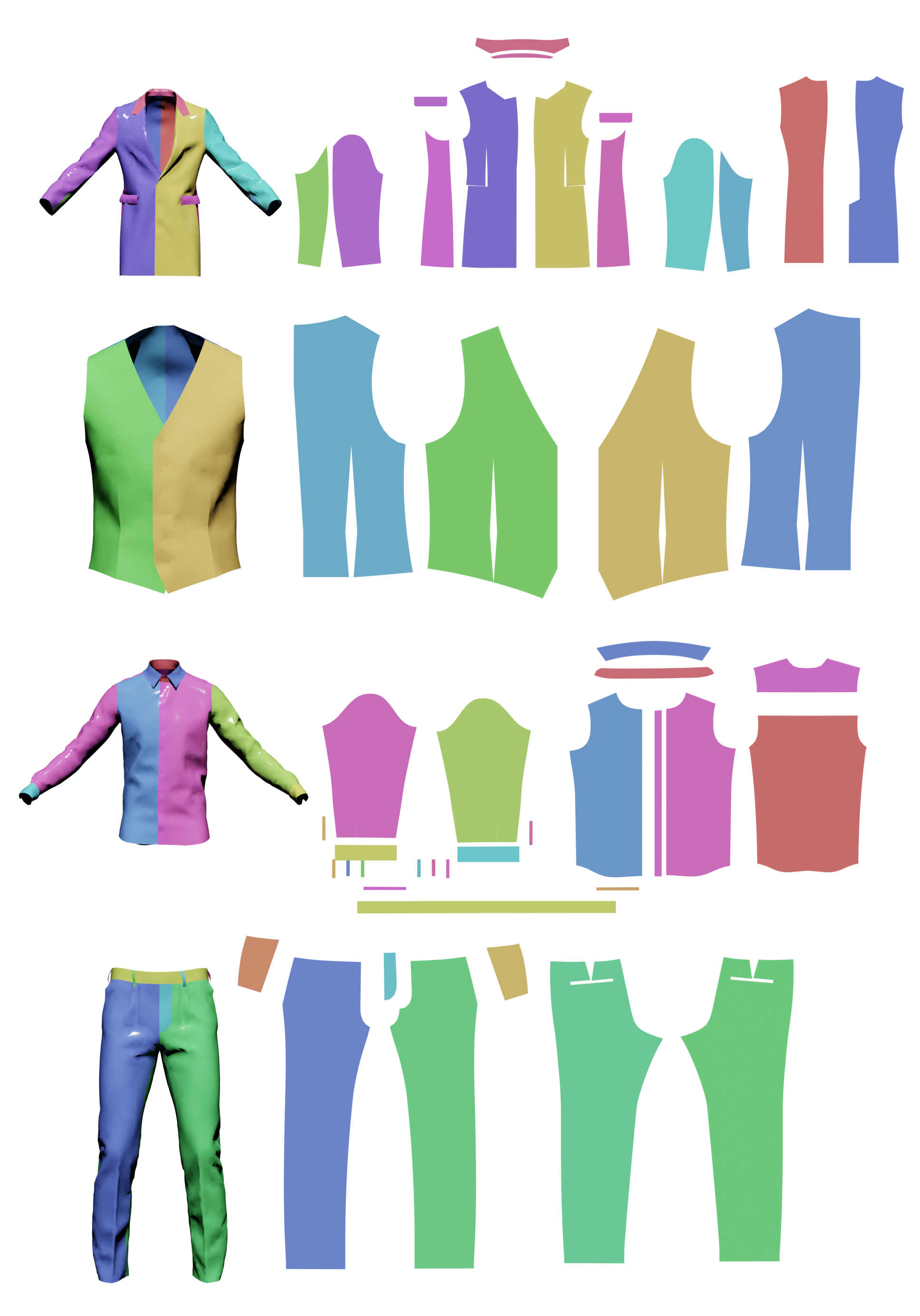}
     &
    \includegraphics[width=\imagewidth, align=c]{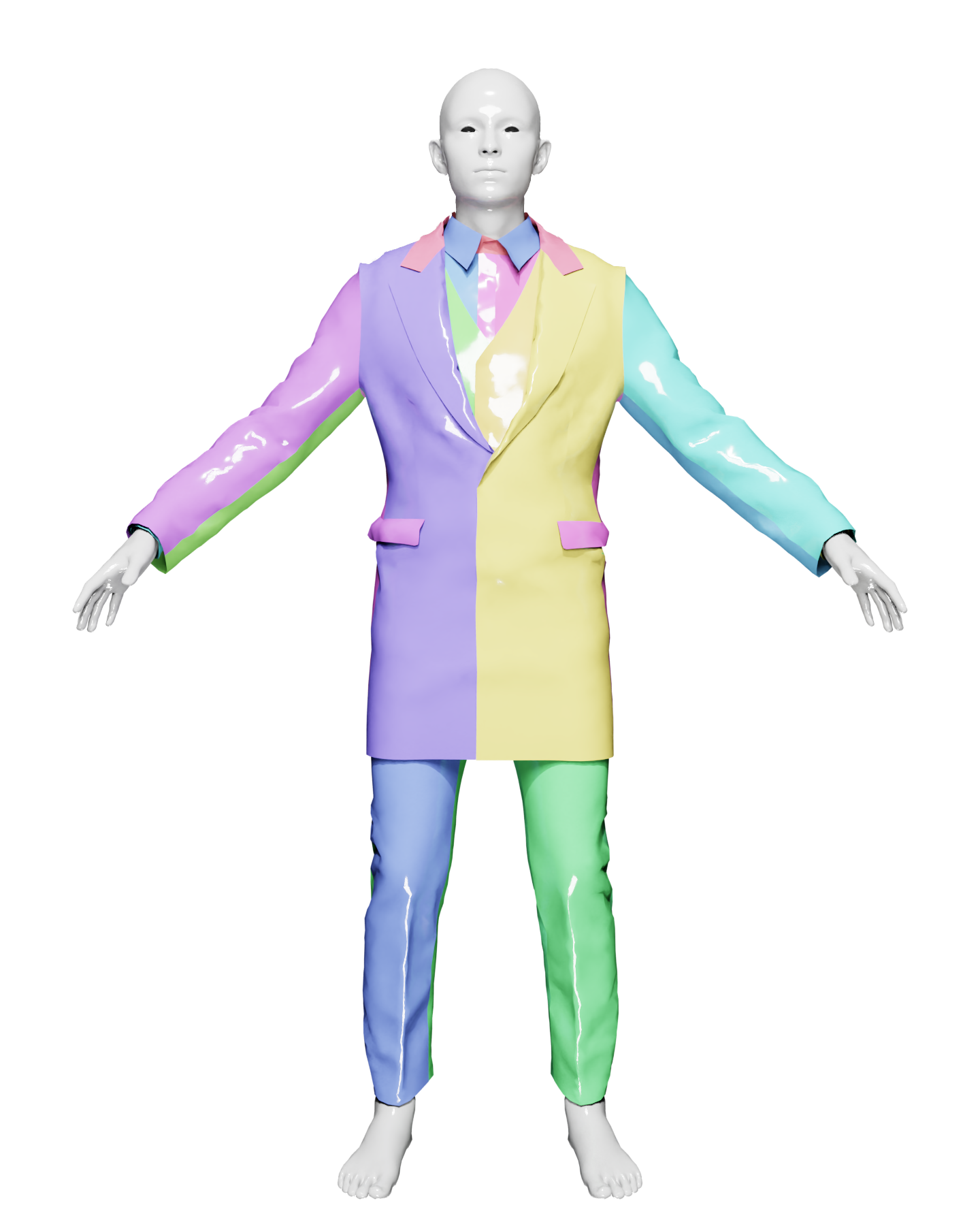}
     &
    \includegraphics[width=\imagewidth, align=c]{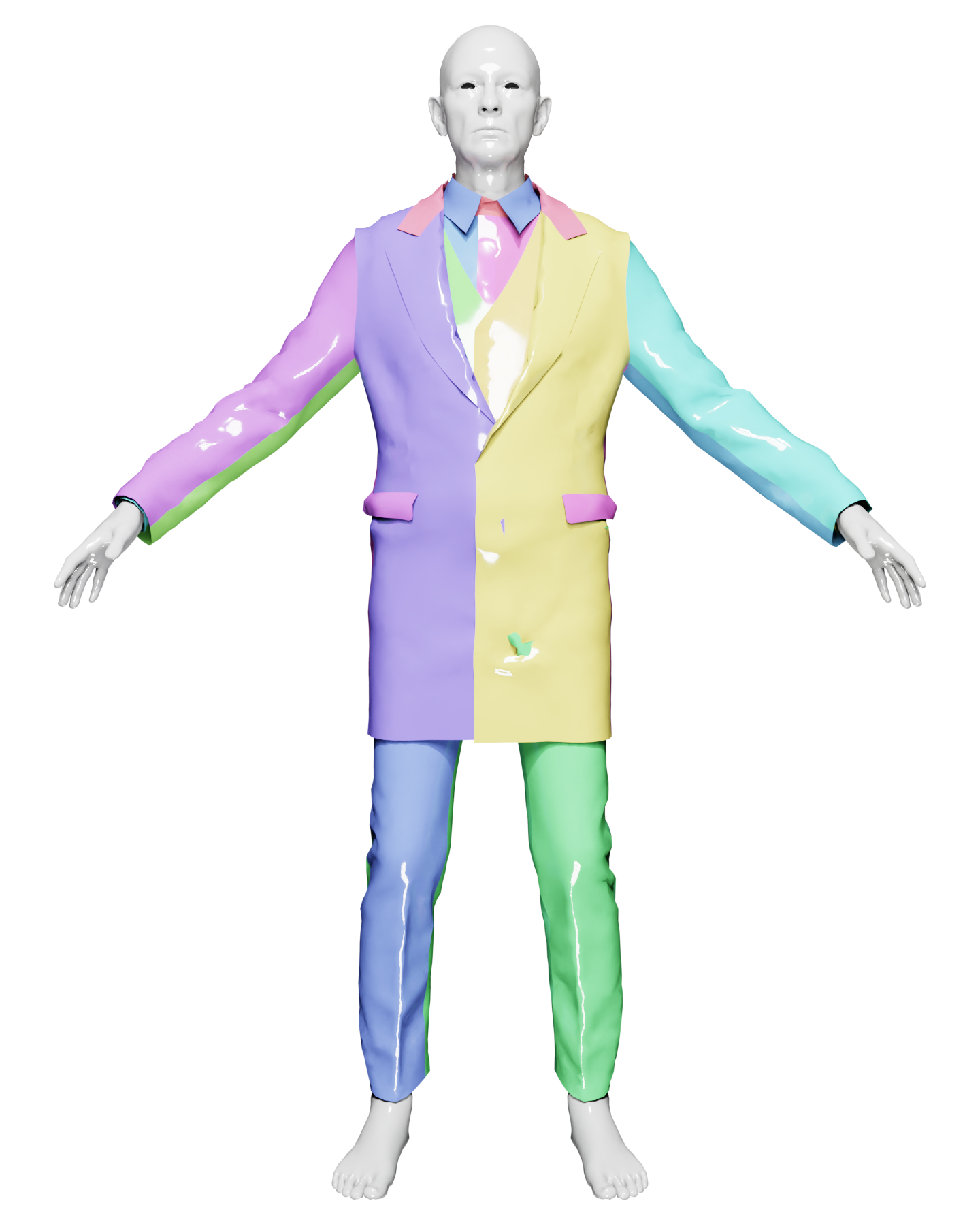}
     &
    \includegraphics[width=\imagewidth, align=c]{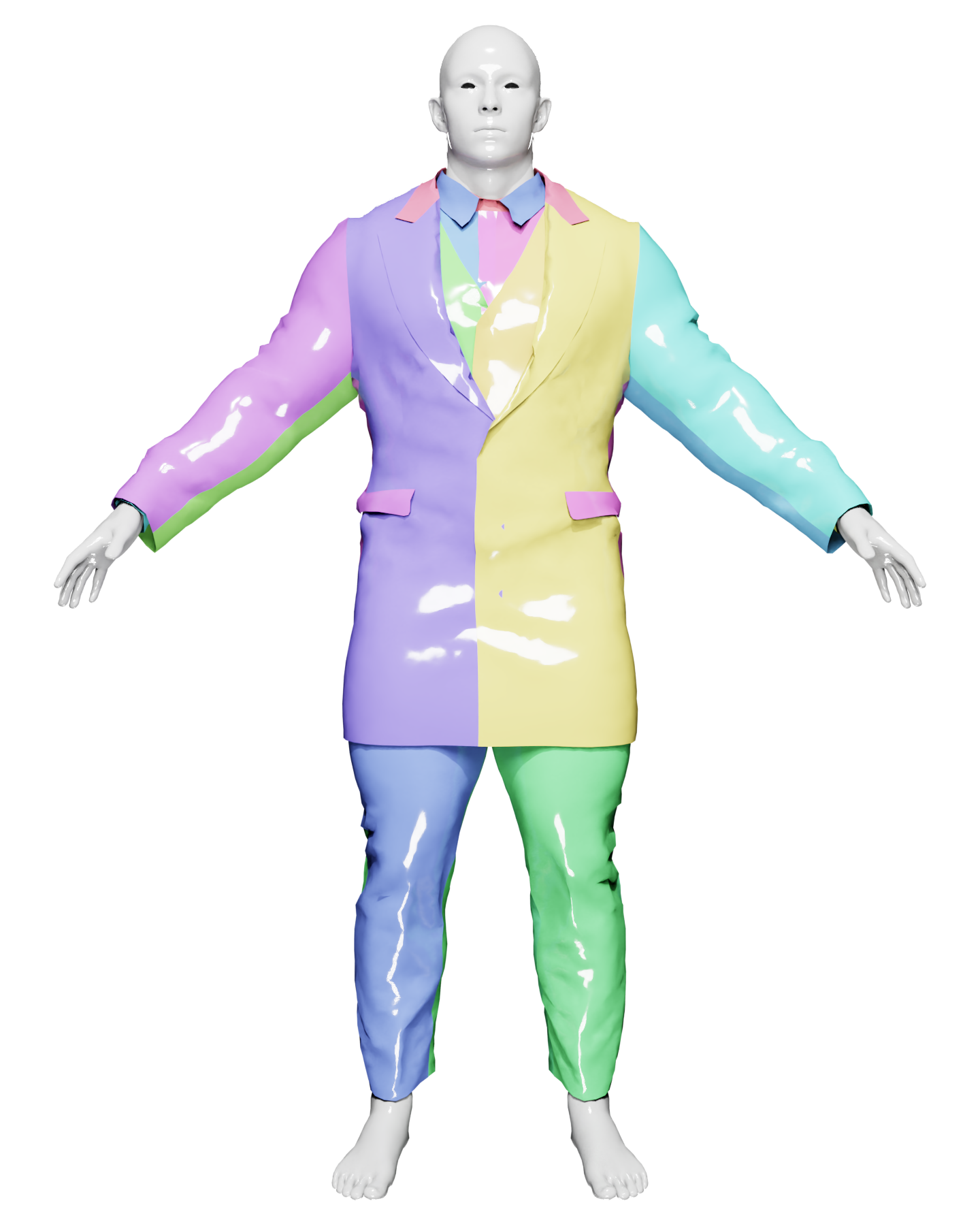}
     &
    \includegraphics[width=\imagewidth, align=c]{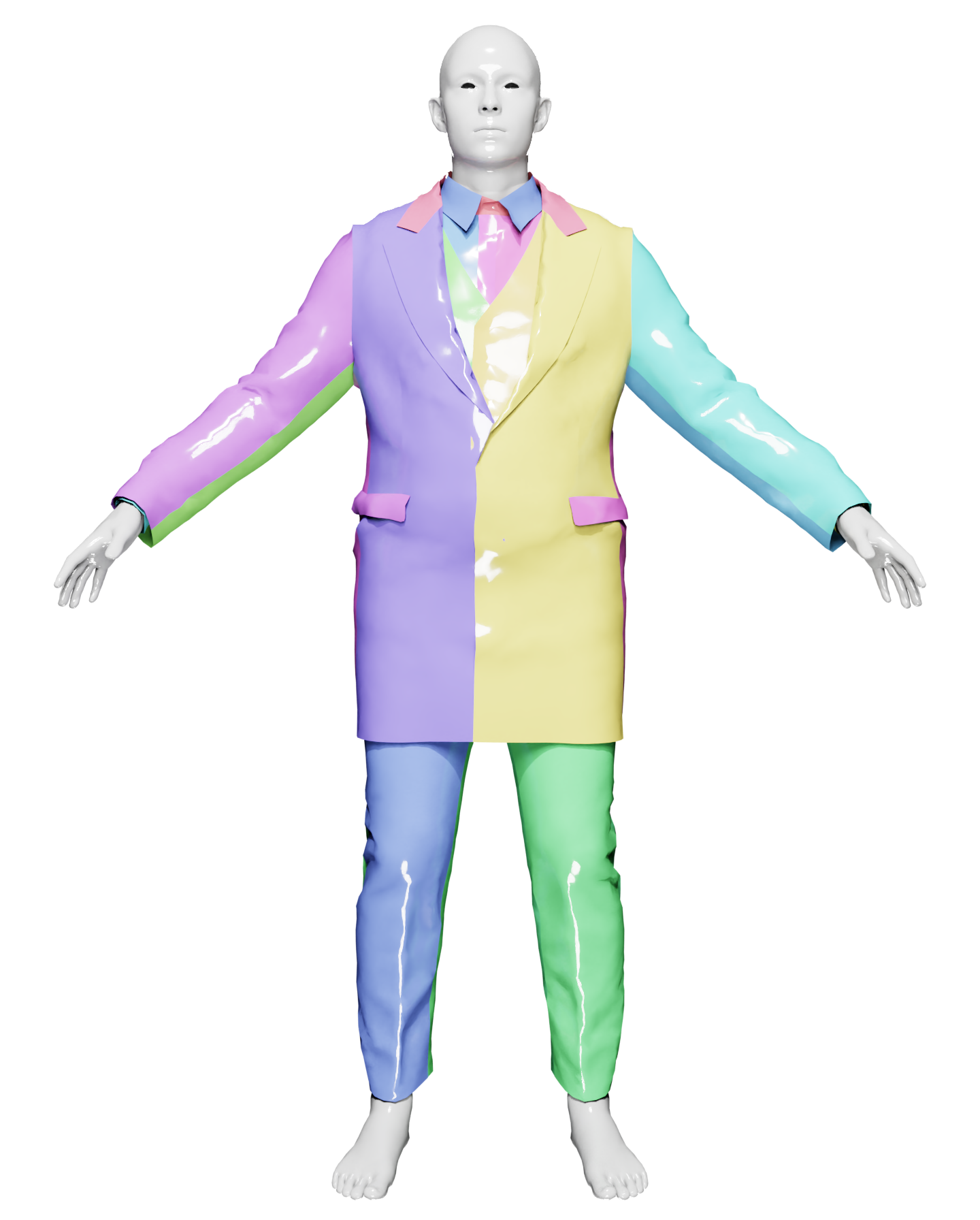}
     &
    \includegraphics[width=\imagewidth, align=c]{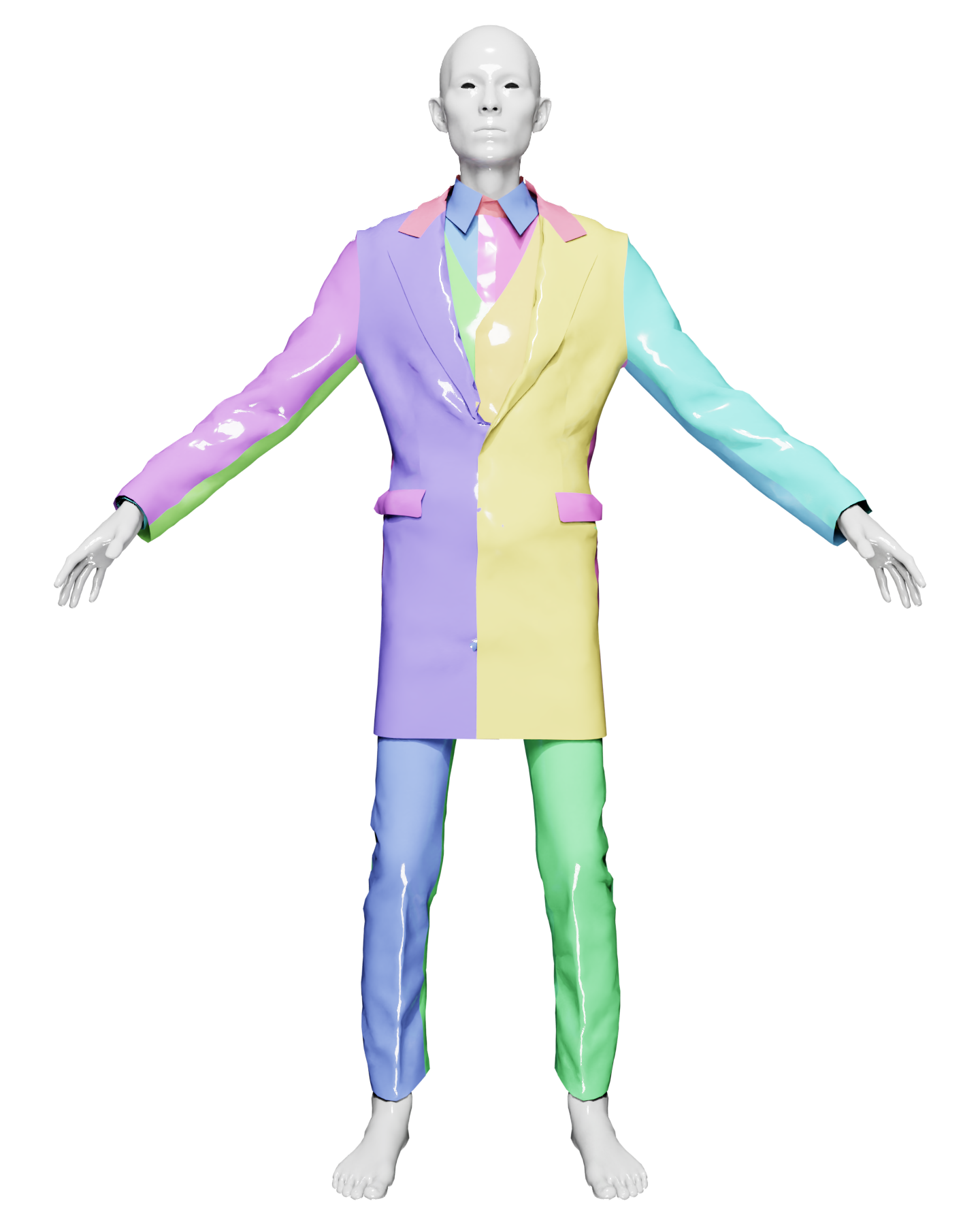}

    \\

  \end{tblr}
  \caption{We demonstrate the end result of refitting and draping outfits composed of diverse garments onto a variety of male characters using the Bolt pipeline.}
  \label{tab:male_outfits_grid}
\end{table*}

\def\bodyA{char_female_age16_body_mid_model_published}
\def\bodyB{char_female_age65_body_mid_model_published}
\def\bodyC{char_female_bodybuilder_body_mid_model_published}
\def\bodyD{char_female_obesityClassI_body_mid_model_published}
\def\bodyE{char_female_atrophied_body_mid_model_published}

\def\outfitA{outfit_female_casual_0001}
\def\outfitB{outfit_female_formal_0001}
\def\outfitC{outfit_female_formal_0004}
\def\outfitD{outfit_unisex_casual_0007}

\begin{table*}
  \begin{tblr}{
      colspec = {X[c,m] | X[c,m] X[c,m] X[c,m] X[c,m] X[c,m]},
    }

     &
    \includegraphics[width=\imagewidth, align=c]{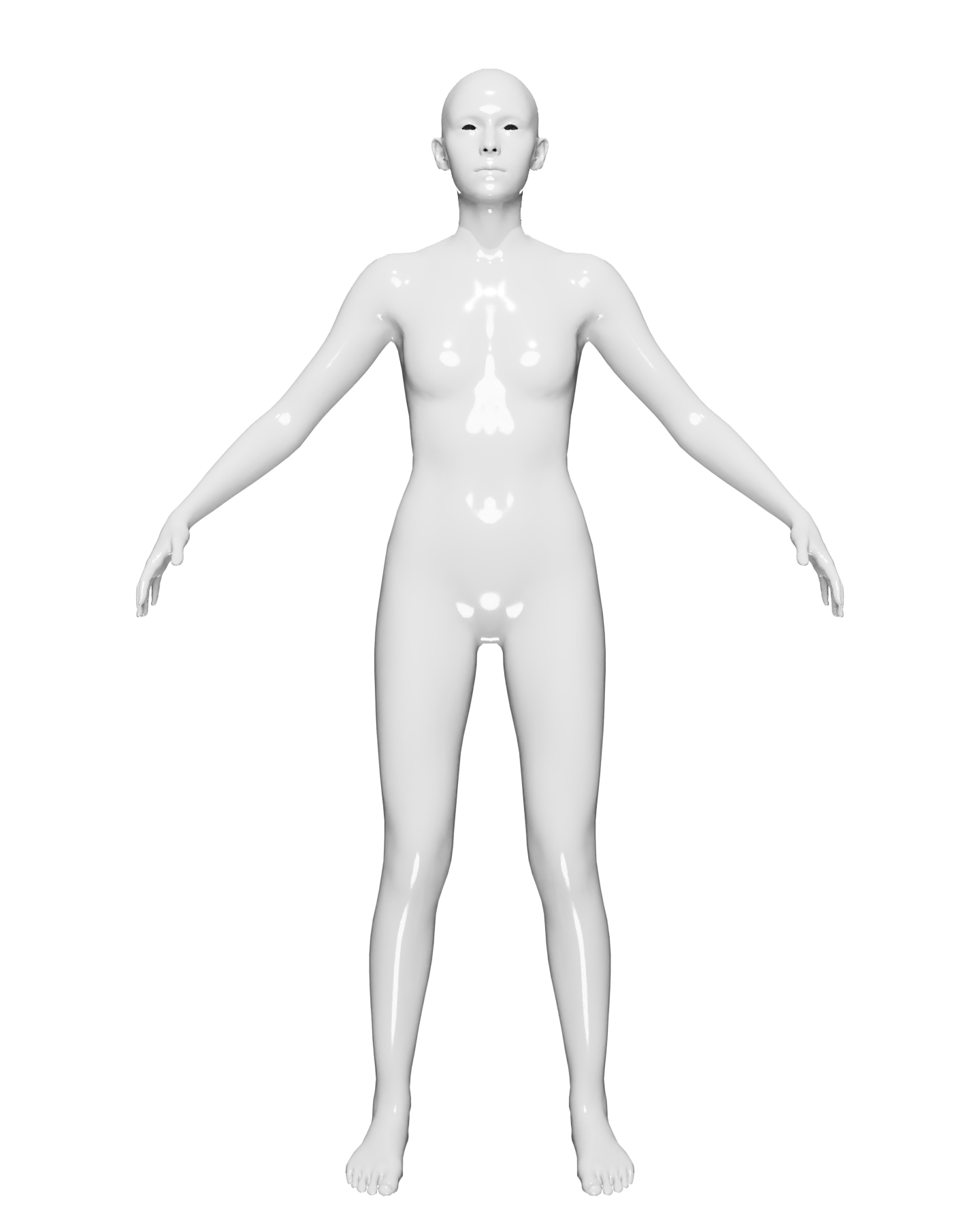}
     &
    \includegraphics[width=\imagewidth, align=c]{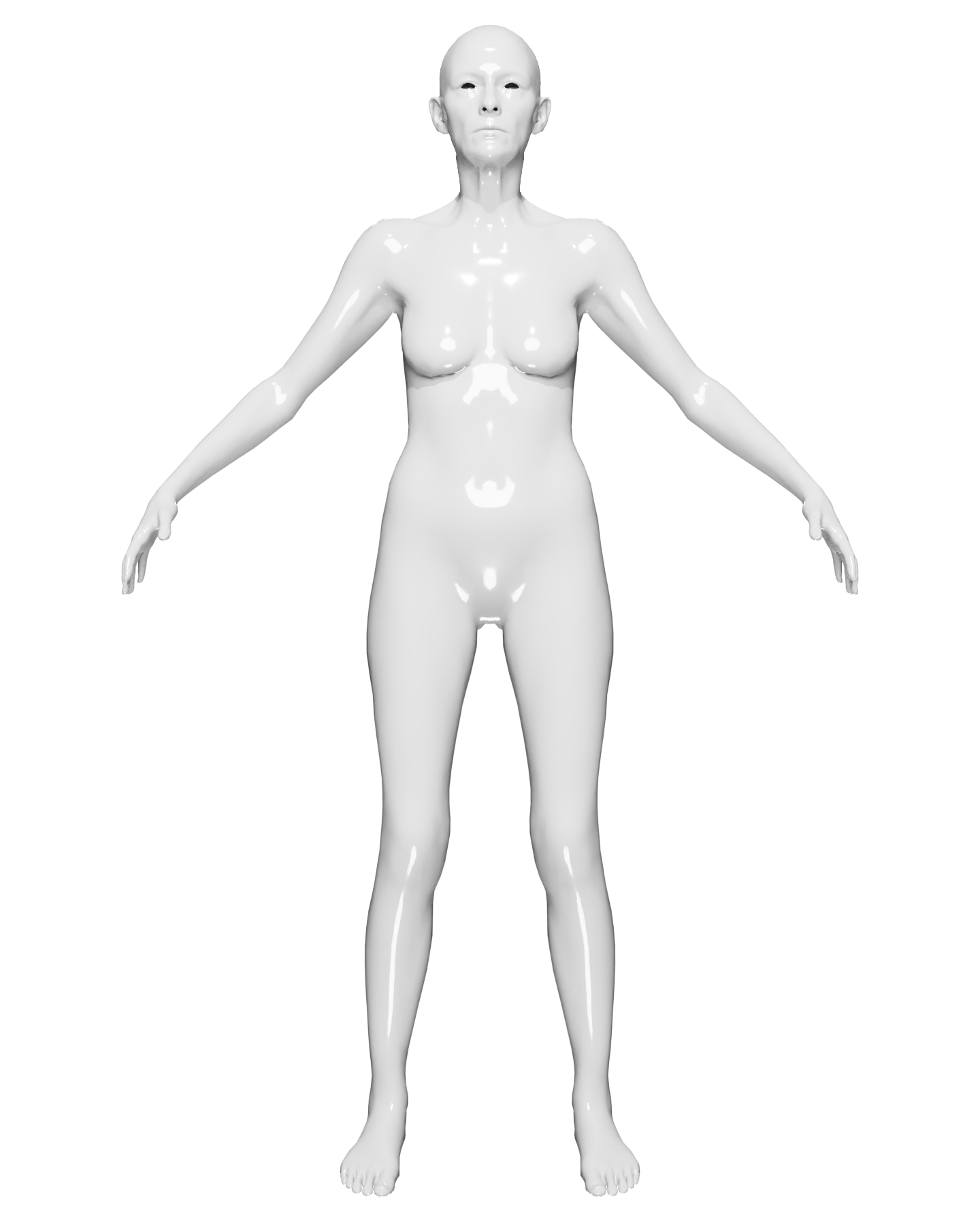}
     &
    \includegraphics[width=\imagewidth, align=c]{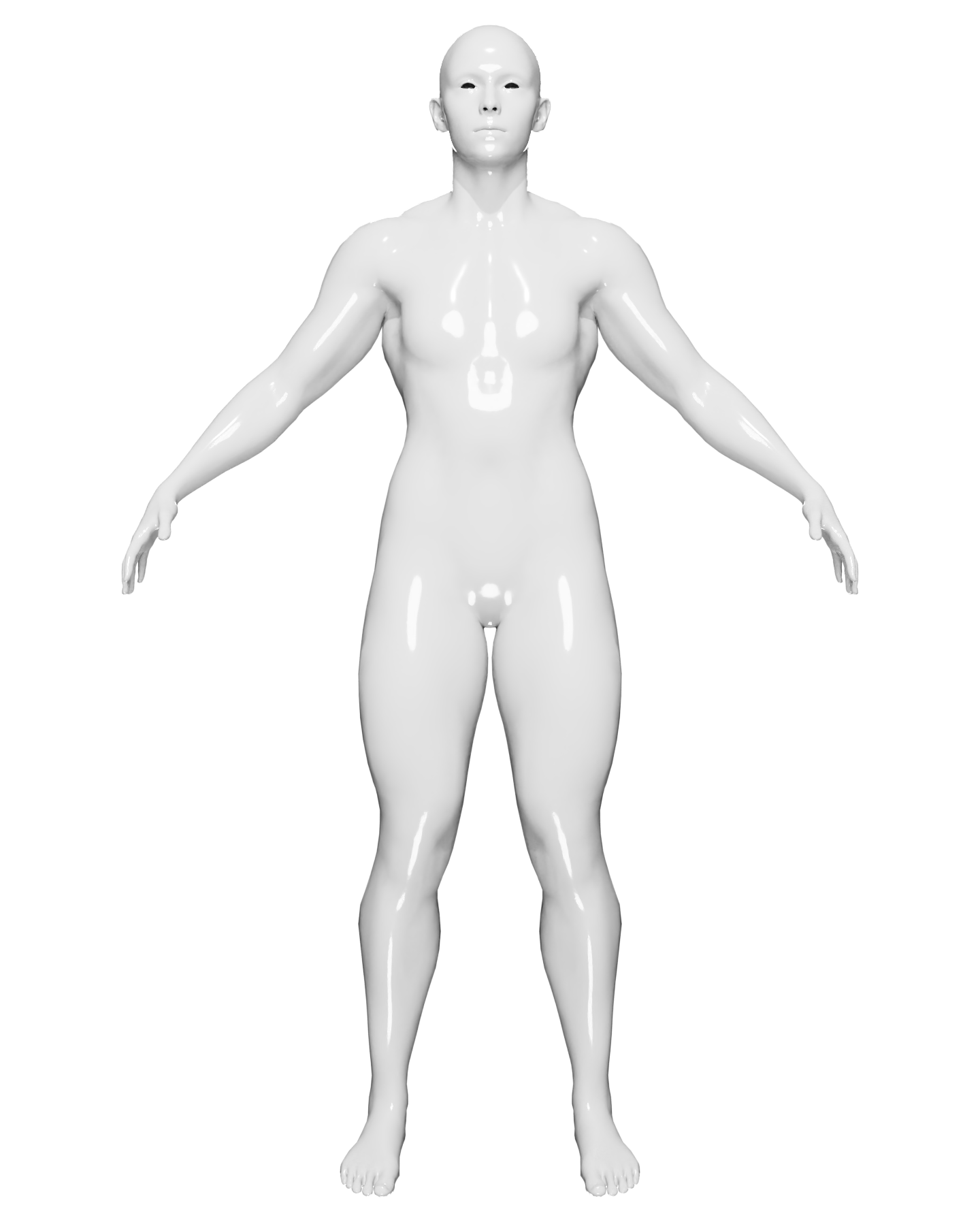}
     &
    \includegraphics[width=\imagewidth, align=c]{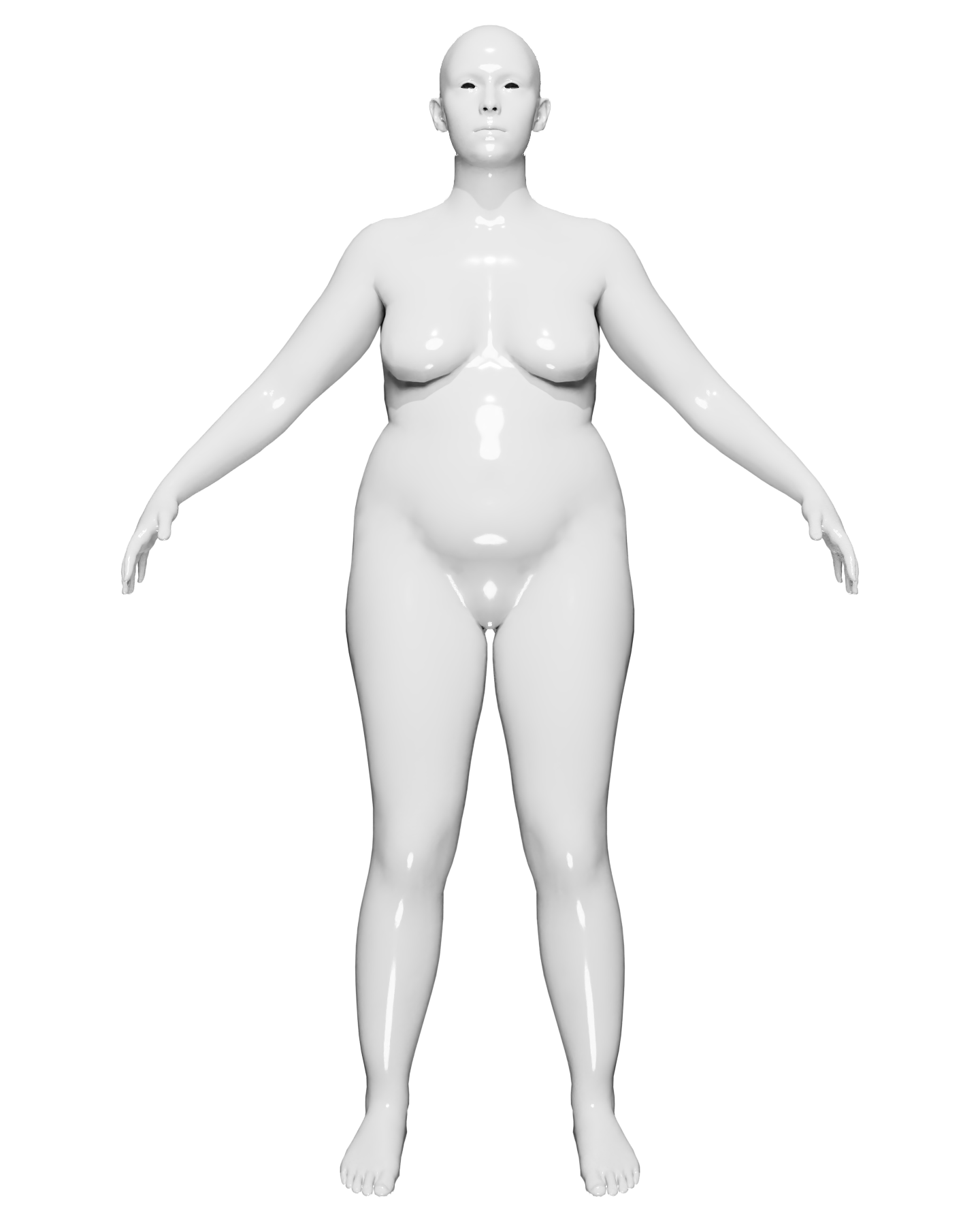}
     &
    \includegraphics[width=\imagewidth, align=c]{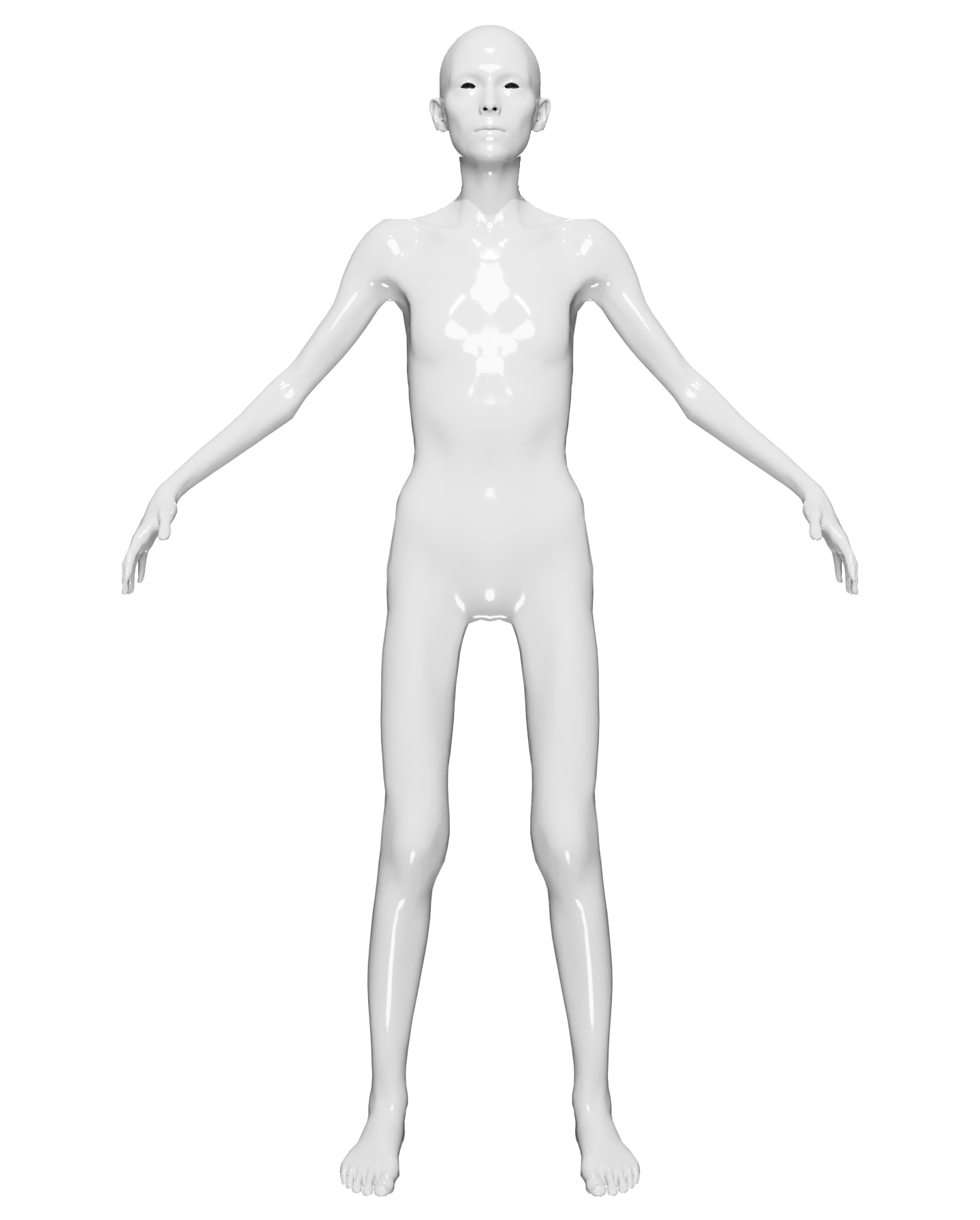}

    \\

    \hline

    \includegraphics[width=\imagewidth, align=c]{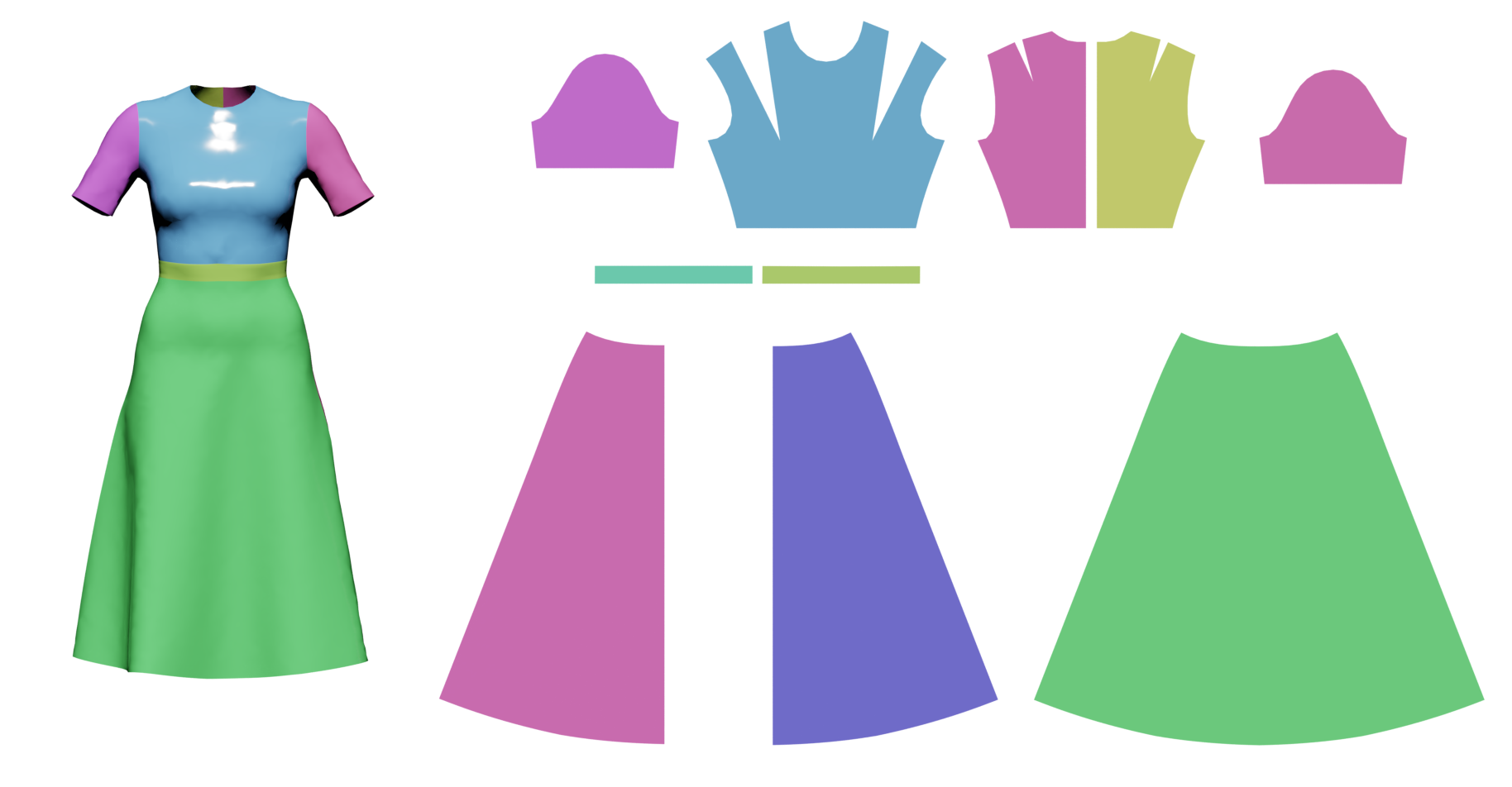}
     &
    \includegraphics[width=\imagewidth, align=c]{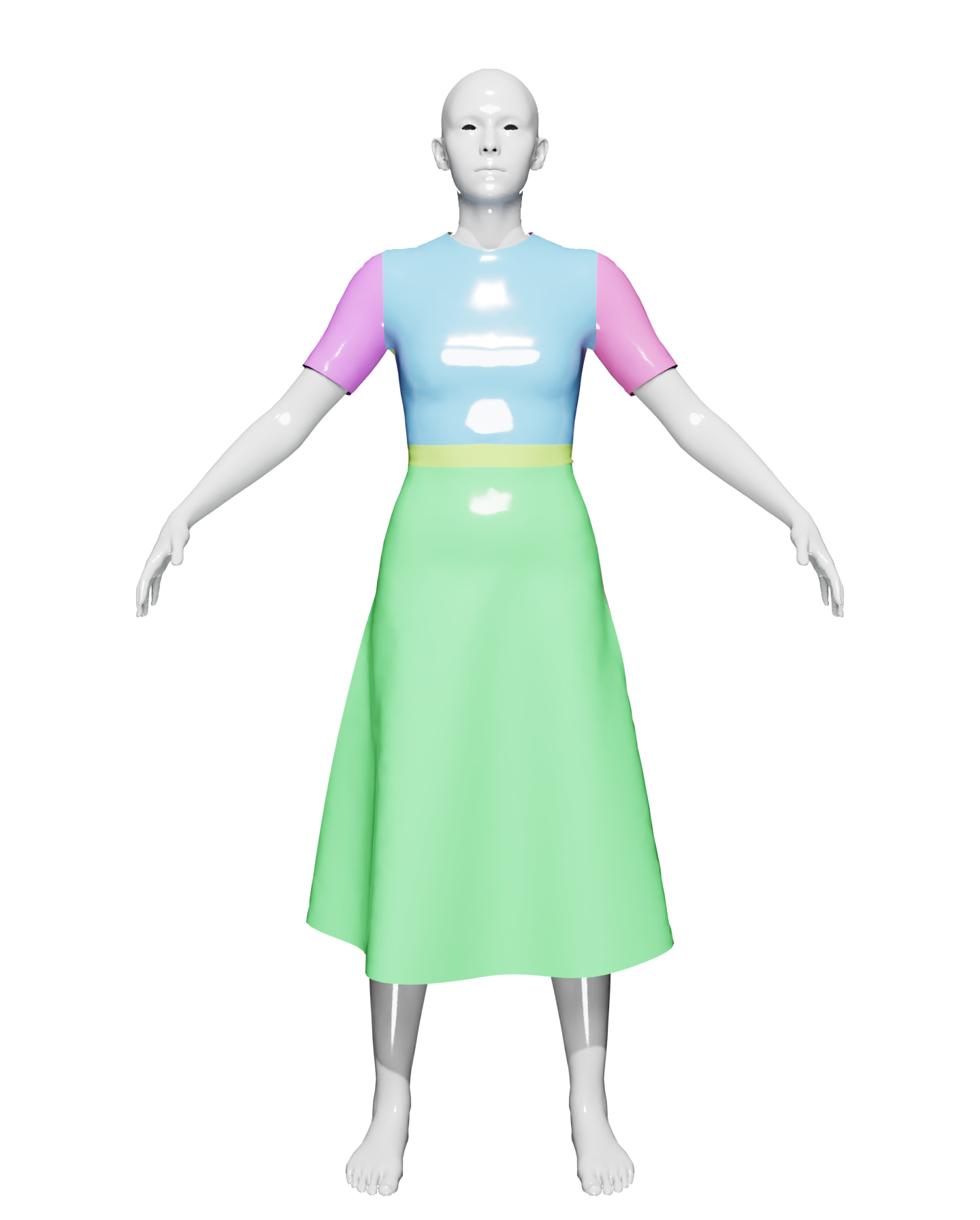}
     &
    \includegraphics[width=\imagewidth, align=c]{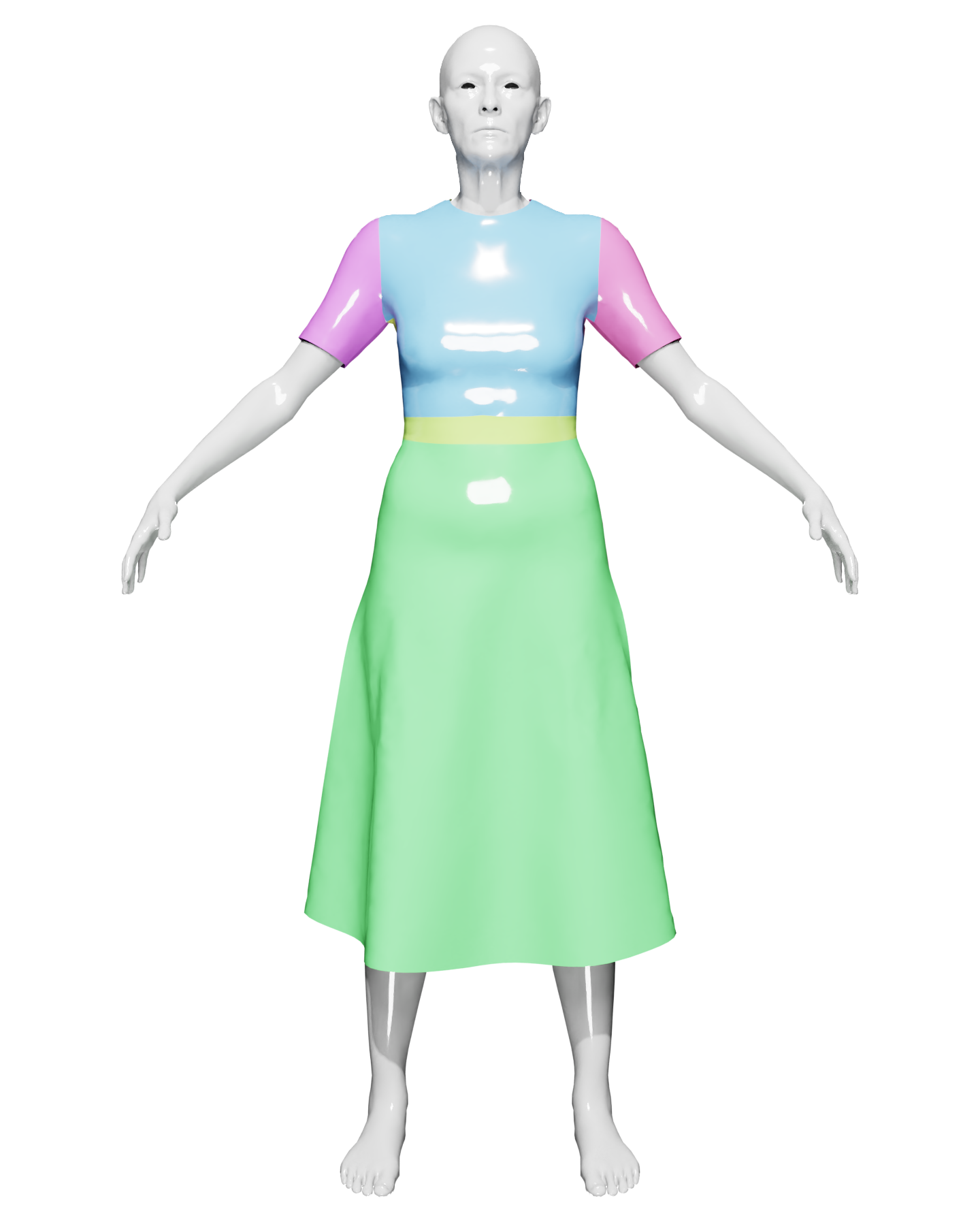}
     &
    \includegraphics[width=\imagewidth, align=c]{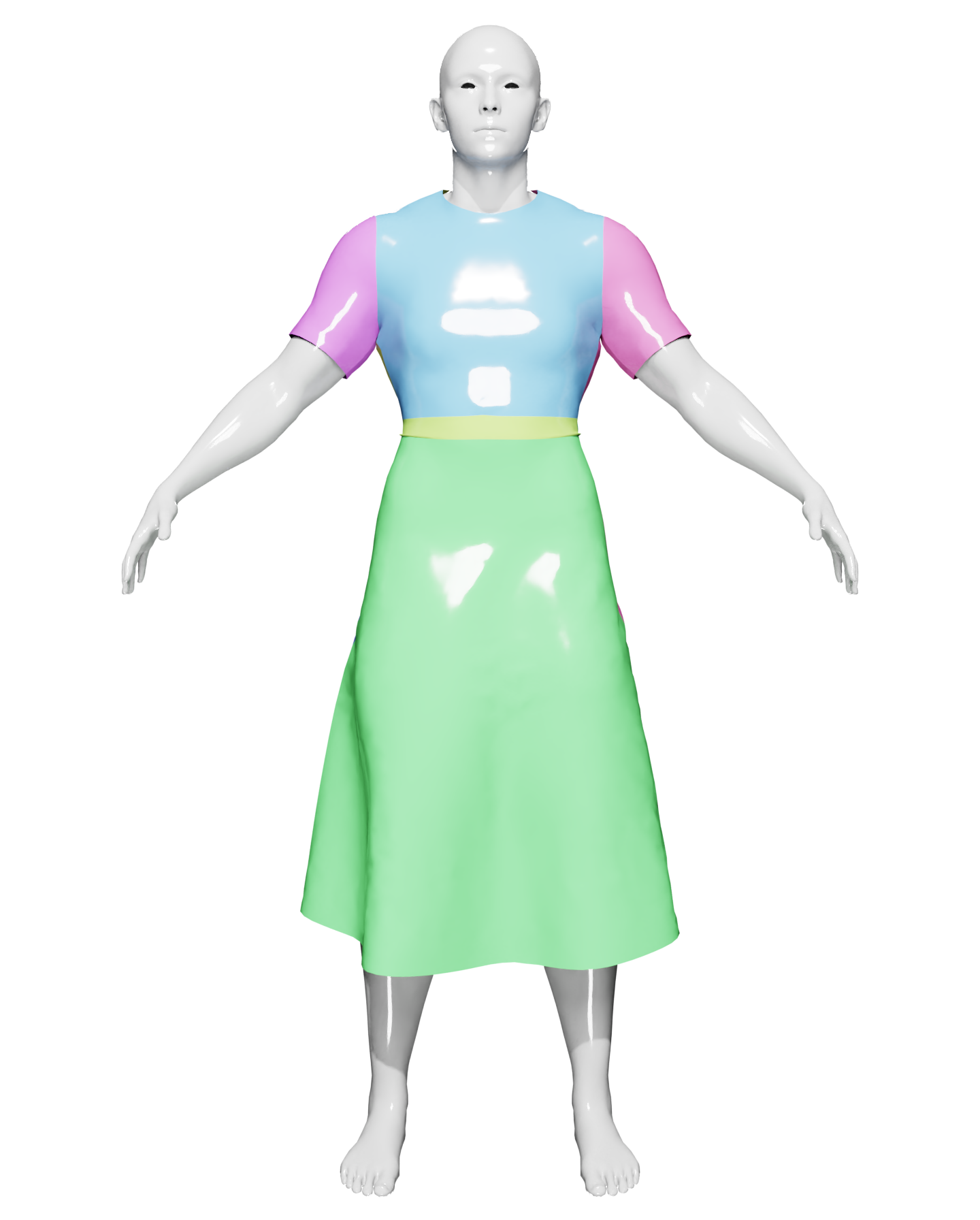}
     &
    \includegraphics[width=\imagewidth, align=c]{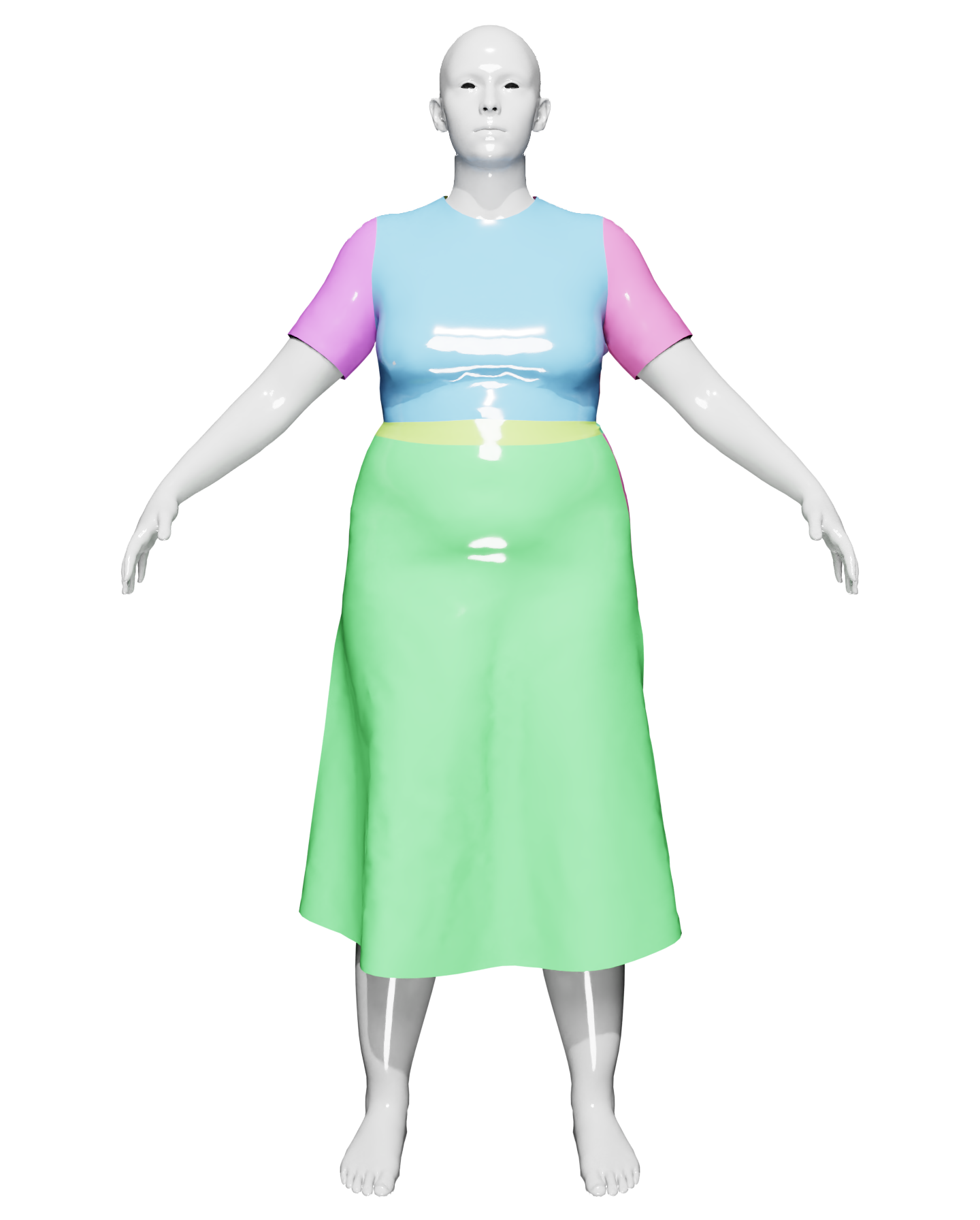}
     &
    \includegraphics[width=\imagewidth, align=c]{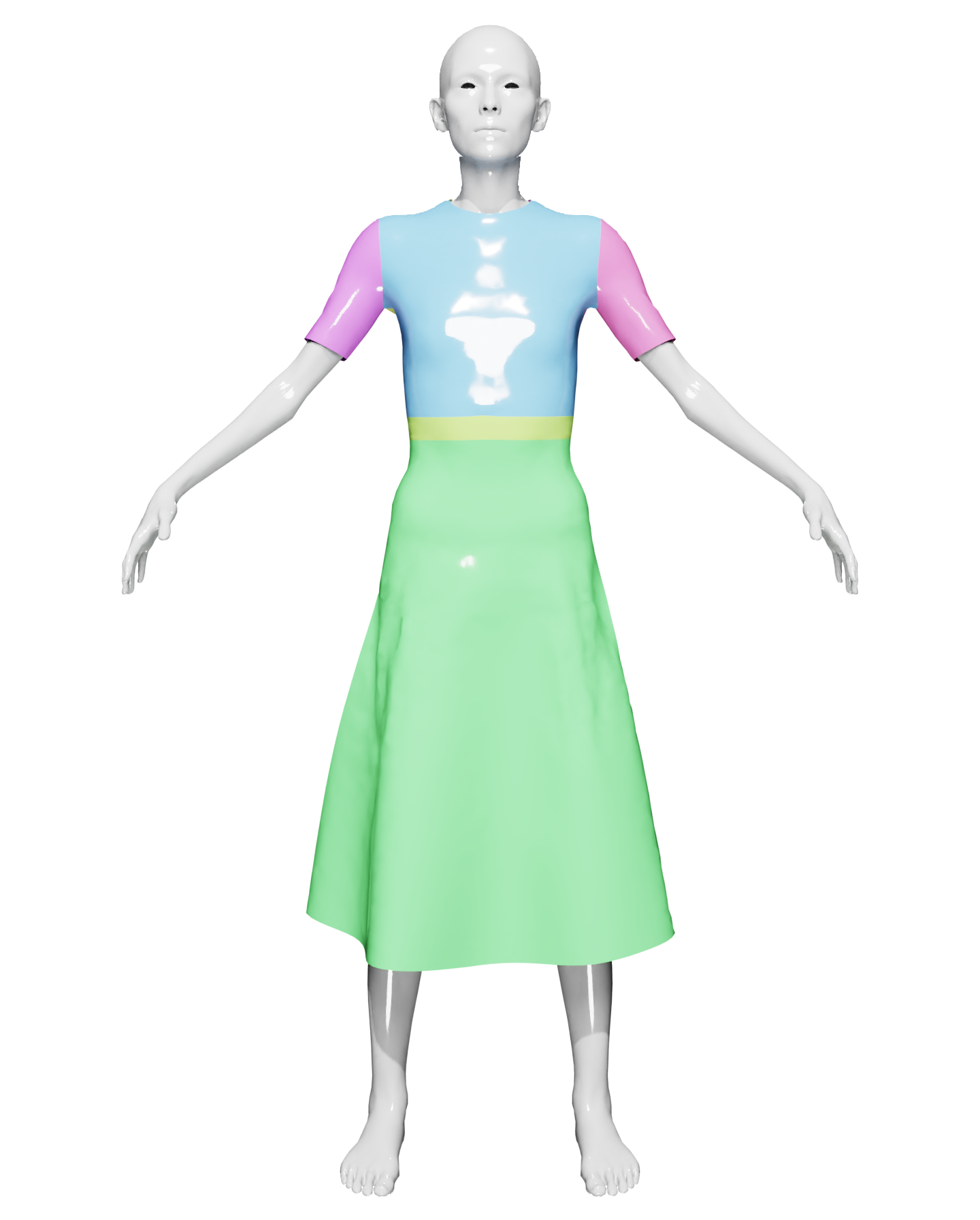}

    \\

    \includegraphics[width=\imagewidth, align=c]{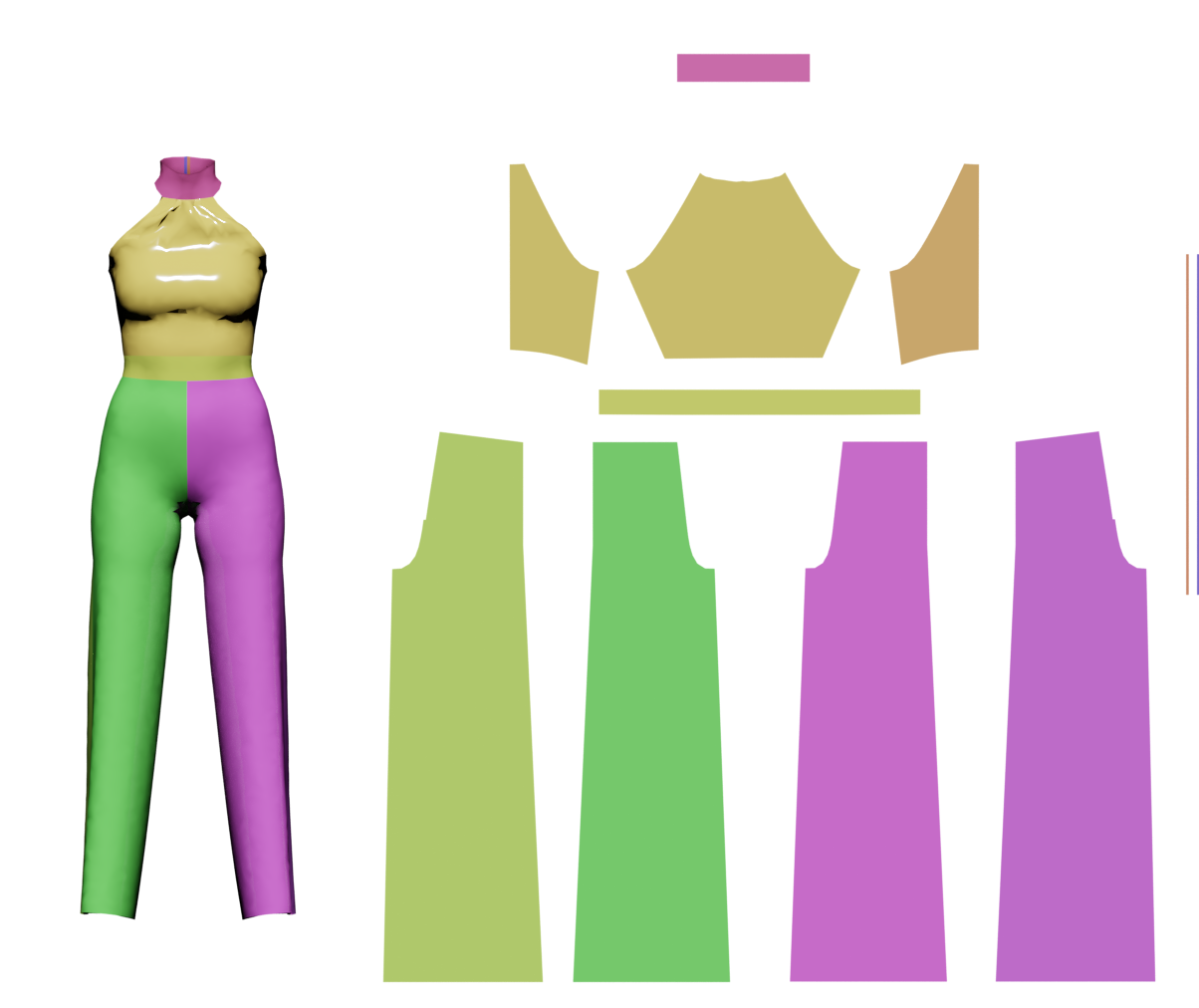}
     &
    \includegraphics[width=\imagewidth, align=c]{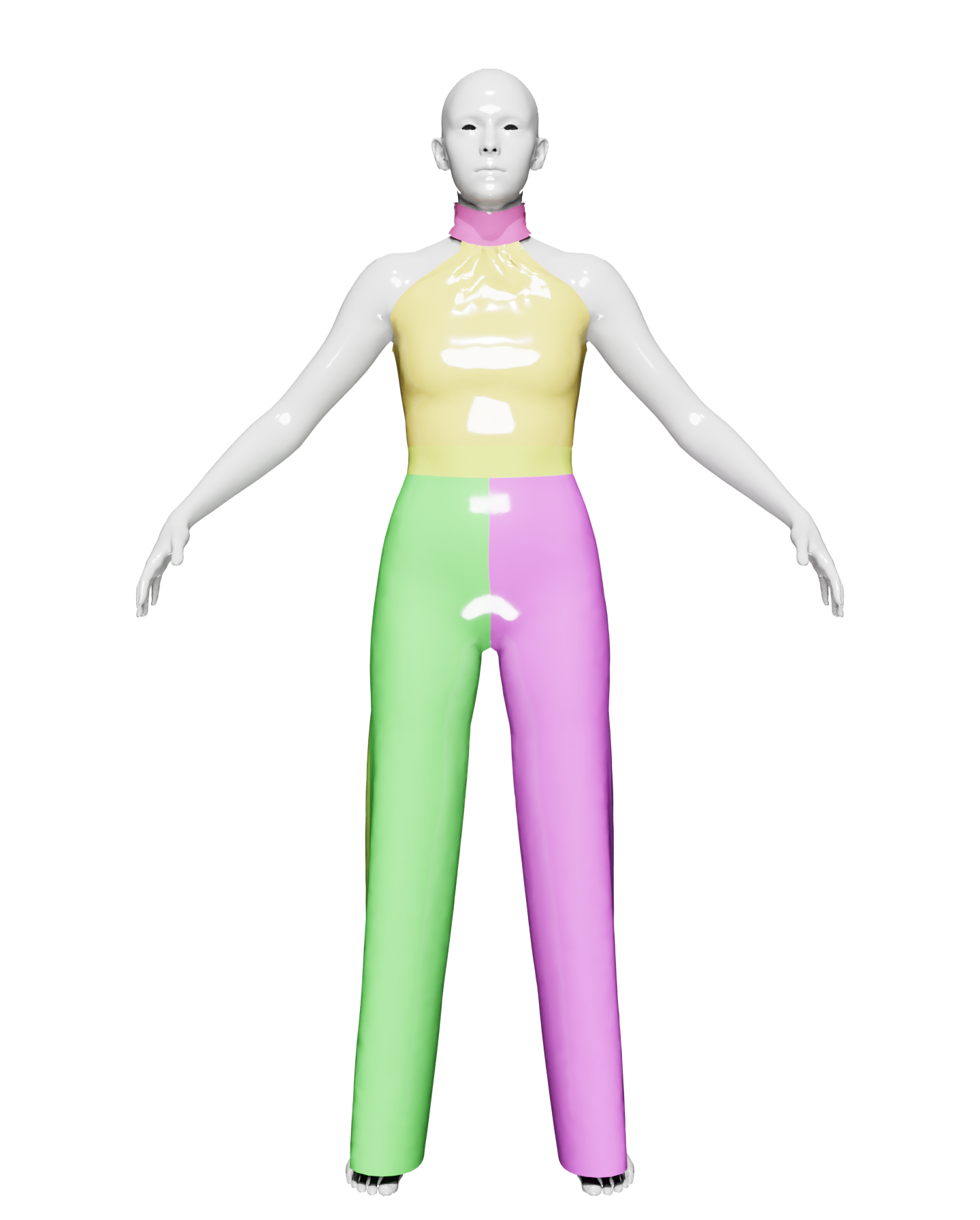}
     &
    \includegraphics[width=\imagewidth, align=c]{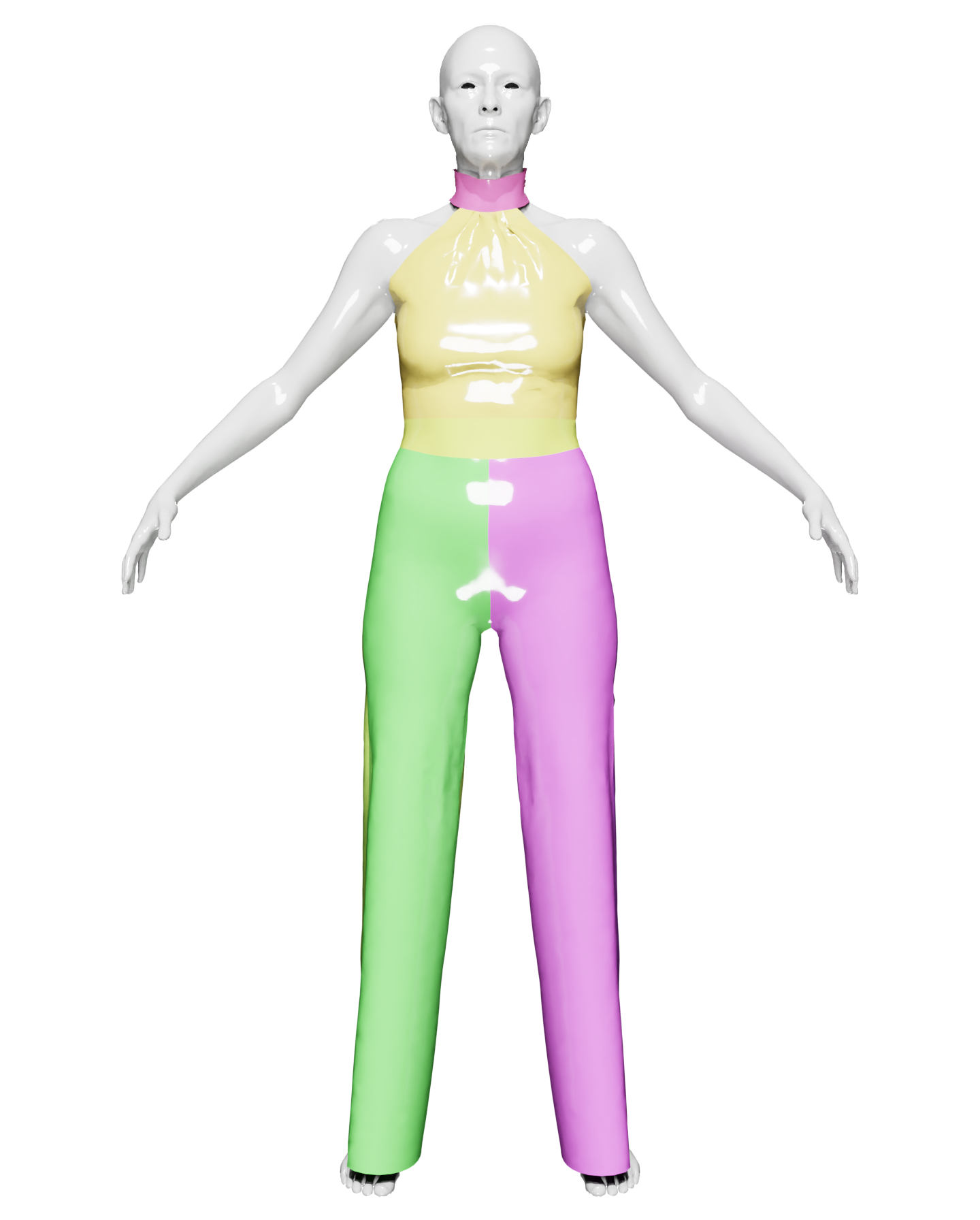}
     &
    \includegraphics[width=\imagewidth, align=c]{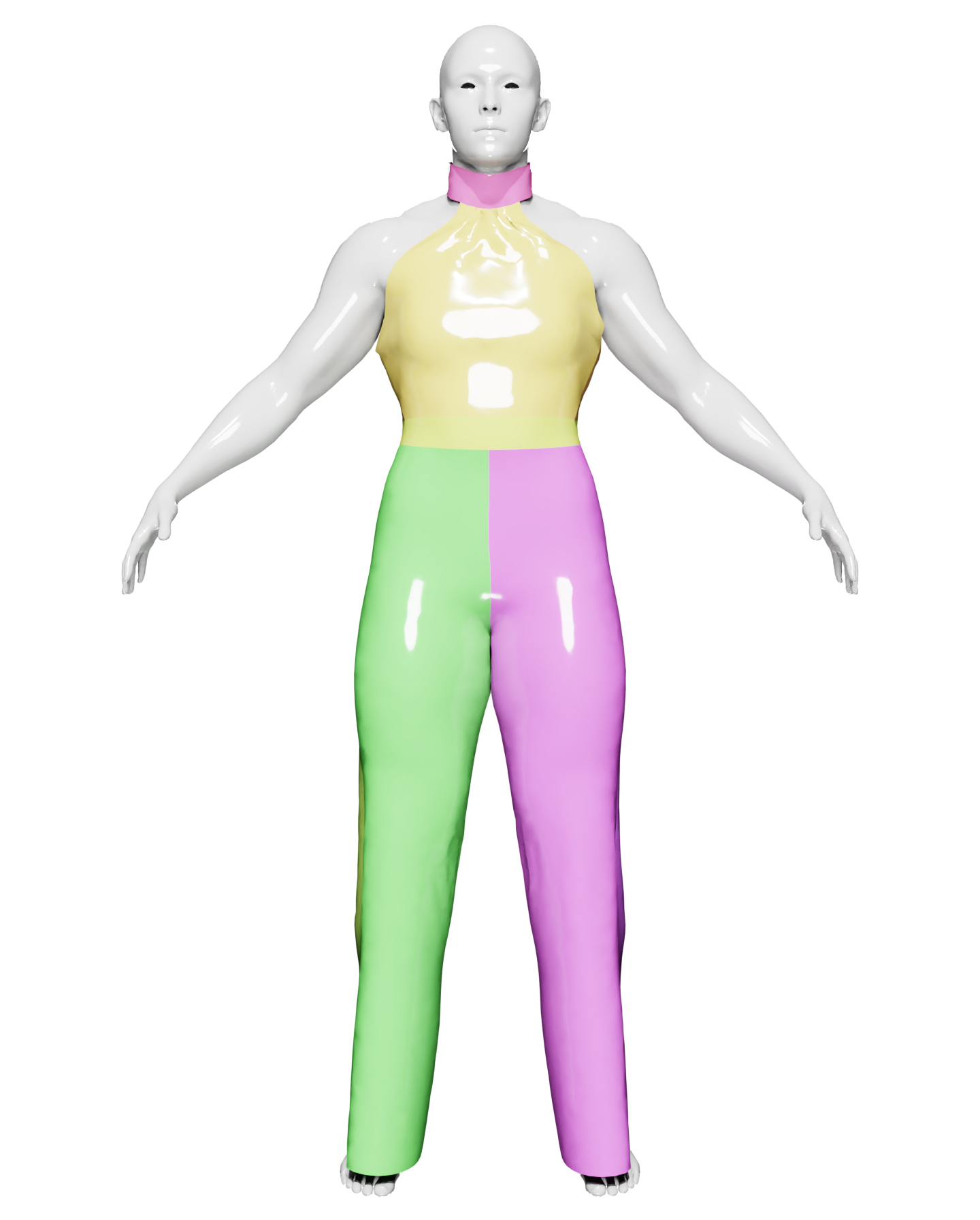}
     &
    \includegraphics[width=\imagewidth, align=c]{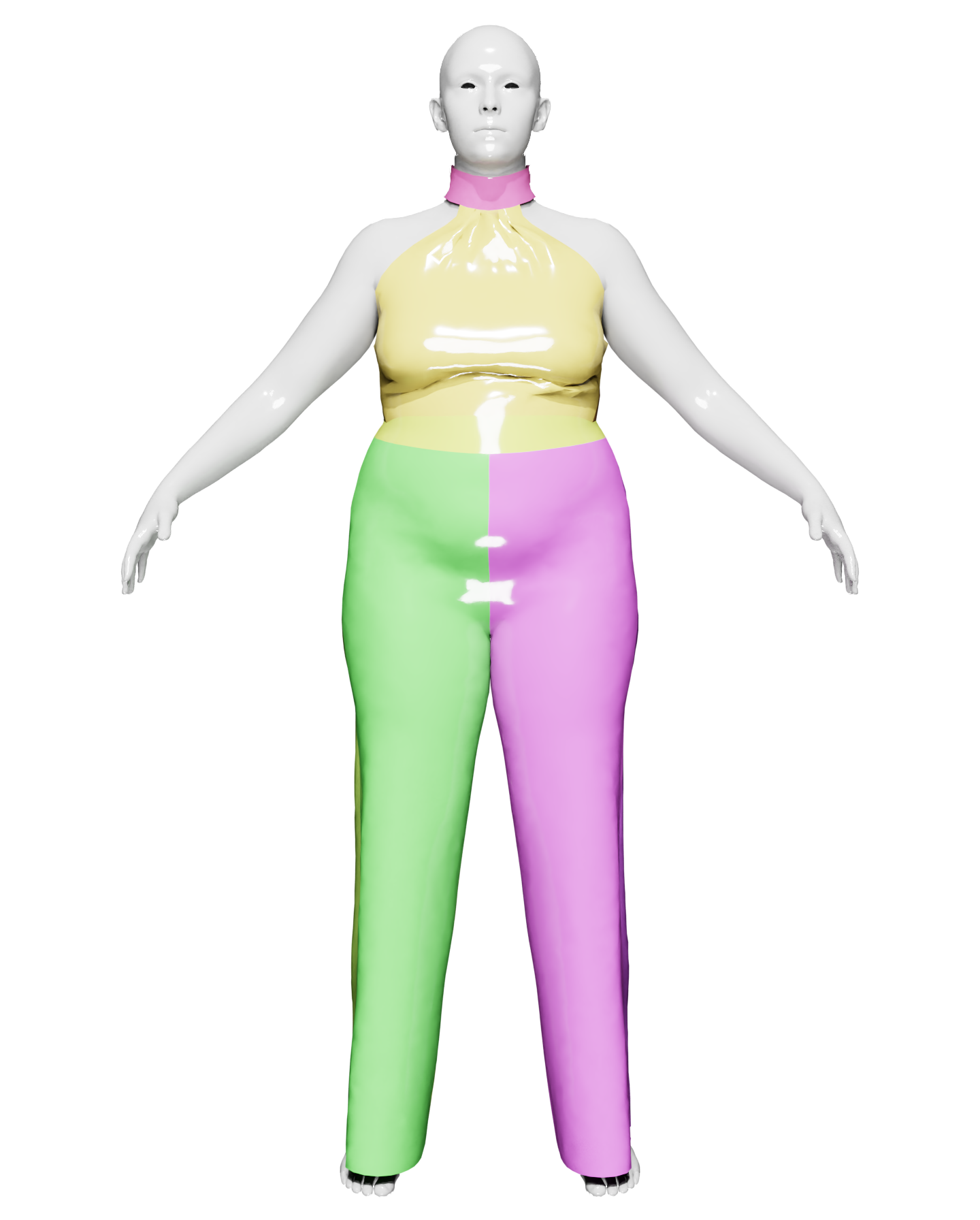}
     &
    \includegraphics[width=\imagewidth, align=c]{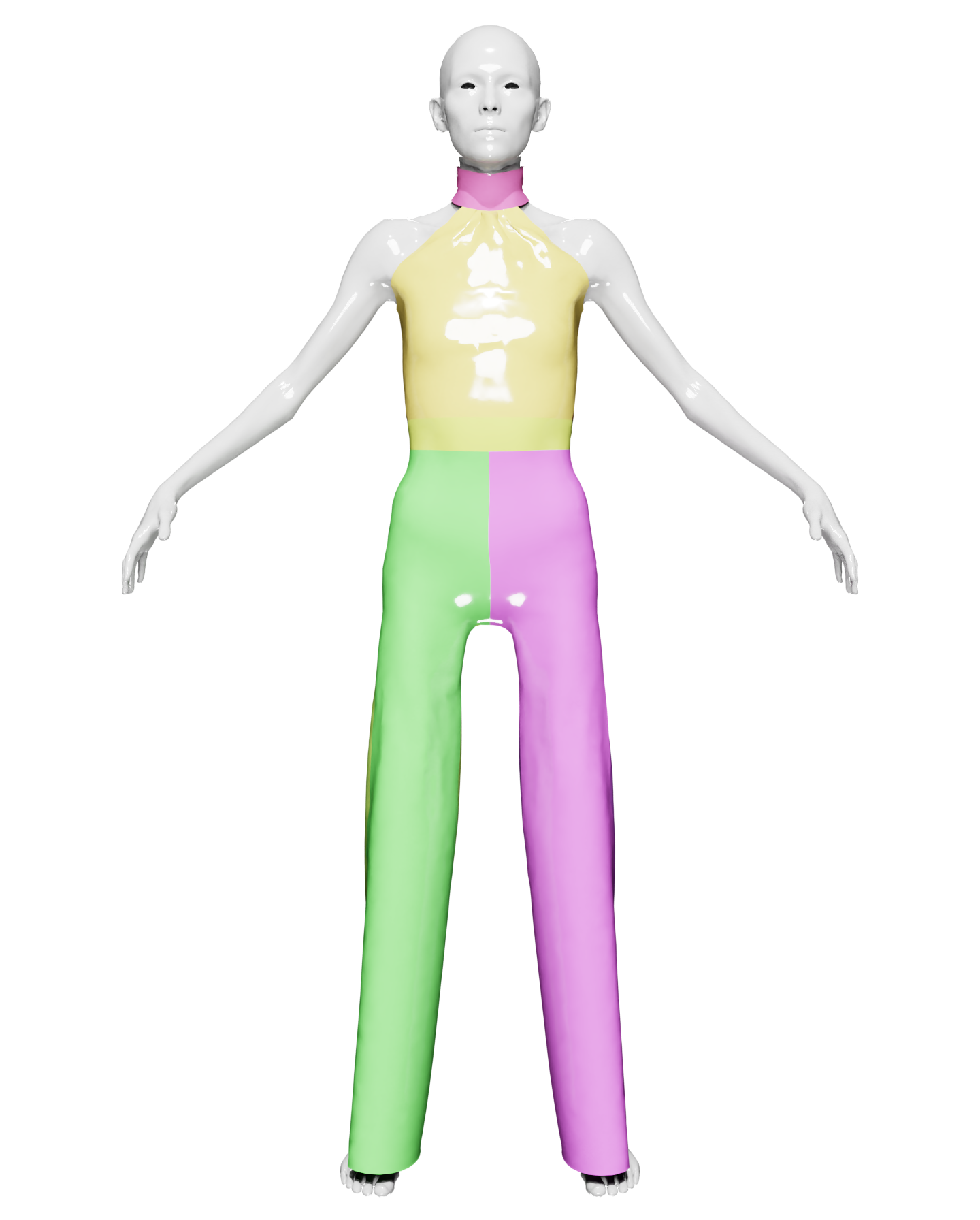}

    \\

    \includegraphics[width=\imagewidth, align=c]{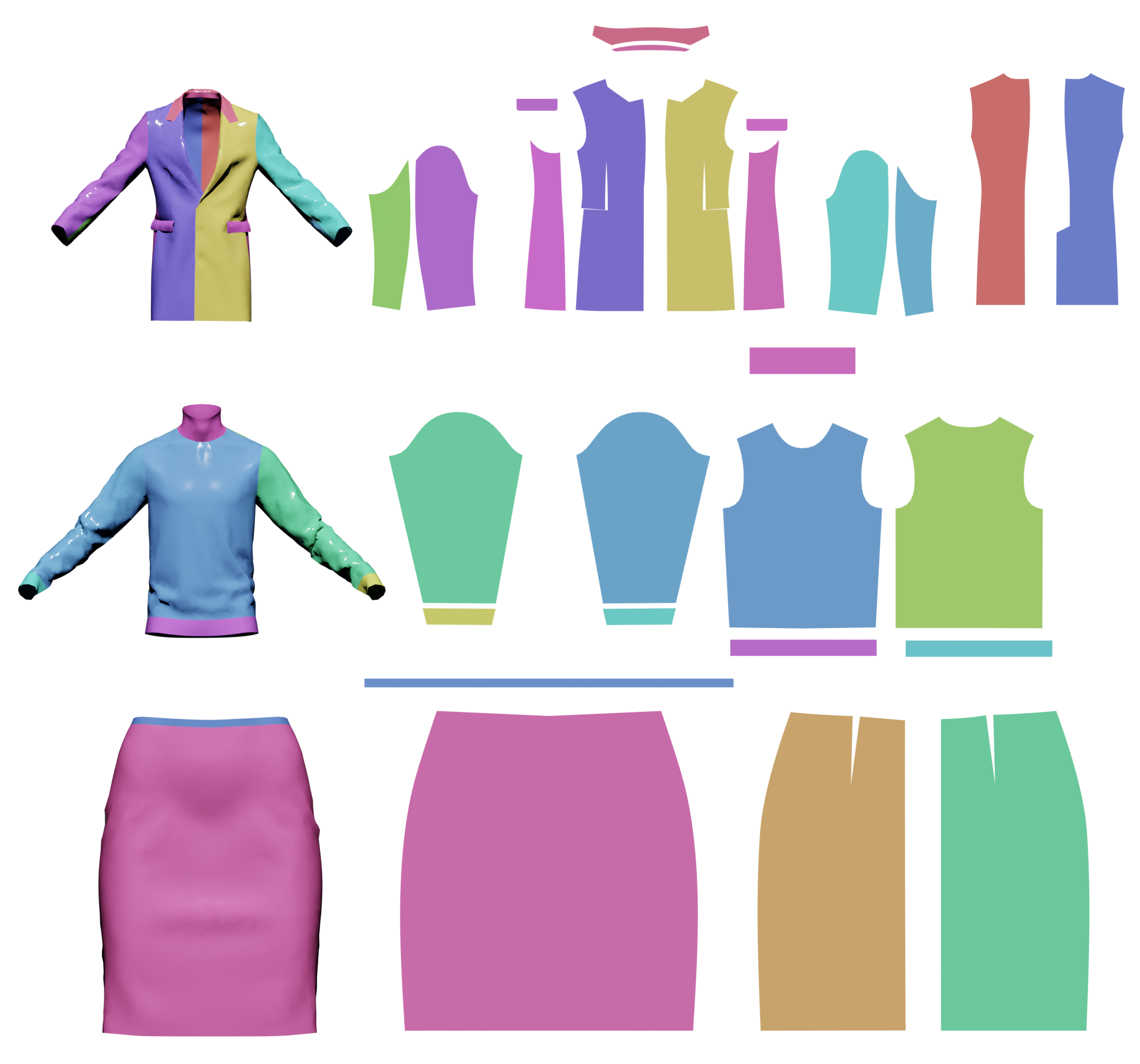}
     &
    \includegraphics[width=\imagewidth, align=c]{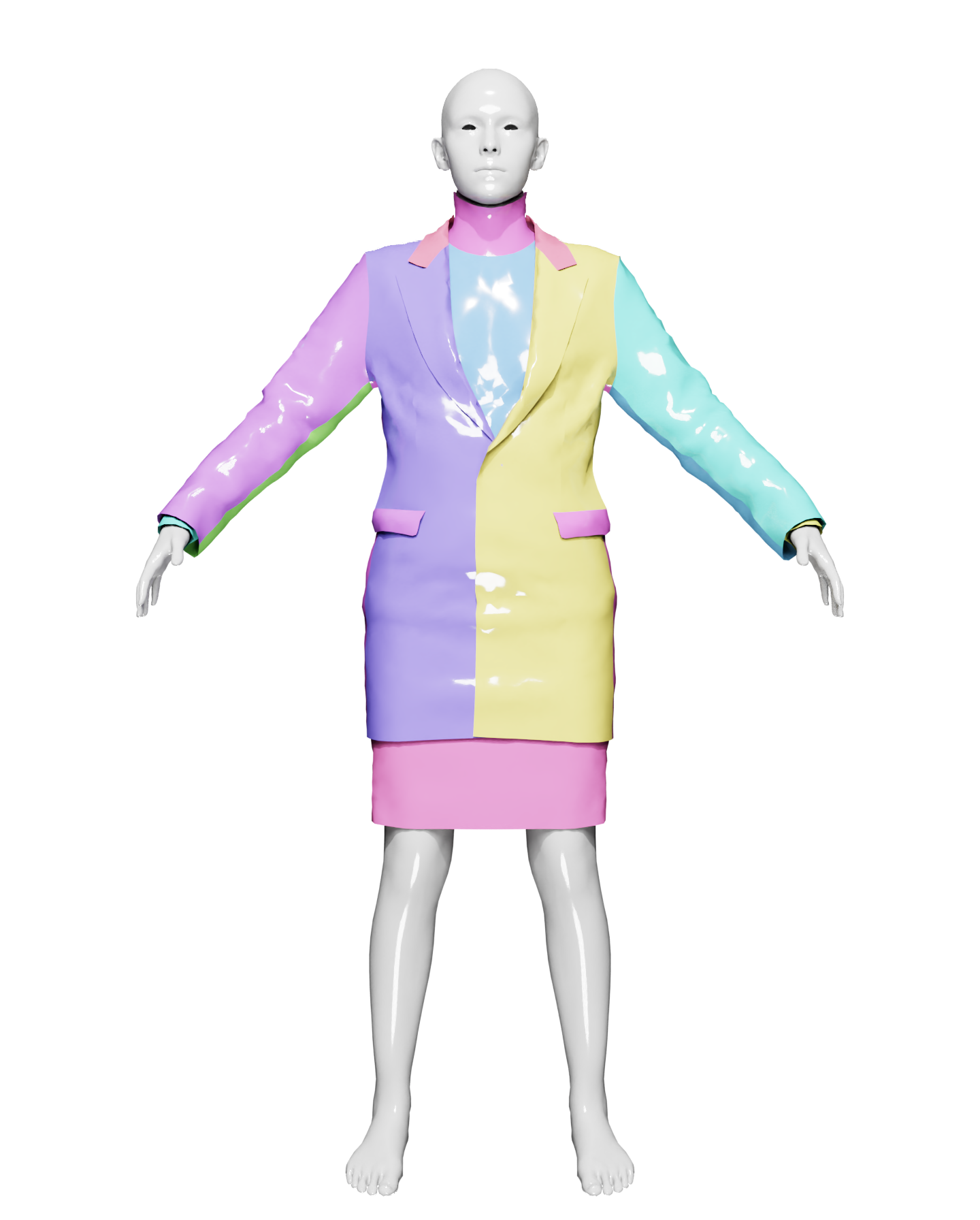}
     &
    \includegraphics[width=\imagewidth, align=c]{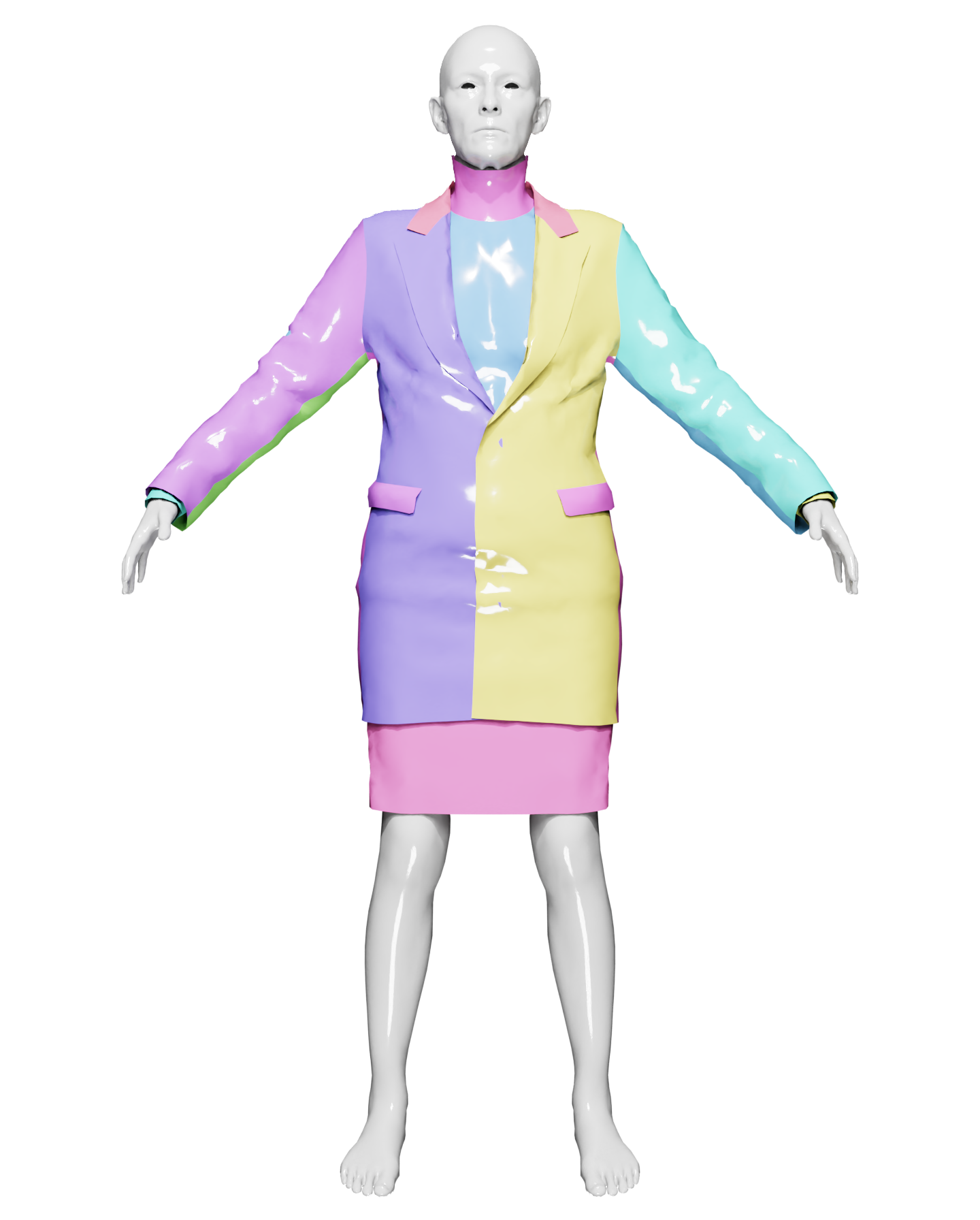}
     &
    \includegraphics[width=\imagewidth, align=c]{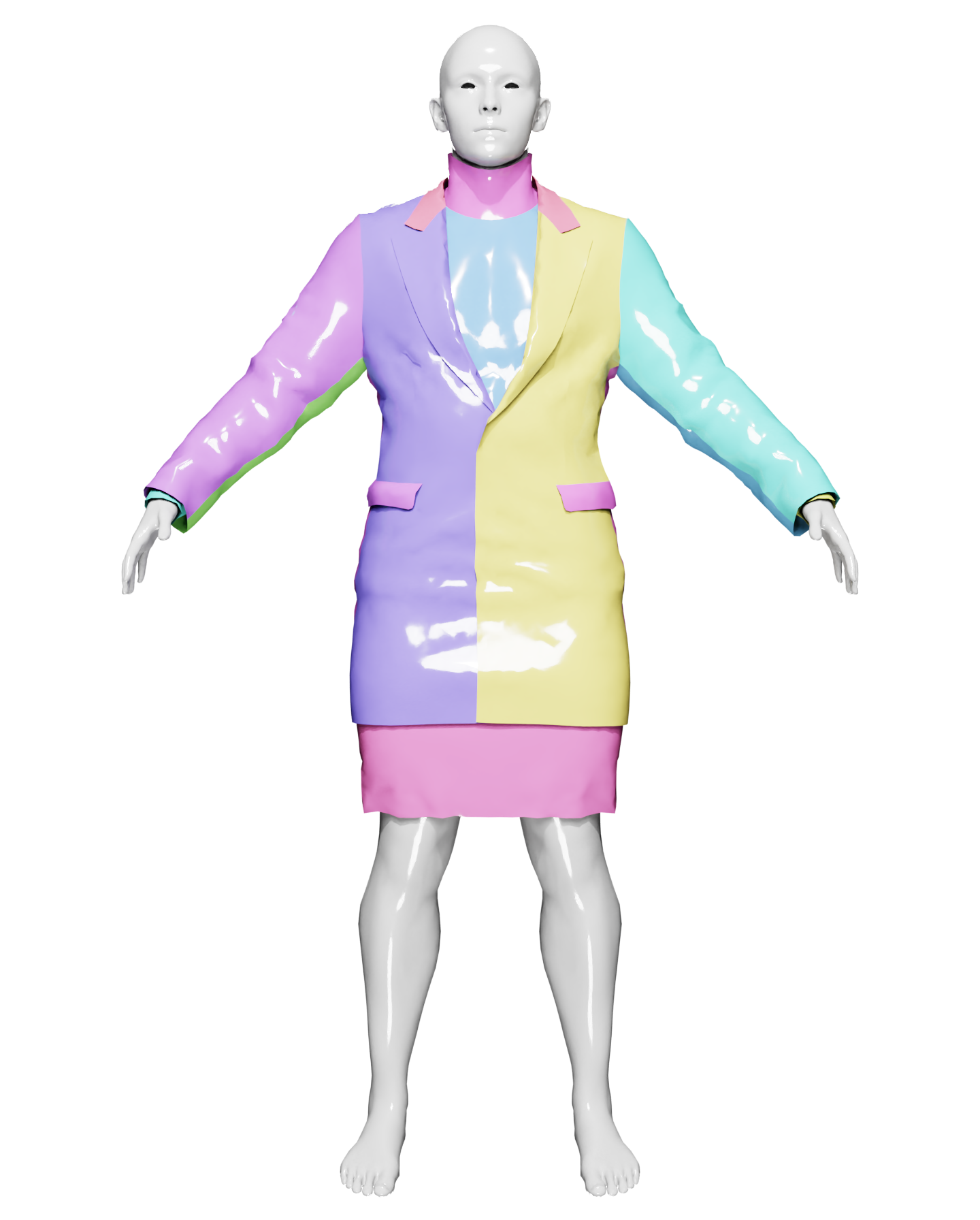}
     &
    \includegraphics[width=\imagewidth, align=c]{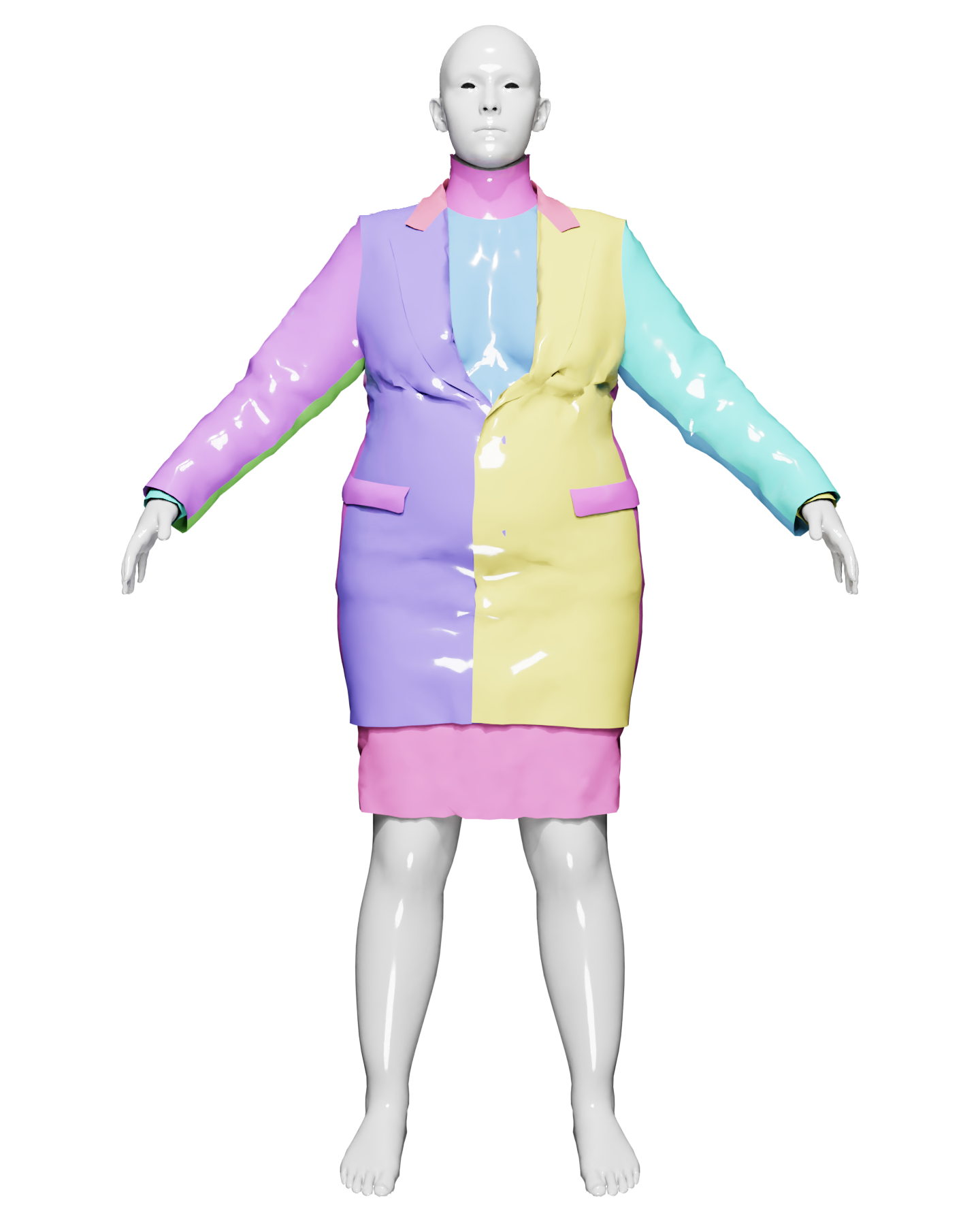}
     &
    \includegraphics[width=\imagewidth, align=c]{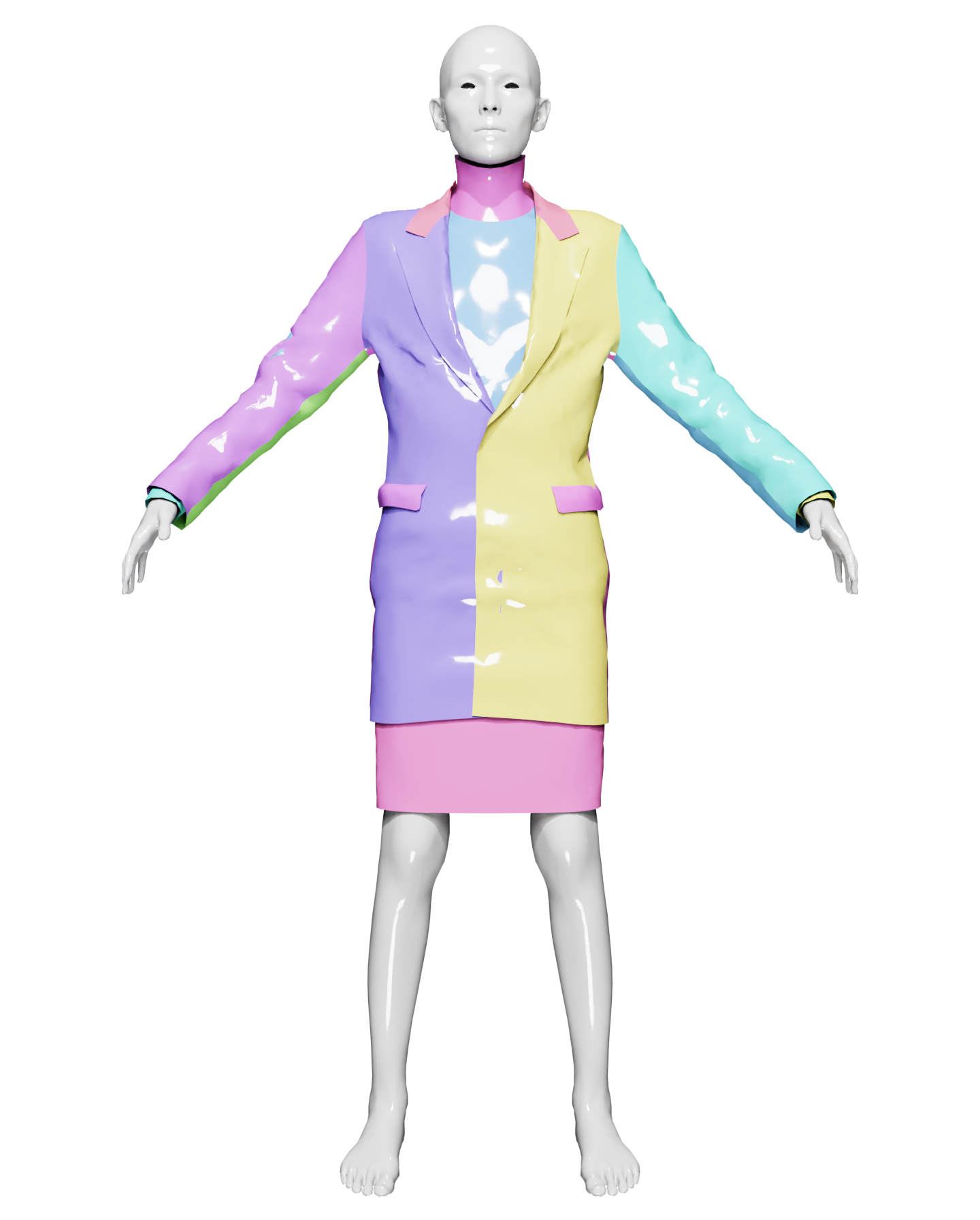}

    \\

    \includegraphics[width=\imagewidth, align=c]{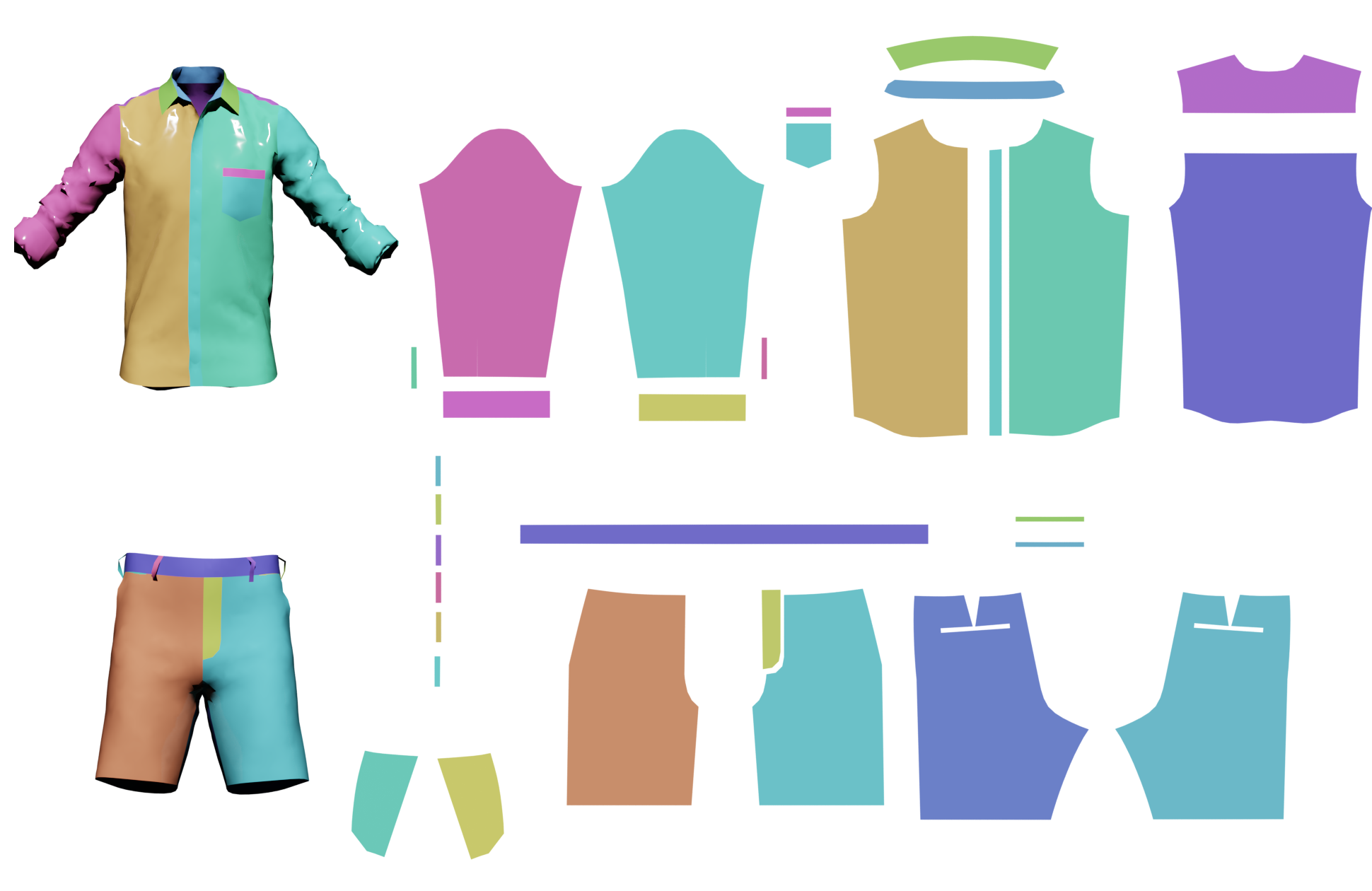}
     &
    \includegraphics[width=\imagewidth, align=c]{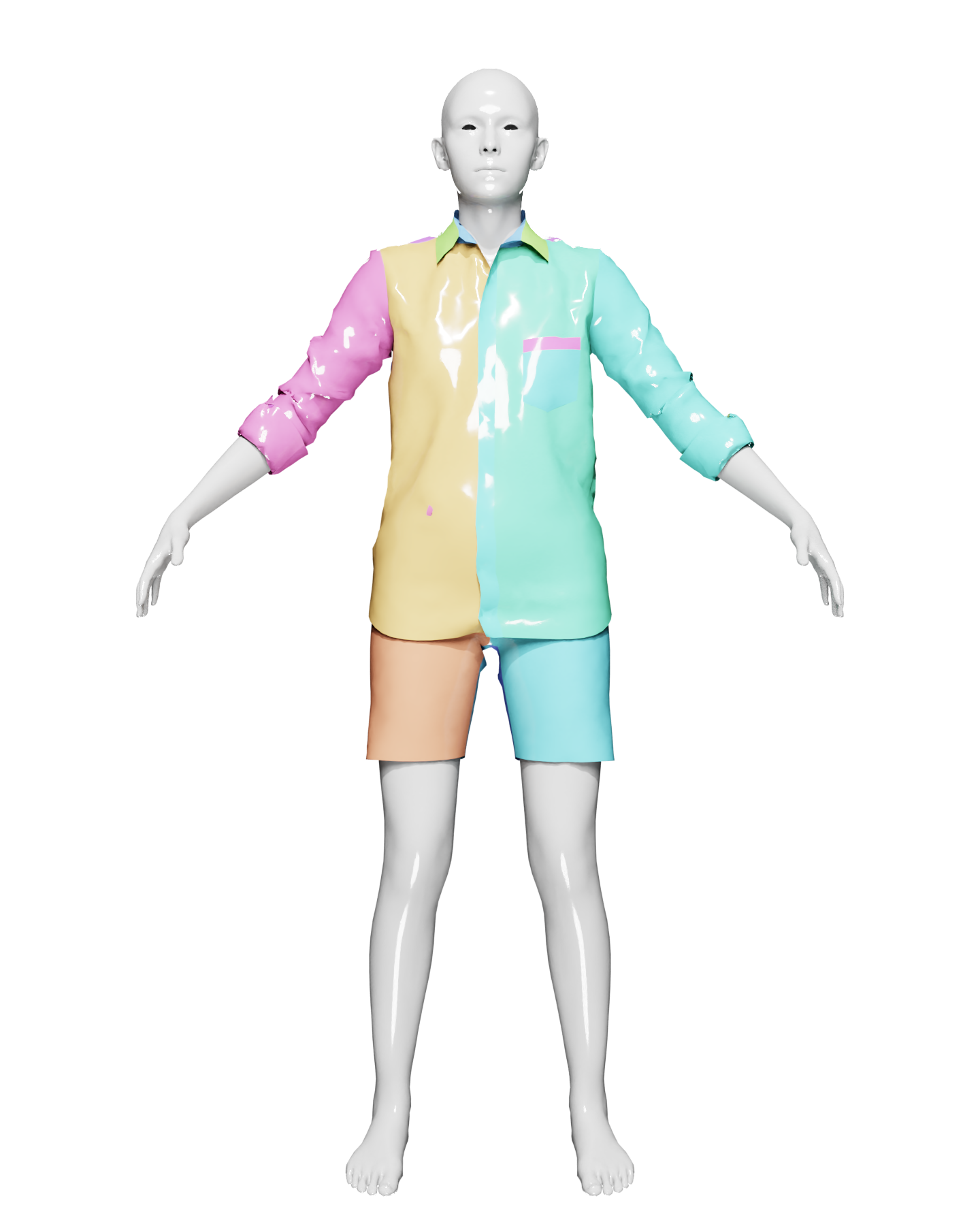}
     &
    \includegraphics[width=\imagewidth, align=c]{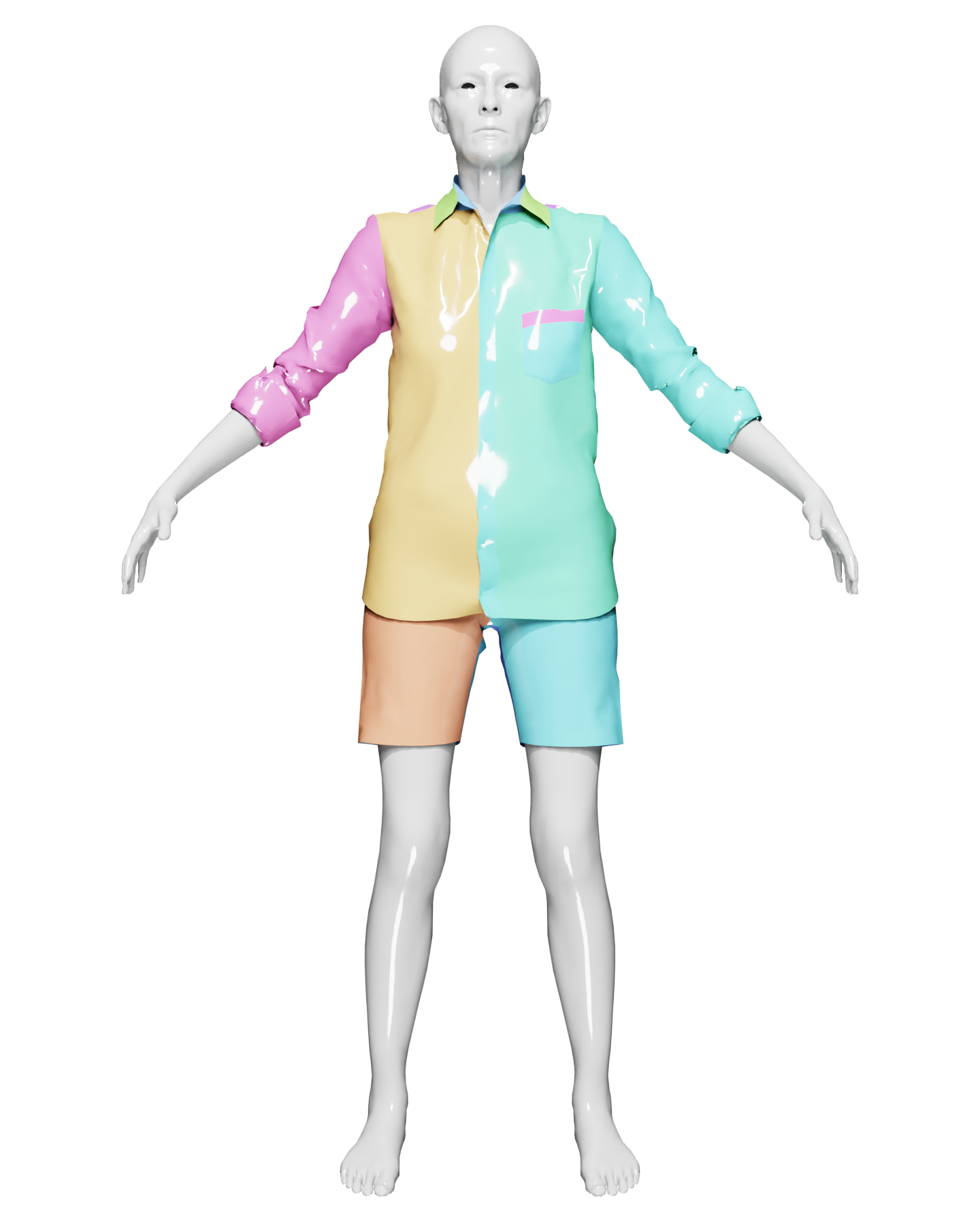}
     &
    \includegraphics[width=\imagewidth, align=c]{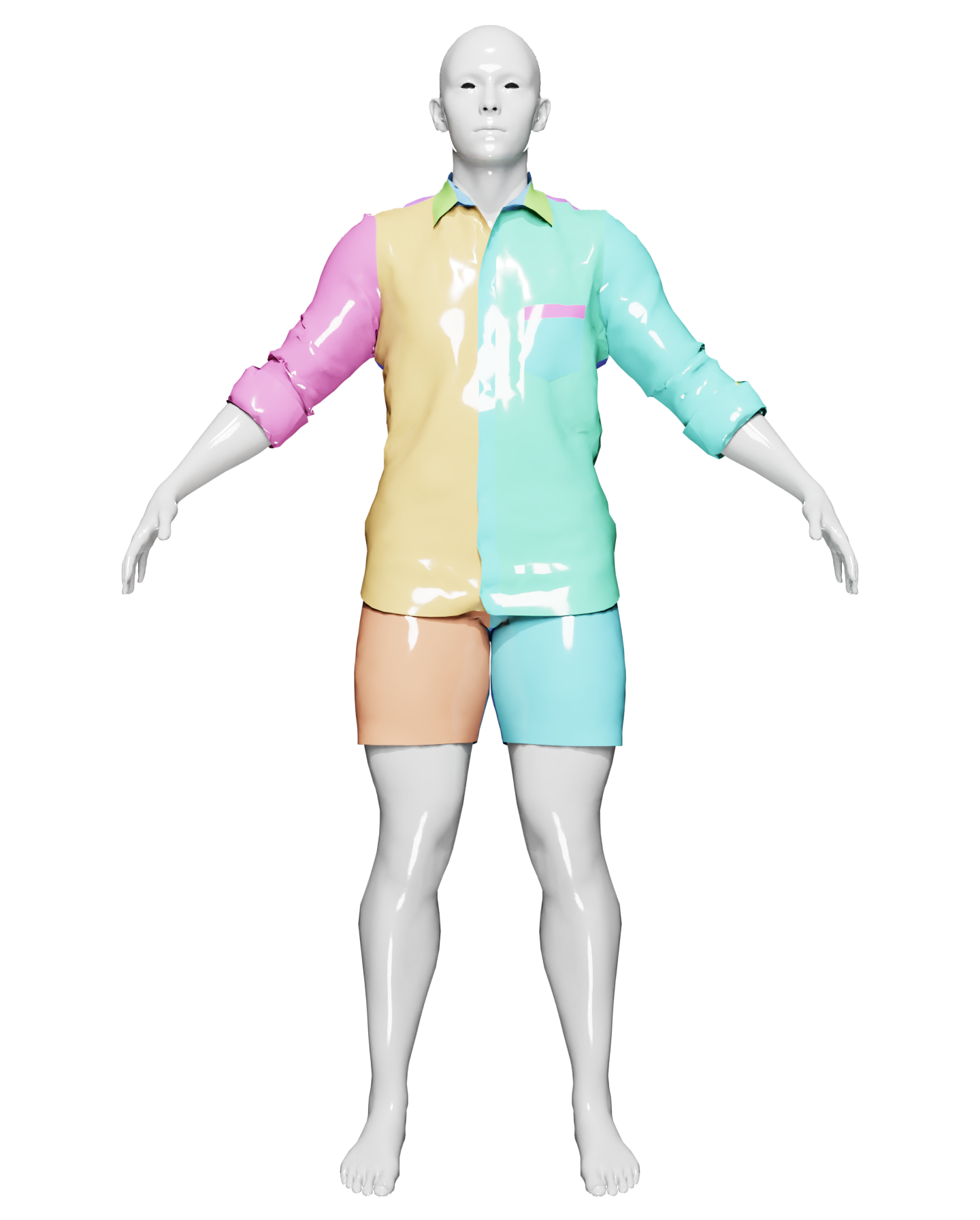}
     &
    \includegraphics[width=\imagewidth, align=c]{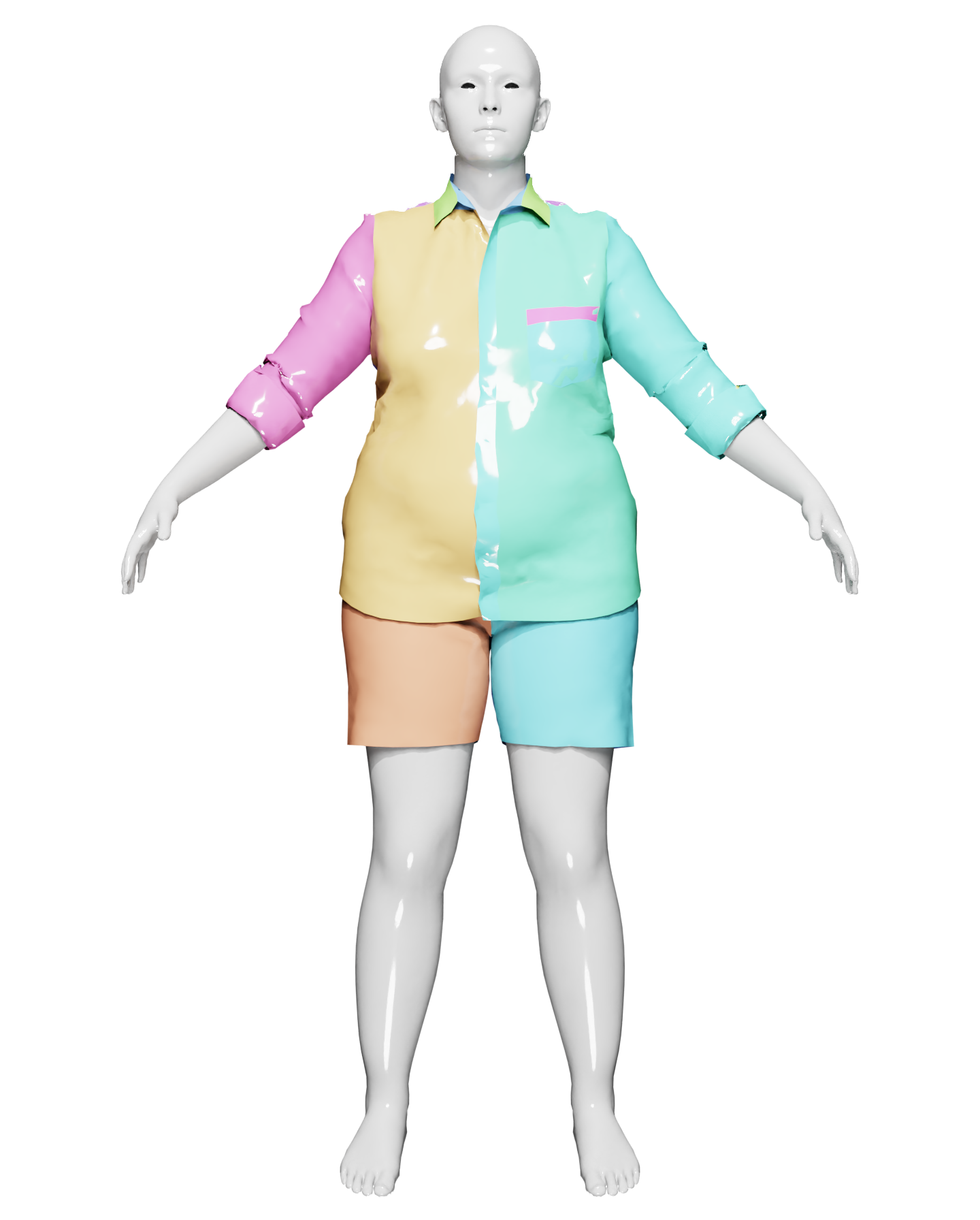}
     &
    \includegraphics[width=\imagewidth, align=c]{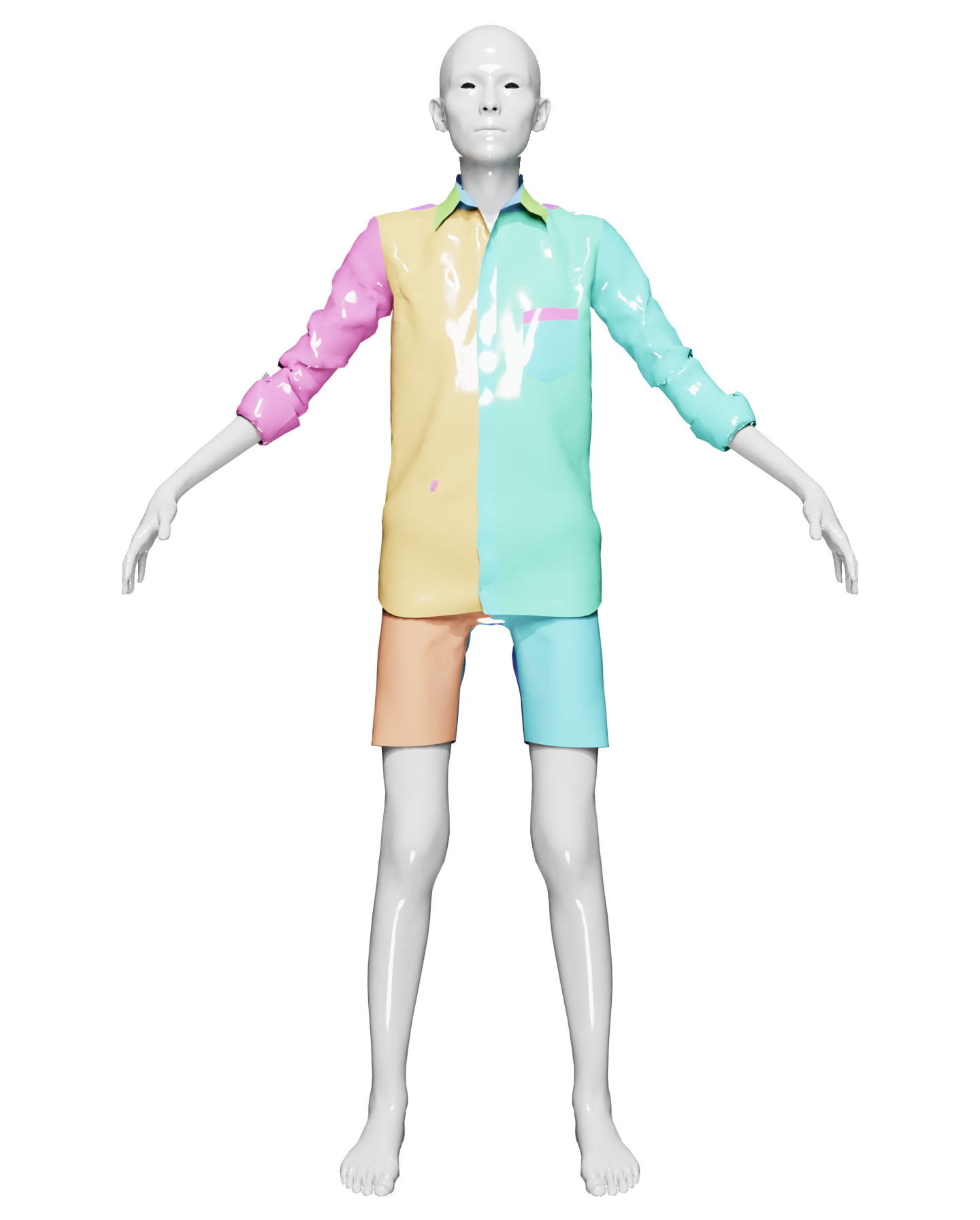}

    \\

  \end{tblr}
  \caption{A variety of outfits composed of different garments are refitted and draped onto a range of female characters using Bolt.}
  \label{tab:female_outfits_grid}
\end{table*}

\end{document}